\documentclass[11pt,a4paper]{article}
\pdfoutput=1
\usepackage{jheppub}
\usepackage{array,hhline,dcolumn}
\usepackage{color}
\usepackage{array}
\usepackage{amsmath,amssymb,ulem,slashed,wasysym}
\usepackage{subfigure,epsfig,pstricks,hyperref}
\usepackage[onehalfspacing]{setspace}

\usepackage{ifthen} 
\newboolean{pdflatex}
\setboolean{pdflatex}{false} 

\newboolean{articletitles}
\setboolean{articletitles}{true} 

\newboolean{uprightparticles}
\setboolean{uprightparticles}{false} 

\newboolean{inbibliography}
\setboolean{inbibliography}{false} 
\def \beq{\begin{equation}}
\def \eeq{\end{equation}}

\def \beqa{\begin{eqnarray}}
\def \eeqa{\end{eqnarray}}

\title{Feasibility of top quark measurements at LHCb and constraints on the large-$x$ gluon PDF}
\author[1]{Rhorry Gauld}
\affiliation[1]{Department of Physics, University of Oxford, Oxford, OX1 3RH}

\emailAdd{r.gauld1@physics.ox.ac.uk}

\abstract{The forward LHCb acceptance opens interesting possibilities of studying precision Standard Model hard processes in a kinematical region beyond the reach of ATLAS and CMS. In this paper we perform a feasibility study for cross-section measurements of top quark pairs with the LHCb detector, with an analysis of signal and background rates for selected final states, and determine the potential precision achievable at $\sqrt{s} =$ 7 and 14~TeV. We then study the dependence of theoretical uncertainties on the pseudorapidity distribution of top quarks produced in pair production at NLO, and observe that a cross-section measurement at high pseudorapidity has enhanced sensitivity to probe the high-$x$ gluon PDF as compared to measurements in the central-region. Based on simulated pseudodata, the impact of a 14~TeV cross-section measurement on the gluon PDF and charge asymmetry is quantified.}

\keywords{Top physics, QCD Phenomenology, Forward physics}
\arxivnumber{1310.1810}

\usepackage{mciteplus}
\begin{document}
\maketitle

\section{Introduction}
Top quark measurements at LHCb were initially proposed in Ref.~\cite{Kagan:2011yx}. It was demonstrated that $t\bar{t}$ production can be probed at high pseudorapidity by partially reconstructing ($\mu^{\pm}b$-jet) the total system. It was also proposed that this method of partial reconstruction can be used to measure the pair production charge asymmetry (charge asymmetry) by comparing the rate of top ($\mu^+$-tagged) to anti-top ($\mu^-$-tagged) events as a function of lepton pseudorapidity within the LHCb acceptance. The main motivation being that at a proton-proton collider, the rapidity of the heavy quarks written in terms of the incoming partons at leading order (LO) is,
\beqa
x_{1,(2)} = \frac{m_T}{\sqrt{\hat{s}}}( e^{(-)y_3} + e^{(-)y_4} ) \\
\mathrm{where } \hspace{0.2cm} m_T= \sqrt{(m^2 + p_T^2)} \hspace{0.2cm} \mathrm{,} \hspace{0.2cm} \hat{s} = 2m^2_T(1 + cosh\Delta y).\nonumber
\eeqa

\begin{figure}[h]
\centering
\begin{tabular}{cc}
\includegraphics[width=.48\linewidth]{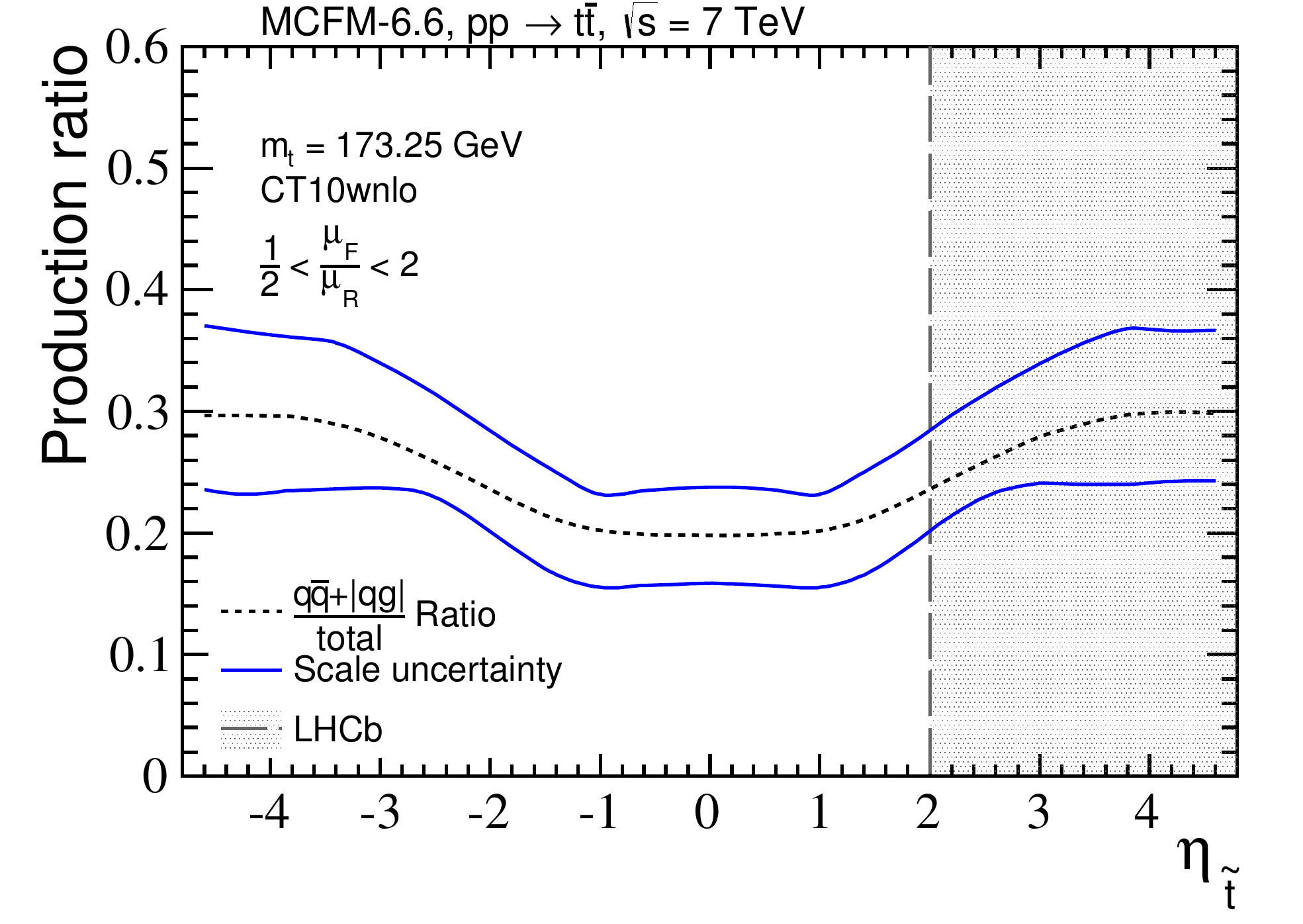} &
\includegraphics[width=.48\linewidth]{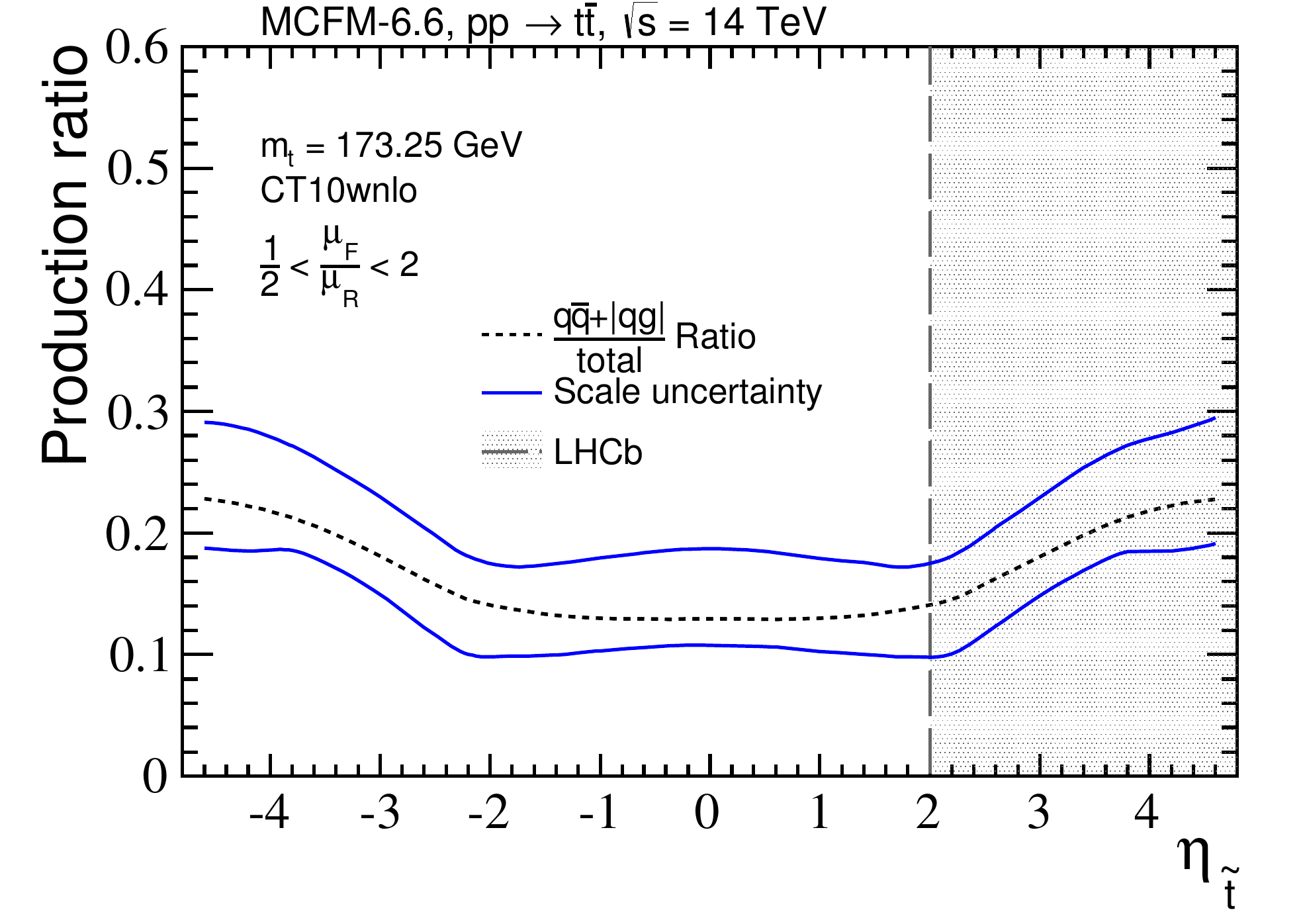}
\end{tabular}
\caption{Ratio of production mechanisms of pseudotop as a function of pseudorapidity at 7 (left) and 14 TeV (right). The blue band corresponds to the uncertainty associated to scale variation.}
\label{prodratio}
\end{figure}

\noindent Meaning that, at momentum exchange scales required to produce pairs of top quarks, top quarks produced in the forward region have a high probability of having come from a high-$x_1$ incoming parton where the ratio of quark to gluon parton distribution functions (PDF) is larger - a consequence of the valence content of the proton. This results in less dilution to the charge asymmetry which arises from the colour structure of interfering diagrams, $qX\rightarrow t\bar{t}Y$, with quarks in the initial state~\cite{Kuhn:1998kw}. However, this also means that forwardly produced top quarks from $gg$-scattering processes are produced from incoming gluons at high-$x_1$. This is presented in Fig.~\ref{prodratio} where the ratio of production mechanisms, ($q\bar{q}+|qg|$)/total, contributing to $t\bar{t}$ production is presented as a function of the arithmetic mean of pseudorapidity distributions of $t$ and $\bar{t}$ (pseudotop - $\tilde{t}$) for 7 (left) and 14 TeV (right) centre of mass energies. Note that the contribution from $gg$-scattering is dominant across the entire range of phase space for both centre of mass energies.

There have been large efforts in the QCD community to improve the precision of top quark pair production predictions. In particular, the completion of the full next-to-next-to-leading order (NNLO) calculation~\cite{Czakon:2013goa,Baernreuther:2012ws,Czakon:2012zr,Czakon:2012pz} as well as resummation of soft gluon emissions to next-to-next-to-leading log (NNLL) accuracy~\cite{Czakon:2009zw,Beneke:2009rj,Cacciari:2011hy}. The reduced scale uncertainty in these predictions is crucial to gaining information on other sources of theoretical uncertainty such as the high-$x$ gluon PDF, $\alpha_s$ and the top mass. A recent study of the impact of these uncertainties on the inclusive cross-section at NNLO+NNLL accuracy can be found in Ref.~\cite{Czakon:2013tha}, where it is observed that such a measurement, with minimal scale uncertainties, has the potential to strongly constrain the gluon PDF. It is clear that a differential result to the same accuracy is highly desirable and will be available in the not-too-distant future. In fact, differential cross-section results and studies using approximate NNLO calculations and resummation techniques have already been obtained in~\cite{Kidonakis:2010dk,Ahrens:2011mw,Kidonakis:2011ca,Beneke:2012wb,Guzzi:2013noa}. To this end, we demonstrate the increased sensitivity of pair production cross-section measurements at high rapidity to the gluon PDF at NLO accuracy.

\section{LHCb analysis at 7, 14~TeV}\label{LHCbanalysis}
This section aims to provide an estimate of the potential statistical precision of a cross-section measurement achievable with the current 7~TeV data ($\int\mathcal{L}dt =$ 1fb$^{-1}$) as well as the projected 14~TeV data sample after 1 year of running ($\int\mathcal{L}dt =$ 5fb$^{-1}$). As pointed out in Ref.~\cite{Kagan:2011yx}, top quarks can be identified through their decay $t \rightarrow (W \rightarrow \mu \nu_\mu) b$, where the muon and the $b$ are registered by the detector. Indeed, in the full $t\bar{t}$ decay it is also possible to reconstruct a $b$,$\mu$ along with $W$ decay products, radiated jets (which tend to be forward) or $b$ quarks which do not come from the same parent top - as demonstrated in Ref.~\cite{Gauld:1557385}. In the following analysis we will consider both $\mu b$  and $\mu bj$ final states. Using multiple final states, requiring a different number of $b$-tags, is a crucial cross check of the background modelling, in particular the $W+$($b$)jets processes. Given that top pairs are produced asymmetrically beyond LO, we introduce the `pseudotop' object where;
\beq
\frac{d \sigma^{\tilde{t}} }{d X} = \frac{1}{2}\left(\frac{d \sigma^t}{d X} + \frac{d \sigma^{\bar{t}}}{d X}\right).
\eeq
\noindent Thus, the $\mu b$  and $\mu bj$ final states are labelled as $\tilde{t}_{\mu b}$, $\tilde{t}_{\mu bj}$. Introducing this definition removes the small bias introduced from the charge asymmetry. Given that the asymmetry in the backgrounds is driven by the quark valence content, which is well constrained by DIS data, the main uncertainty on backgrounds arises from total normalisation and so is not affected by this definition.

\begin{description}
	\item[Signal and background]
\end{description}
The $t\bar{t}$ signal and backgrounds are simulated using \texttt{POWHEG}~\cite{Nason:2004rx,Frixione:2007vw,Alioli:2010xd,Frixione:2007nw,Alioli:2010qp,Alioli:2009je,Frederix:2012dh,Oleari:2011ey} with the central CT10wnlo~\cite{Lai:2010vv} PDF set and then matched to \texttt{Pythia8176}~\cite{Sjostrand:2007gs}, with the exception of $Z+b$jets where the matrix element is produced using \texttt{MadGraph5}~\cite{Alwall:2011uj} with CTEQ6l1~\cite{Pumplin:2002vw}. The $t\bar{t}$ signal is found by fixing the reference factorisation and renormalisation scales equal to the top mass ($m_t$ = 173.25~GeV). It is found that the difference between the parton level \texttt{POWHEG}$\rightarrow$\texttt{Pythia8176} and \texttt{MCFM}~\cite{Campbell:2010ff} pseudotop pseudorapidity distributions is negligible\footnote{Provided the switch \texttt{SpaceShower:phiIntAsym} is turned off during the shower. This switch introduces un-desired colour-reconnection effects which are already accounted for correctly in the NLO matrix element.}. The main backgrounds are identified as single top, $W+$($b$)jets and $Z+$($b$)jets. The QCD background originating from di-$b$jet production where a secondary muon passes isolation and kinematic cuts has previously been shown to be negligible~\cite{Kagan:2011yx}. Given a di-$b$jet background rejection of $\mathcal{O}$(10$^{-5}$), and that the relative increase in the ratio of $(pp\rightarrow t\bar{t})/(pp\rightarrow b\bar{b})$ from 7$\rightarrow$14~TeV is $\approx$ 3, this will also be ignored for the 14~TeV analysis. The $t$-channel single top process is modelled in both 4 and 5 flavour (ST, tch) schemes, the 4-flavour cross-section is normalised to that found in the case of 5 flavours and the average of these distributions is plotted with a systematic error associated to the envelope between the two descriptions\footnote{The top decay is not included in the matrix element in the 4-flavour scheme.}. There is a small combined (below 10$\%$ of $t\bar{t}$ signal) contribution from $tW$ and $s$-channel single top which is not included. The $Z+$($b$)jets background arises from leptonic decay of a $Z$ where only one of the leptons is detected in association with either a correctly identified (in $Zb/Zg\rightarrow b\bar{b}$) or mis-tagged $b$-jet. The $W+$($b$)jets background is separated into $W$jets, where jet is a light flavour (u,d,s,c,g) which is mis-tagged as a $b$-jet, and $W$$b$jets (where $g\rightarrow b\bar{b}$) with a correctly identified $b$-jet.

\begin{description}
	\item[Selection and reconstruction] 
\end{description}
Jet objects are defined to have a jet parameter R\footnote{R $= \sqrt{y^2 + \phi^2}$} of 0.5, $p_{T} > $15~GeV, and to be built with the anti-kt algorithm using \texttt{FastJet3}~\cite{Cacciari:2011ma} software. In this analysis, $b$-jets are defined to be jets which are matched to a parton $b$-quark from the hard process (within R). It is also required that charged leptons are isolated ($\Delta$ R ($\mu$, jet) $\geq$ R) which is necessary to suppress QCD background. It was found that reconstructed jets with an R parameter of 0.7 better match parton energy as well as increasing the $b$-matching efficiency. However, the combination of the lepton isolation requirement and background reduction favours an R parameter of 0.5. Kinematic cuts of $p_T >$ 60~GeV on the leading $b$-jet $p_T$ and a cut of $p_T >$ 20~GeV on the muon and sub-leading jet are imposed which dramatically reduces background whilst having a comparatively small effect on signal. Jets and muons are also required to be within the pseudorapidity range of $2 < \eta < 4.5$. An efficiency of 75$\%$ is applied to muons, which is estimated from the combined trigger efficiency of $\approx 75\%$ and identification/tracking $> 95\%$ for high $p_T$ ($p_T >$~20~GeV) muons from Ref.~\cite{Aaij:2012vn}. A $b$-tagging efficiency of 70$\%$ with a corresponding mis-tag rate for light jets of 1$\%$ is assumed. It is noted that a more detailed study could separate $c$-jet background processes and apply an appropriate charm mis-tag rate, this is left for the data analysis.

\begin{figure}
\centering
\begin{tabular}{cc}
\includegraphics[width=.48\linewidth]{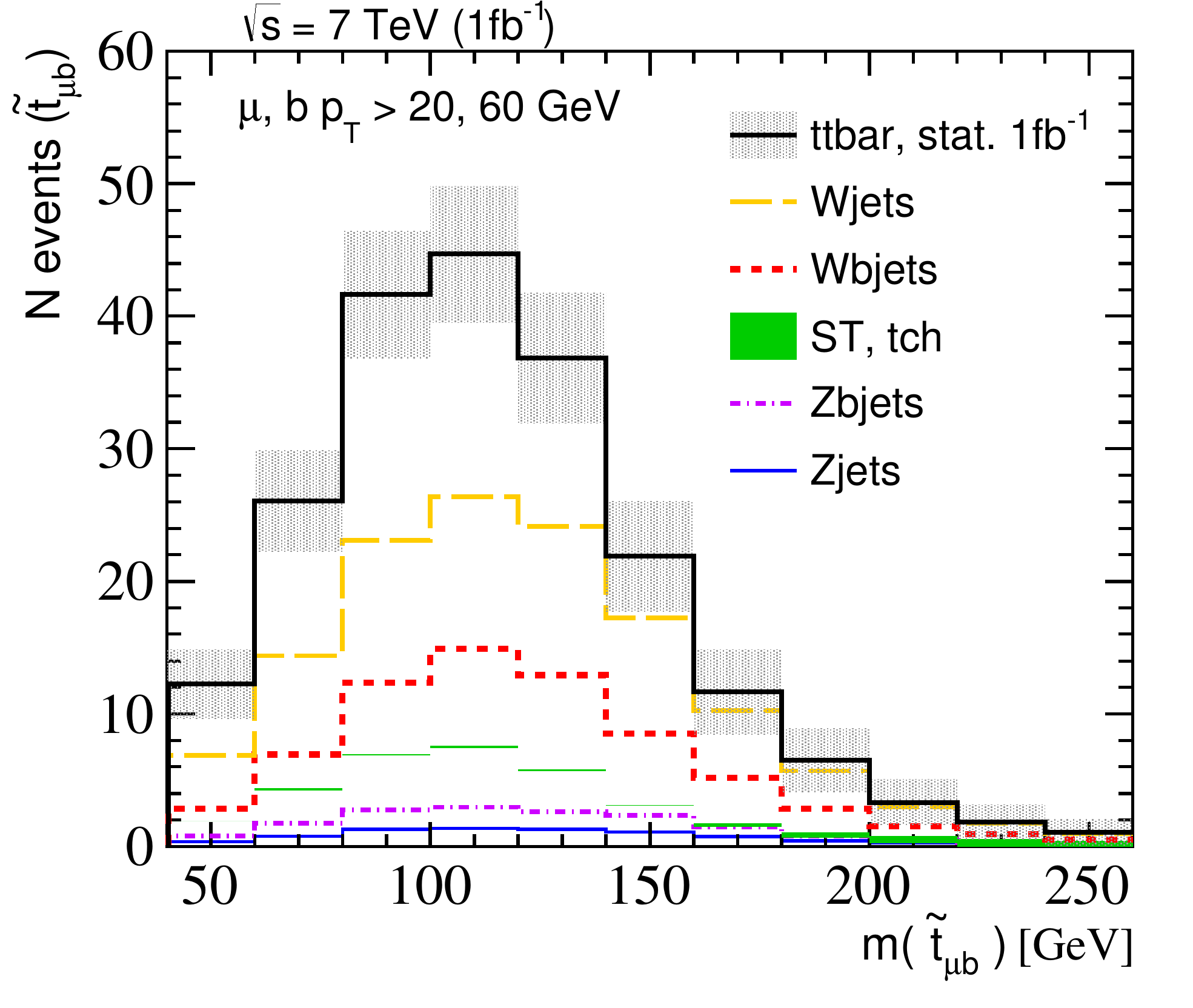} &
\includegraphics[width=.48\linewidth]{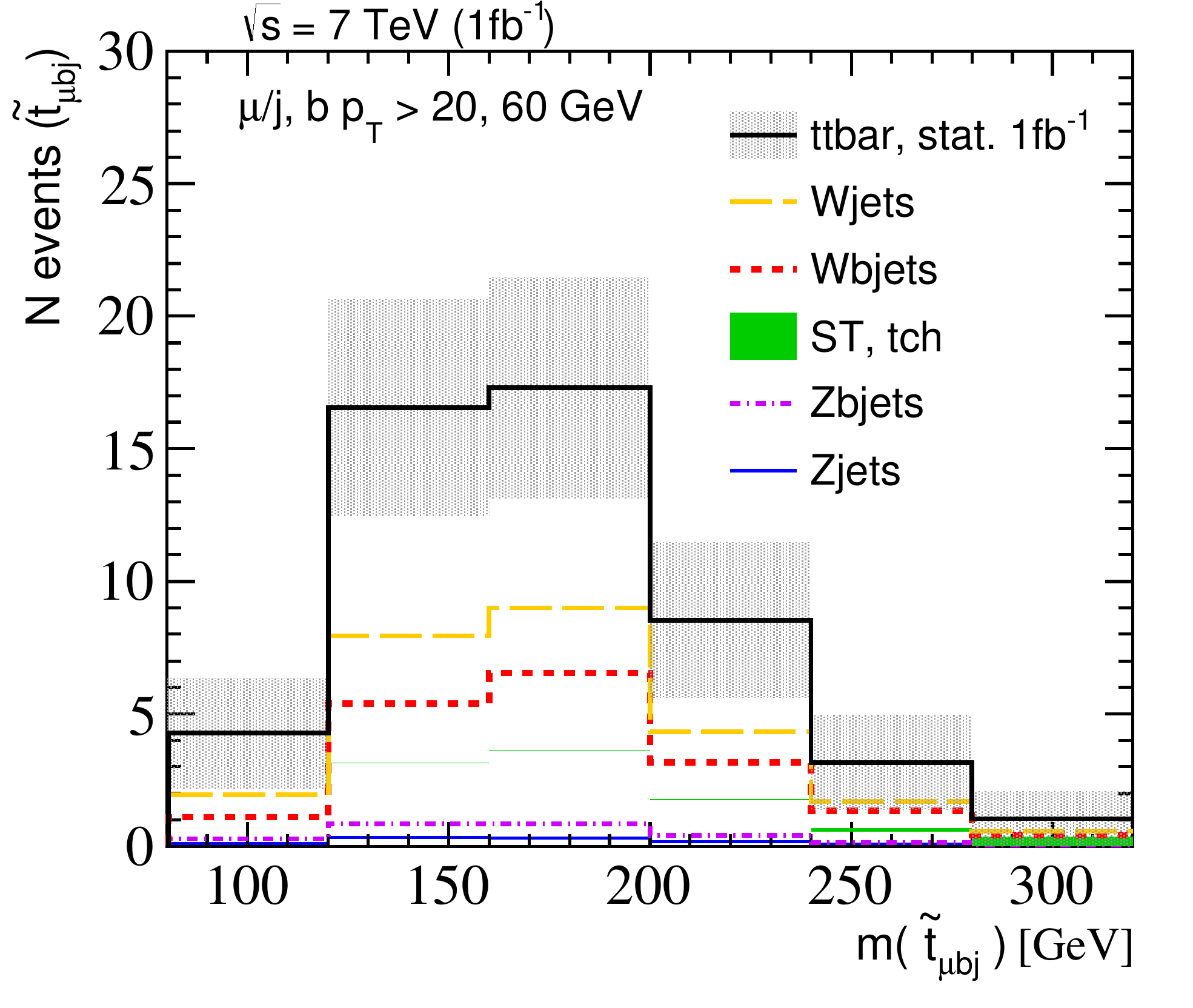} \\
\includegraphics[width=.48\linewidth]{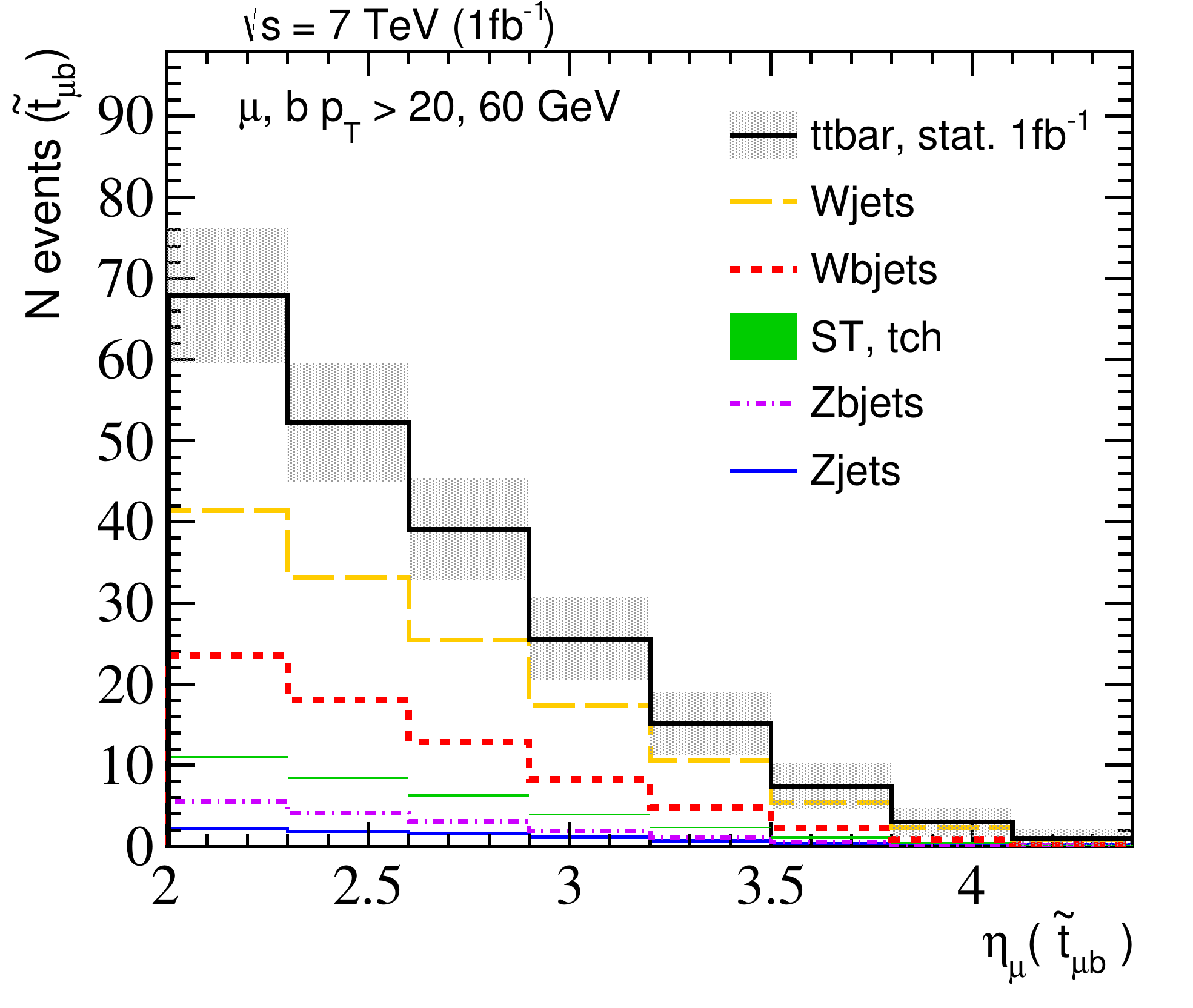} &
\includegraphics[width=.48\linewidth]{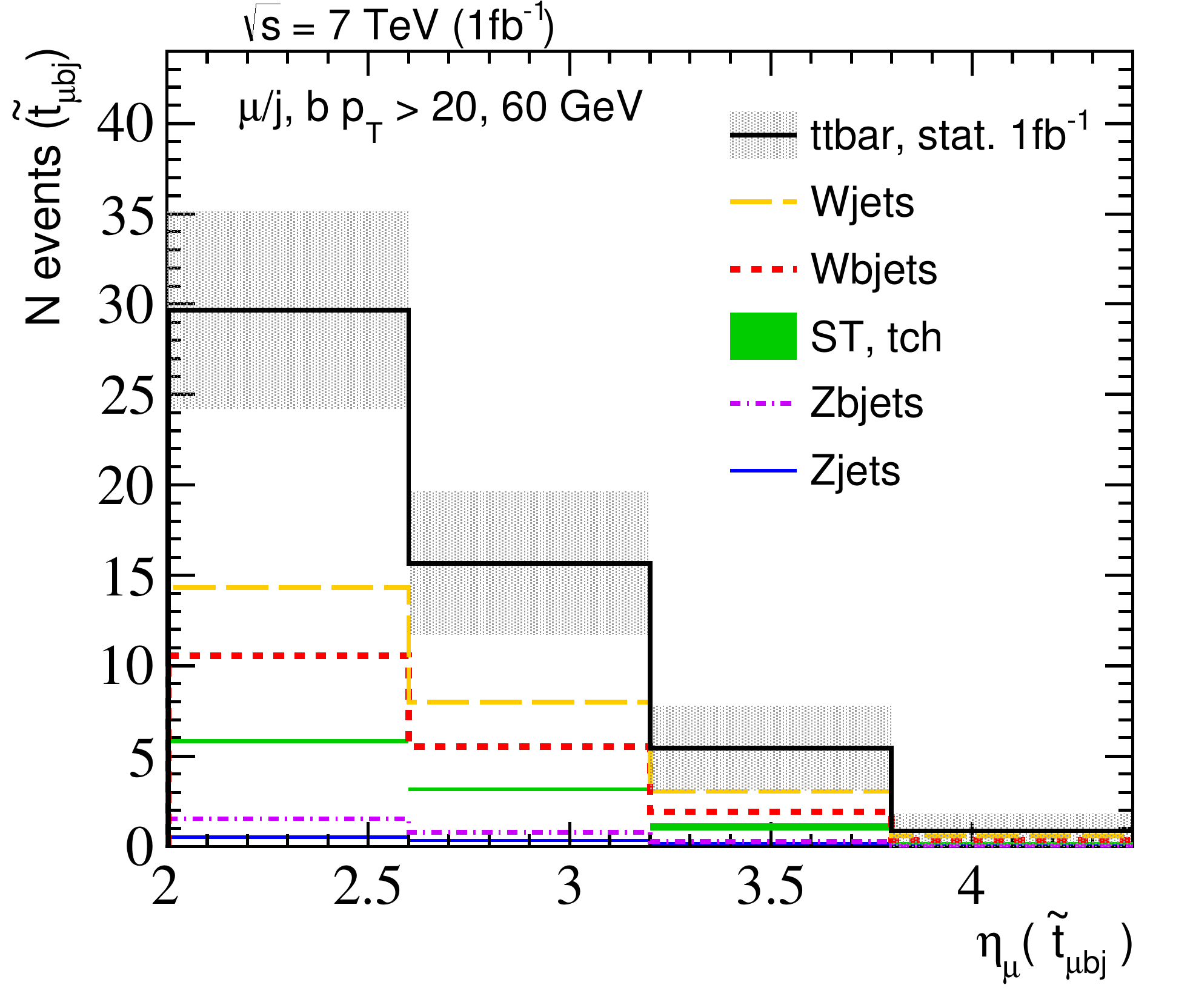} \\
\end{tabular}
\caption{Expected signal and background as a function of pseudotop ($\tilde{t}$) invariant mass (upper) and muon pseudorapidity (lower) at $\sqrt{s}$ = 7~TeV with $\int \mathcal{L} dt$ = 1fb$^{-1}$. Cuts on the $b$-jet and muon/light-jet $p_T$ of 60, 20~GeV are applied to the $\tilde{t}_{\mu b}$ (left) and $\tilde{t}_{\mu bj}$ (right) selections.}\label{plot7}
\end{figure}

The expected number of pseudotop events in 1~fb$^{-1}$ as a function of reconstructed invariant mass (upper) and muon pseudorapidity (lower) for both $\mu b$ (left) and $\mu bj$ (right) channels at 7~TeV are plotted in Fig.~\ref{plot7}. The background and signal is stacked and the resultant uncertainty band corresponds to the statistical error for the given choice of binning. The single top (ST, tch) distribution is the envelope of the 4 and 5-flavour predictions as previously mentioned. The 4-flavour differential cross-section tends to be slightly larger for high invariant masses due to a larger number of events where the co-linear spectator $b$-quark and lepton are reconstructed together within the acceptance - in particular beyond $m_t$. The $\eta_{\mu}$ distributions (lower) have overwhelming statistical uncertainties in the pseudorapidity bins beyond $\eta =$ 3. This is region of phase space where the asymmetry in the $t$/$\bar{t}$ pseudorapidity distributions in pair production is largest and indicates that more data is required for a statistically meaningful differential charge asymmetry measurement at LHCb.

\begin{figure}
\centering
\begin{tabular}{cc}
\includegraphics[width=.48\linewidth]{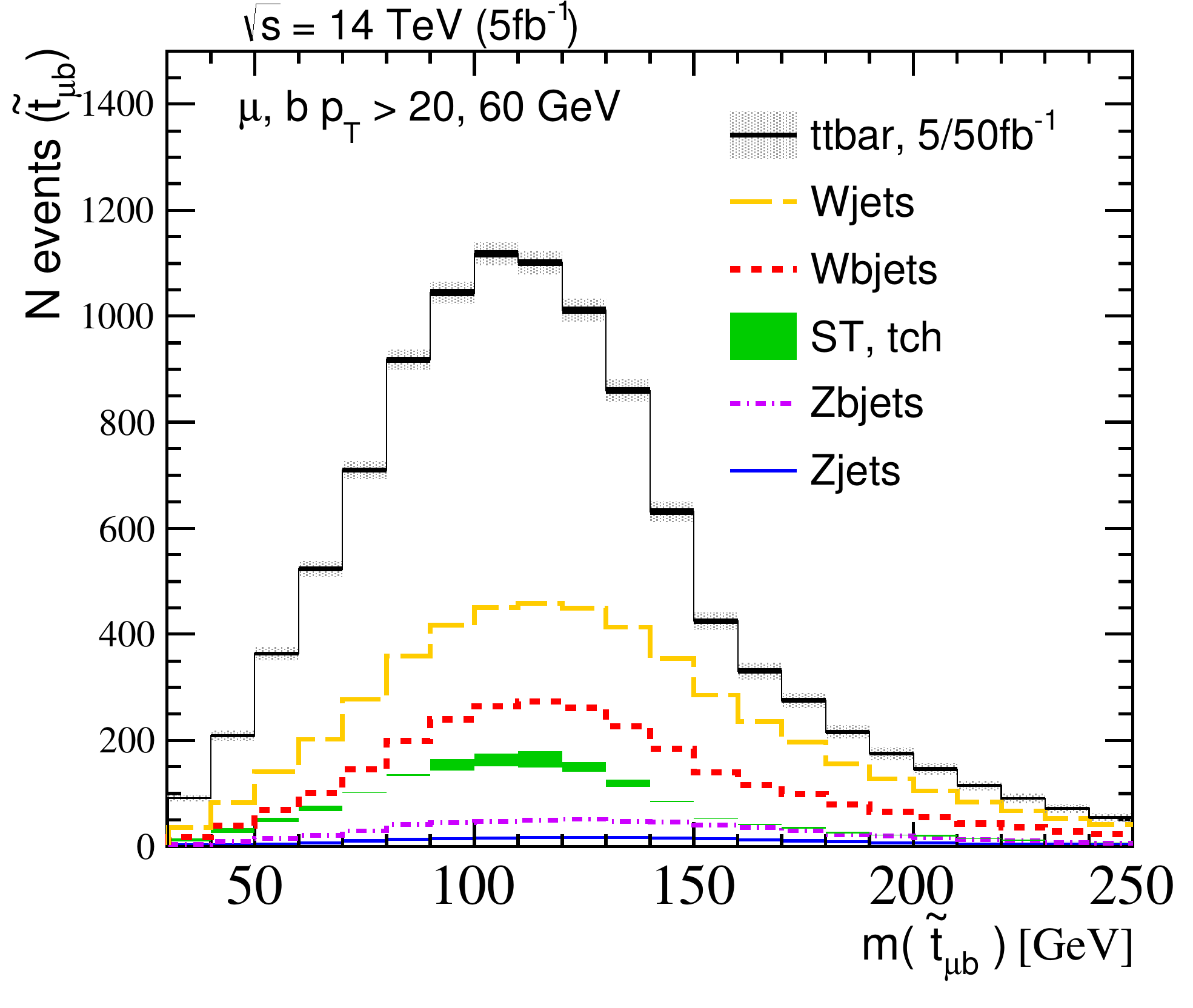} &
\includegraphics[width=.48\linewidth]{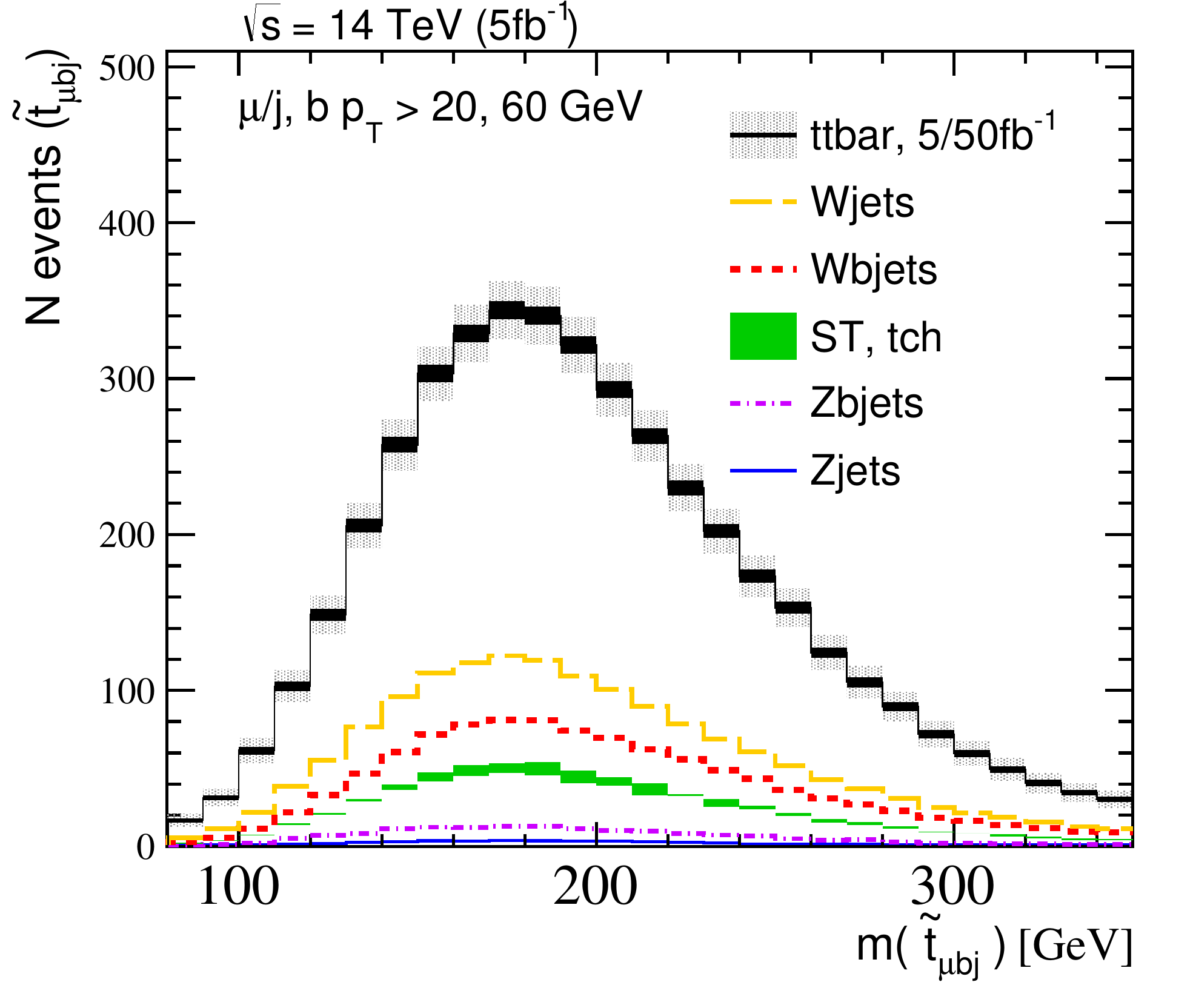} \\
\includegraphics[width=.48\linewidth]{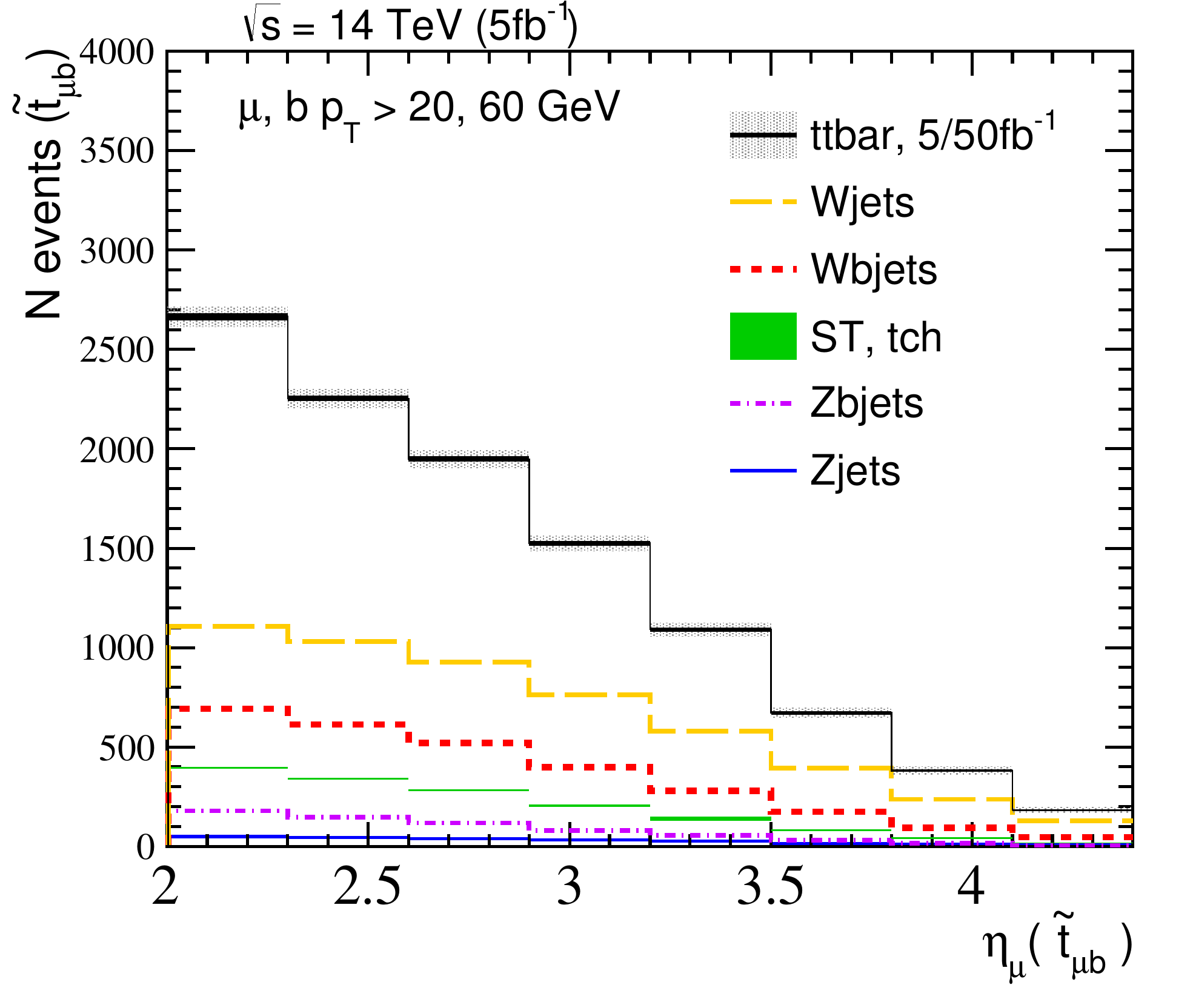} &
\includegraphics[width=.48\linewidth]{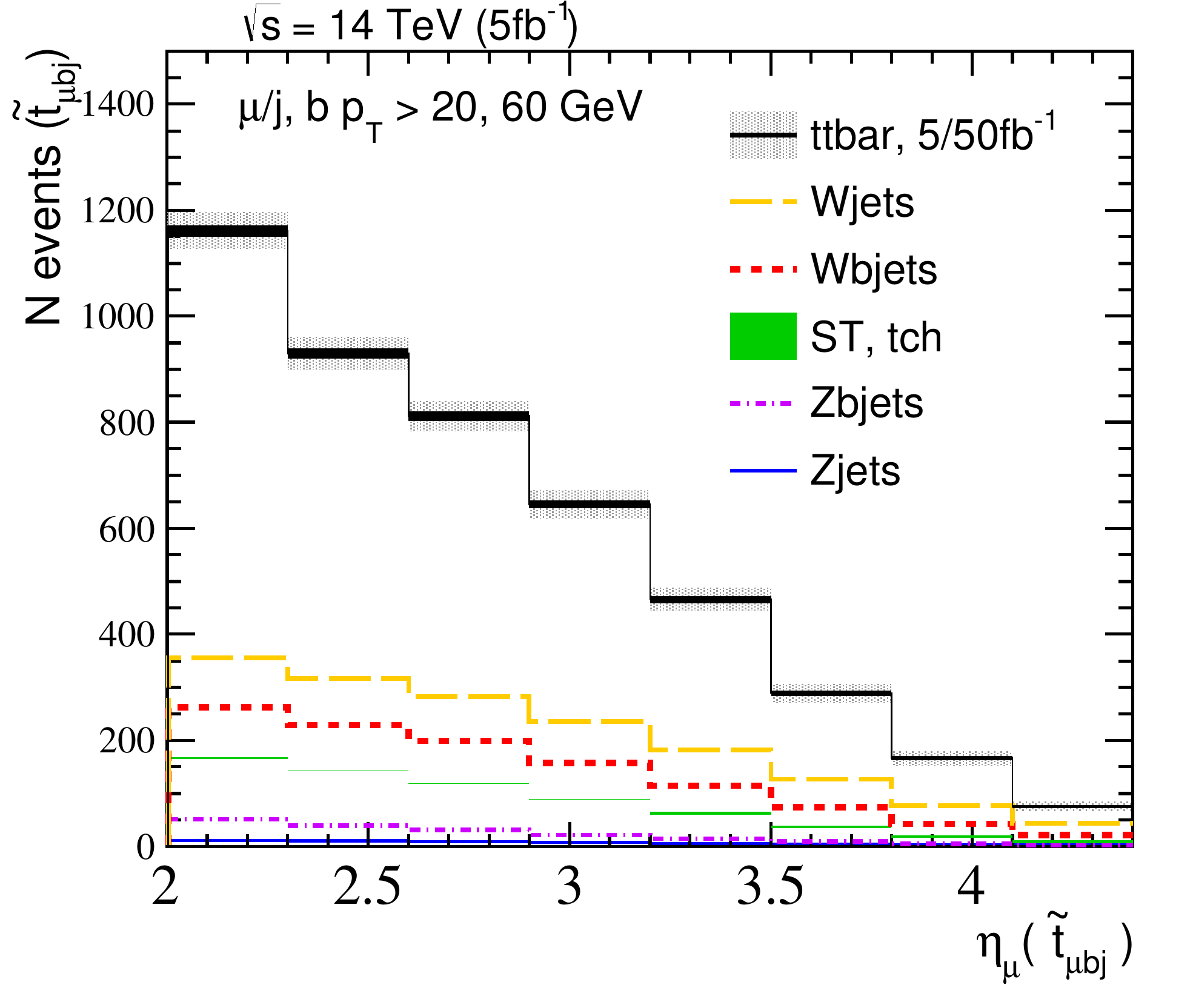} \\
\end{tabular}
\caption{Expected signal and background as a function of pseudotop ($\tilde{t}$) invariant mass (upper) and muon pseudorapidity (lower) at $\sqrt{s}$ = 14~TeV with $\int \mathcal{L} dt$ = 5fb$^{-1}$. Cuts on the $b$-jet and muon/light-jet $p_T$ of 60, 20~GeV are applied to the $\tilde{t}_{\mu b}$ (left) and $\tilde{t}_{\mu bj}$ (right) selections. The solid internal fill corresponds the statistical error expected with $\int \mathcal{L} dt$ = 50fb$^{-1}$ }\label{plot14}
\end{figure}

The same analysis of signal and background at 14~TeV is also presented. The expected number of pseudotop events in 5~fb$^{-1}$ as a function of reconstructed invariant mass (upper) and of muon pseudorapidity (lower) for both $\mu b$ (left) and $\mu bj$ (right) channels at 14~TeV are shown in Fig.~\ref{plot14}. The larger data sample size and dramatic increase in $t\bar{t}$ cross-section at 14~TeV suggest that high statistical precision will be achievable for several high multiplicity pseudotop final states. The wider grey fill on the $t\bar{t}$ signal corresponds to the statistical precision expected with 5~fb$^{-1}$, while the black band corresponds to 50~fb$^{-1}$ (achievable after $\approx$~10 years of running). It is clear that high-statistical precision ($< 2\%$) can be obtained across the entire acceptance in $\eta-$phase space even for a fine choice of binning (/0.3$\eta$).

\section{Pair production cross-section}\label{parton}
Given the promising signal yield and observability at 14 and 7~TeV, we study the theoretical uncertainties on the signal at the parton level within the LHCb acceptance. The parton level NLO results are produced with \texttt{MCFM} and compared to the inclusive NNLO+NNLL (NNLO$^*$) results presented in Ref.~\cite{Czakon:2013tha}. The LHCb cross-section for pseudotop production is,
\begin{eqnarray}
\sigma^{\mathrm{LHCb}} = \int_{\eta = 2} \frac{d \sigma^{\tilde{t}}}{d \eta}.
\end{eqnarray}

\noindent In accordance with Ref.~\cite{Czakon:2013tha}, the theoretical uncertainties are obtained in the following way,

\begin{description}
	\item [Top mass]
\end{description}
The central top quark pole mass is assumed to be 173.25~GeV. The dependence on the cross-section from the uncertainty of the top mass, $\delta m_t$, is then found by varying the mass within the range $m_t \in [171.75 - 174.75]$~GeV and then taking the average. This range is in agreement with the current PDG value of $m_t = $ 173.07 $\pm$ 0.52 $\pm$ 0.72 GeV~\cite{Beringer:1900zz} and latest LHC combination of $m_t = $ 173.29 $\pm$ 0.23 $\pm$ 0.792 GeV~\cite{ATLAS-CONF-2013-102} from direct measurements.

\begin{description}
	\item [PDF]
\end{description}
The following NLO PDF sets are studied; ABM11(5flv)\cite{Alekhin:2012ig}, CT10wnlo, HERAPDF1.5~\cite{CooperSarkar:2011aa}, MSTW08nlo68cl~\cite{Martin:2009iq} and NNPDF2.3nlo~\cite{Ball:2011mu}, where the central value of $\alpha_s (M_z) =$ 0.118, 0.118, 0.1176, 0.120 and 0.119 is chosen for each set respectively. Asymmetric/symmetric uncertainties are found in the usual way as;
\beq
\begin{aligned}
\Delta X^+ &= \sqrt{ \sum_{i=1}^N ( max[X^+_i - X_0, X^-_i - X_0, 0 ] )^2}, \\
\Delta X^- &= \sqrt{ \sum_{i=1}^N ( max[X_0 - X^+_i, X_0 - X_i^-, 0 ] )^2}, \\
\Delta X &= \frac{1}{2} \sqrt{ \sum_{i=1}^N ( X^+_i - X_i^- )^2}.
\end{aligned}
\eeq

\noindent $X^{\pm}_i$ represents the observable calculated from eigenvector member $\mathcal{S}^{\pm}_i$. The uncertainties obtained for each PDF collaboration are quoted at 1$\sigma$ confidence level (CL), where the CT10 uncertainties provided at 90\% CL have been scaled down by a factor of 1.645. The PDFs are accessed through the LHADPF interface~\cite{Whalley:2005nh}.

\begin{figure}[h]
\centering
\begin{tabular}{c}
\includegraphics[width=0.6\linewidth]{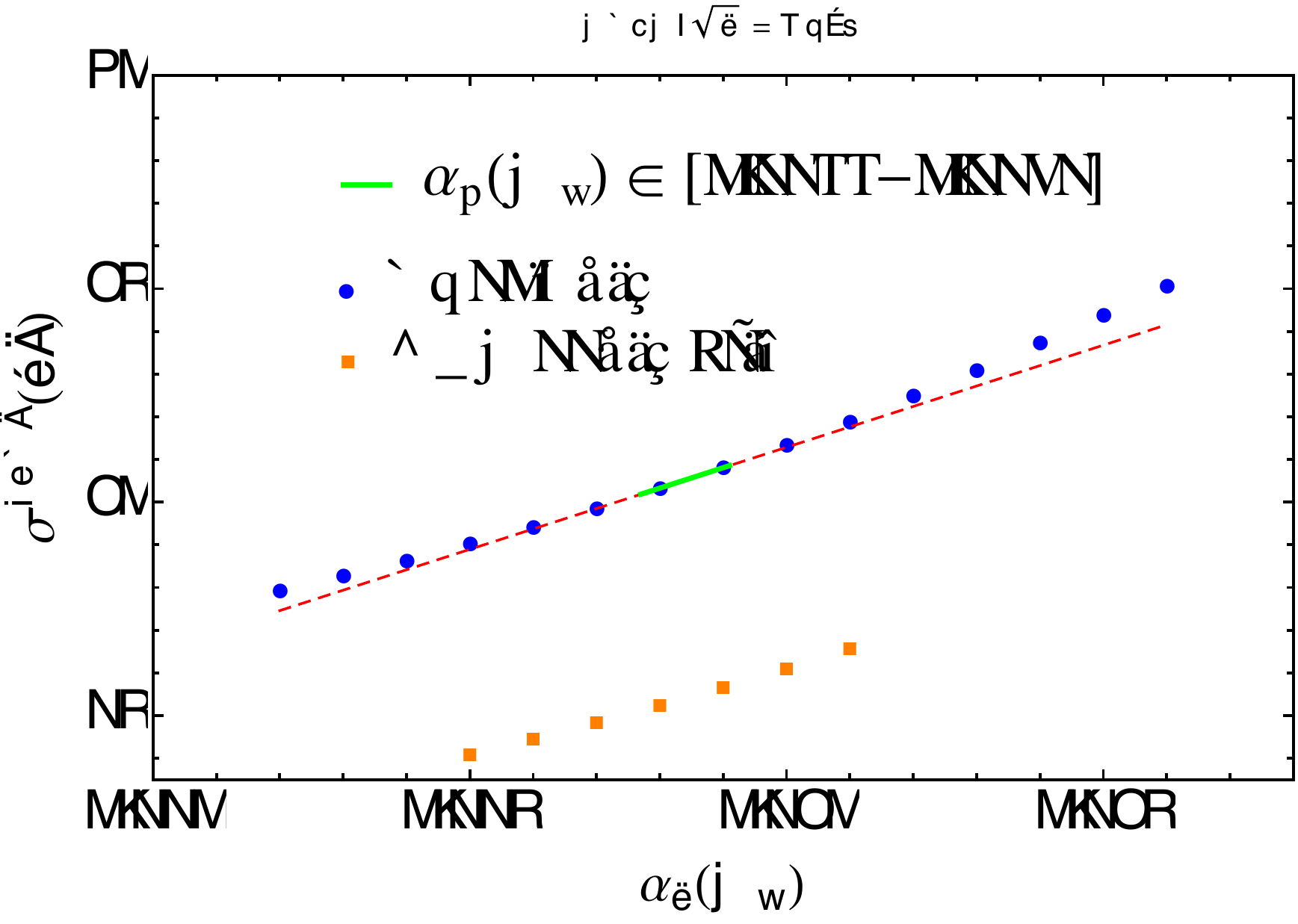}
\end{tabular}
\caption{Linear fit of $\alpha_s(M_Z)$ within the range $\alpha_s(M_Z) \in [0.116-0.120]$ at 7~TeV. The 1$\sigma$ uncertainty is then calculated from this - see text for details.}
\label{fig:alphas}
\end{figure}

\begin{description}
	\item [\textbf{$\alpha_s$}]
\end{description}
The strong coupling uncertainty, $\delta_{\alpha_s}$, is found by computing a linear fit for the cross-section for variation of $\alpha_s(M_Z) \in [0.116 - 0.120]$ which spans the preferred central value for all PDF sets. The 1$\sigma$ CL range is then extracted by matching this fit to the current PDG average of $\alpha_s (M_z)$ is 0.1184 $\pm$ 0.0007~\cite{Beringer:1900zz} - see Fig. \ref{fig:alphas}. A linear fit within this range is a good approximation to the leading quadratic behaviour. In the case of ABM PDFs, $\alpha_s(M_Z)$ is included as a parameter in the fit and is therefore already accounted for in the symmetric PDF uncertainty. ABM however do provide a set of PDFs for $\alpha_s(M_Z)$ variation, the results obtained from this variation set is compared to that of CT10 within Fig. \ref{fig:alphas} for reference.

\begin{description}
	\item[Scale] 
\end{description}
The scale uncertainty, $\delta_{\alpha_s}$, is found from varying factorisation and renormalisation scales $\mu_F$ and $\mu_R$ independently by a factor of two in both directions of the top mass - this is done such that the scale ratio ($\mu_F/\mu_R$) is always within this range. The central value is chosen as $\mu_0 = m_t$.

\begin{figure}[h]
\centering
\begin{tabular}{cc}
\includegraphics[width=.48\linewidth]{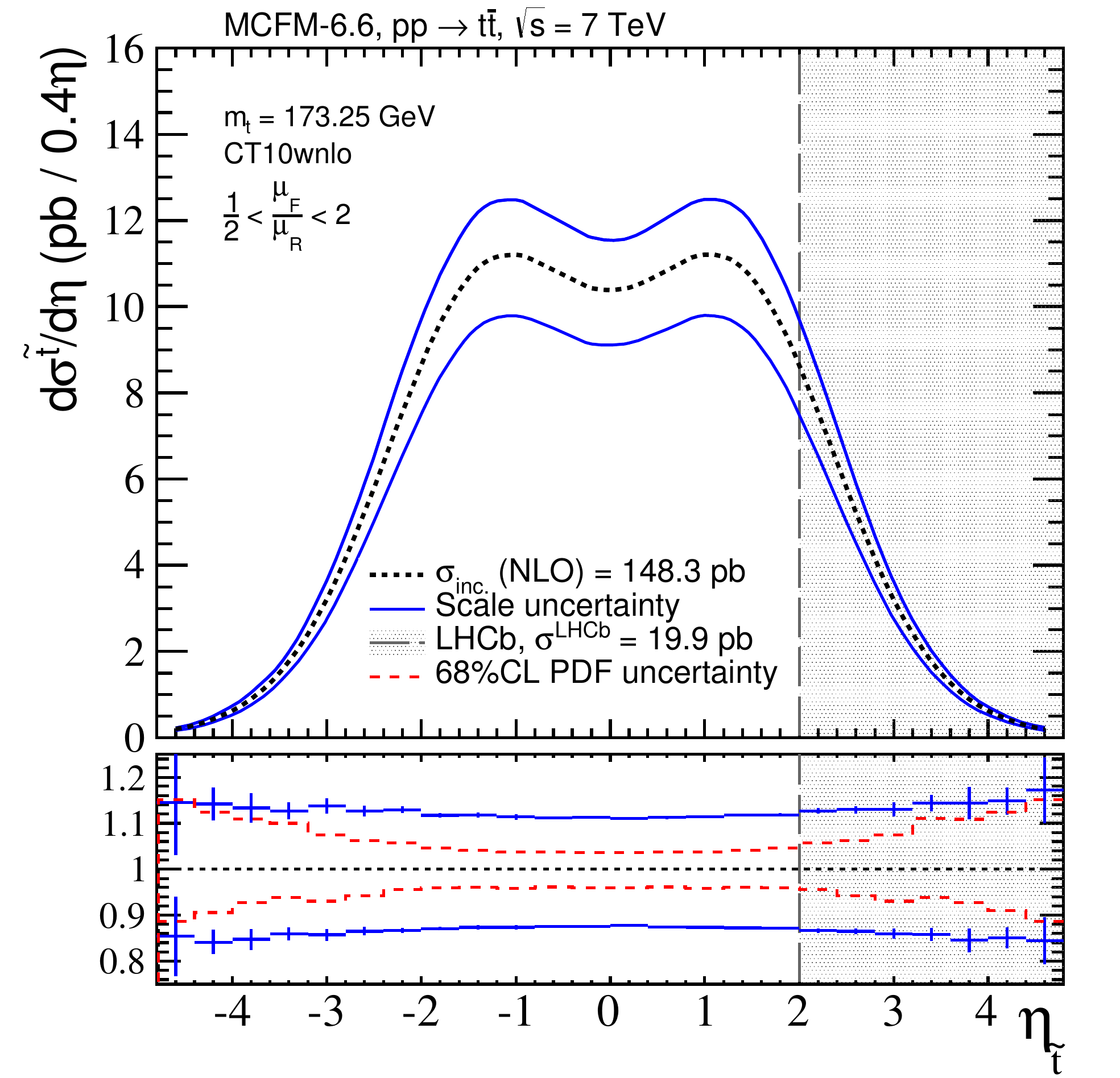} &
\includegraphics[width=.48\linewidth]{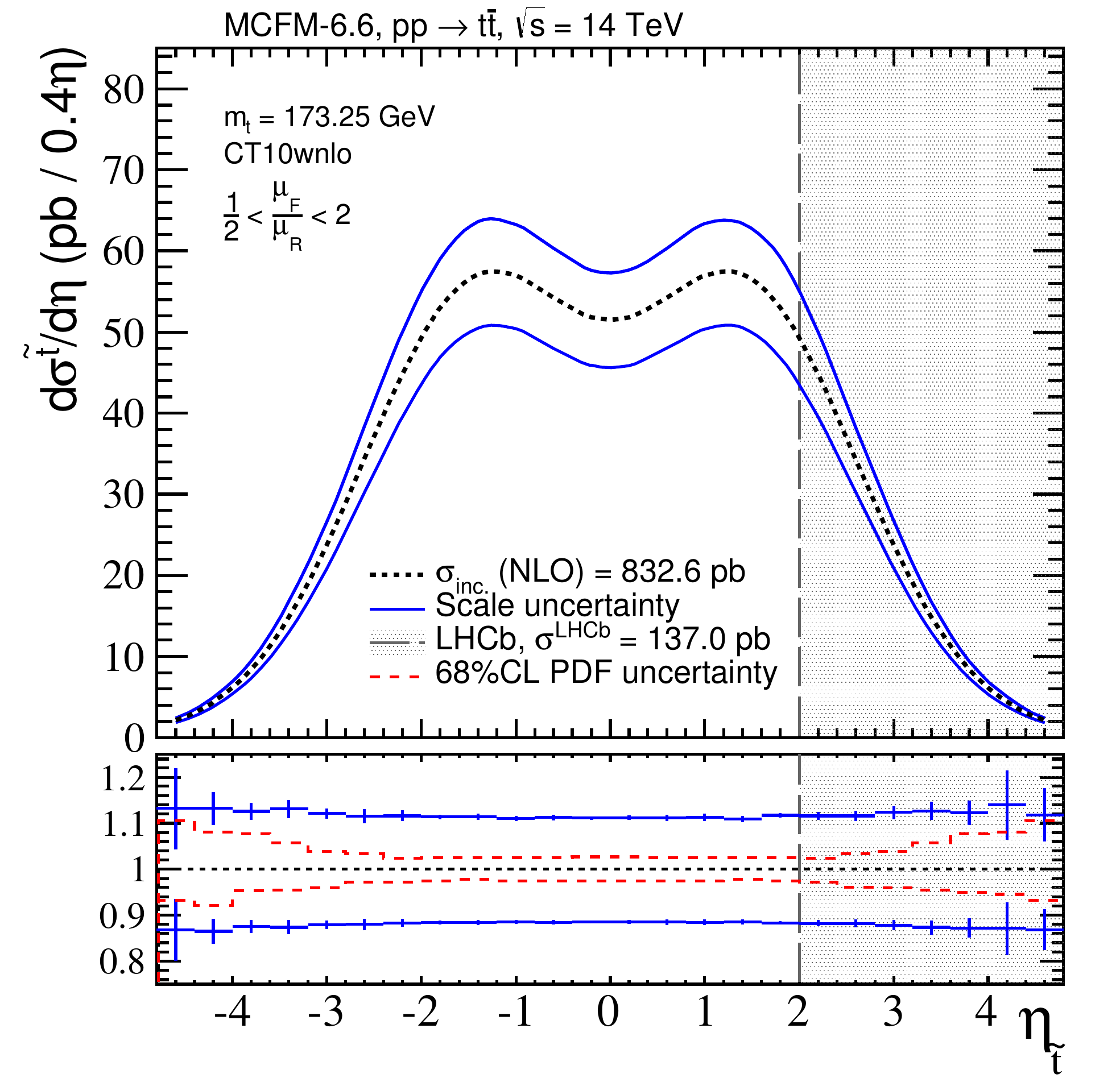}
\end{tabular}
\caption{Differential pseudotop ($\tilde{t}$) cross-section with respect to pseudorapidity at 7 (left) and 14 TeV (right). The central CT10wnlo PDF set is used and the blue uncertainty corresponds to the scale uncertainty - see text for details.}
\label{diffsigma}
\end{figure}

The pseudotop differential cross-section with respect to pseudorapidity is shown in Fig. \ref{diffsigma} for 7 and 14 TeV, the blue uncertainty band is due to scale variation as previously described. The relative 1$\sigma$ PDF uncertainty is plotted alongside the relative scale uncertainty, demonstrating the increase in PDF uncertainty with pseudorapidity. The LHCb acceptance is also highlighted to provide some intuition as to the fraction of events where one of the pair produced top quarks is forward.

The magnitude of $\delta_{\mathrm{PDF}}$ increases with pseudorapidity as this corresponds to events produced from partons at both very high and low-$x$ where the gluon and anti-quark PDFs are respectively not well known. There is also a rapidity dependence of $\delta_{\alpha_s}$ which arises from uncertainty in the gluon PDF indirectly, where an increase in $\alpha_s$ leads to a smaller gluon PDF at lower values of x while momentum sum rules compensate this by increasing the gluon PDF at large x, resulting in a rapidity dependent uncertainty. There is also a small pseudorapidity dependence on scales due to differences in the physical scale, $Q^2$, for forward events.

\begin{figure}[h]
\centering
\begin{tabular}{cc}
\includegraphics[width=0.48\linewidth]{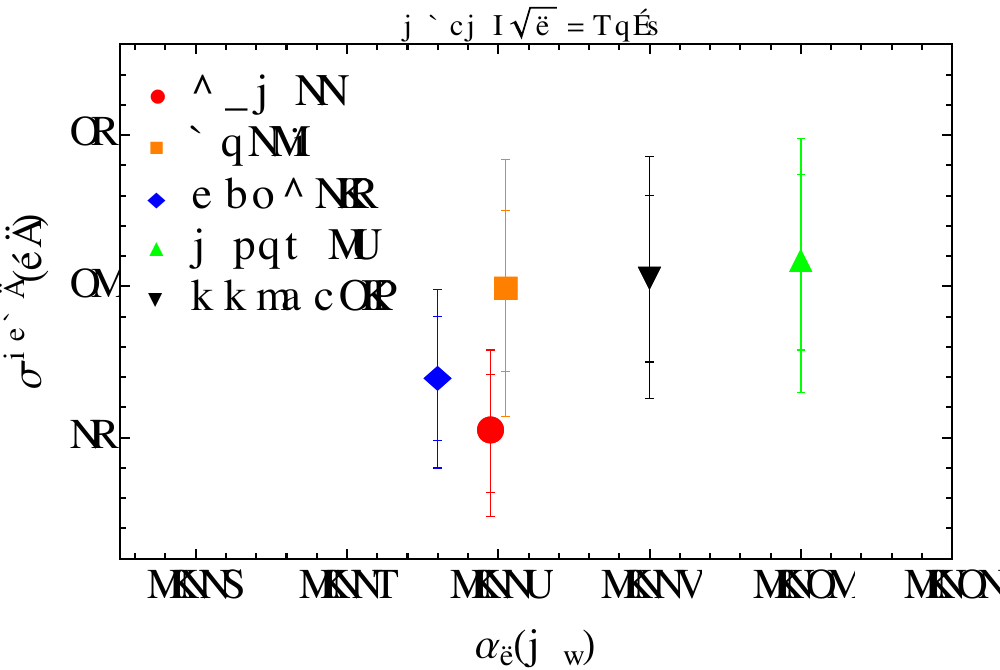} &
\includegraphics[width=0.48\linewidth]{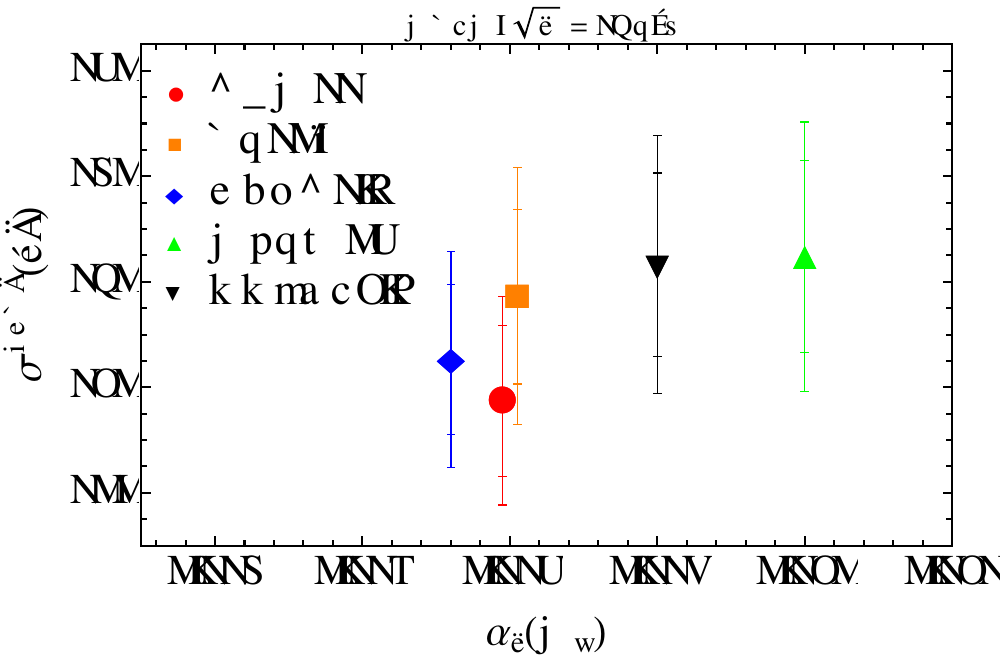}
\end{tabular}
\caption{Summary of cross-section and theoretical uncertainties within the LHCb fiducial region at $\sqrt{s} =$ 7 (left) and 14~TeV (right), plotted with respect to each PDF collaborations preferred value for $\alpha_s(M_Z)$. The inner and outer error bars correspond to the scale and total uncertainties respectively.}
\label{fig:Sigma}
\end{figure}

The contribution from the individual sources of systematic uncertainties to the LHCb cross-section are now evaluated and compared to the inclusive NLO and NNLO$^*$ results - from Ref.~\cite{Czakon:2013tha}. The total uncertainty is found by combining the the individual uncertainties following the recommendation of the Higgs Cross Section Working Group~\cite{Dittmaier:2011ti} as,
\begin{equation}
\delta_{\mathrm{total}} = \delta_{\mathrm{scale}} + (\delta_{\mathrm{PDF}}^2 + \delta_{\alpha_s}^2 + \delta_{m_t}^2)^\frac{1}{2}.
\end{equation}
\noindent The 7 and 14~TeV results are summarized in Table~\ref{Sigma7} and Table~\ref{Sigma14} respectively. A summary plot including both scale and total uncertainties is also provided both energies in Fig.~\ref{fig:Sigma}. The NNLO$^*$ results have been corrected to the chosen top mass range of $m_t \in [ 171.75 - 174.75 ]$, where it is found that a 1~GeV uncertainty on $m_t$ translates into a 3.0, 2.7$\%$ uncertainty on the cross-section at 7 and 14~TeV.

\begin{table}[htdp]
\begin{footnotesize}
\centering
	\begin{tabular}{r | c | c | r c r | r c r | r c r | r c r | r c r }
	Order 	& PDF 	&$\sigma$(pb) 	& \multicolumn{3}{ c | }{$\delta_{\mathrm{scale}}$ (pb)}	& \multicolumn{3}{ c | }{$\delta_{\mathrm{PDF}}$ (pb)}	& \multicolumn{3}{ c | }{$\delta_{\alpha_s}$ (pb)} & \multicolumn{3}{ c | }{$\delta_{m_t}$ (pb)}	& \multicolumn{3}{ c }{$\delta_{\mathrm{total}}$ (pb)}\\ \hline
	
	NNLO$^*$(inc.)	& 	& 135.8		& $^{+ 3.5}_{- 4.2}$ & \hspace{-0.82cm} & $^{(+ 2.6\%)}_{(- 3.1\%)}$	& $^{+ 6.4}_{- 6.4}$ & \hspace{-0.82cm} & $^{(+ 4.7\%)}_{(- 4.7\%)}$		& $^{+ 0.0}_{- 0.0}$ & \hspace{-0.82cm} & $^{(+ 0.0\%)}_{(- 0.0\%)}$ 	& $^{+ 6.5}_{- 6.3}$ & \hspace{-0.82cm} & $^{(+ 4.8\%)}_{(- 4.7\%)}$		& $^{+ 12.7}_{- 13.2}$ & \hspace{-0.82cm} & $^{(+ 9.3\%)}_{(- 9.7\%)}$ \\

	NLO(inc.)	& ABM	& 123.5		& $^{+ 14.6}_{- 16.1}$ & \hspace{-0.82cm} & $^{(+ 11.8\%)}_{(- 13.0\%)}$	& $^{+ 2.3}_{- 2.3}$ & \hspace{-0.82cm} & $^{(+ 1.9\%)}_{(- 1.9\%)}$		& $^{+ 0.0}_{- 0.0}$ & \hspace{-0.82cm} & $^{(+ 0.0\%)}_{(- 0.0\%)}$ 	& $^{+ 5.8}_{- 5.7}$ & \hspace{-0.82cm} & $^{(+ 4.7\%)}_{(- 4.6\%)}$		& $^{+ 20.8}_{- 22.2}$ & \hspace{-0.82cm} & $^{(+ 16.9\%)}_{(- 18.0\%)}$ \\

	NLO(LHCb)	& 	& 15.2		& $^{+ 1.9}_{- 2.0}$ & \hspace{-0.82cm} & $^{(+ 12.4\%)}_{(- 13.0\%)}$	& $^{+ 0.3}_{- 0.3}$ & \hspace{-0.82cm} & $^{(+ 2.1\%)}_{(- 2.1\%)}$		& $^{+ 0.0}_{- 0.0}$ & \hspace{-0.82cm} & $^{(+ 0.0\%)}_{(- 0.0\%)}$ 	& $^{+ 0.8}_{- 0.7}$ & \hspace{-0.82cm} & $^{(+ 5.0\%)}_{(- 4.9\%)}$		& $^{+ 2.7}_{- 2.8}$ & \hspace{-0.82cm} & $^{(+ 17.8\%)}_{(- 18.3\%)}$ \\
	\hline
	
	NNLO$^*$(inc.)	& 	& 172.5		& $^{+ 4.6}_{- 6.0}$ & \hspace{-0.82cm} & $^{(+ 2.7\%)}_{(- 3.5\%)}$	& $^{+ 8.0}_{- 6.5}$ & \hspace{-0.82cm} & $^{(+ 4.6\%)}_{(- 3.8\%)}$		& $^{+ 3.7}_{- 3.7}$ & \hspace{-0.82cm} & $^{(+ 2.2\%)}_{(- 2.2\%)}$ 	& $^{+ 8.0}_{- 7.7}$ & \hspace{-0.82cm} & $^{(+ 4.6\%)}_{(- 4.4\%)}$		& $^{+ 16.5}_{- 16.7}$ & \hspace{-0.82cm} & $^{(+ 9.5\%)}_{(- 9.7\%)}$ \\

	NLO(inc.)	& CT10	& 148.3		& $^{+ 17.6}_{- 19.2}$ & \hspace{-0.82cm} & $^{(+ 11.9\%)}_{(- 13.0\%)}$	& $^{+ 6.6}_{- 6.3}$ & \hspace{-0.82cm} & $^{(+ 4.4\%)}_{(- 4.2\%)}$		& $^{+ 2.0}_{- 2.0}$ & \hspace{-0.82cm} & $^{(+ 1.3\%)}_{(- 1.3\%)}$ 	& $^{+ 6.8}_{- 6.6}$ & \hspace{-0.82cm} & $^{(+ 4.6\%)}_{(- 4.4\%)}$		& $^{+ 27.2}_{- 28.5}$ & \hspace{-0.82cm} & $^{(+ 18.4\%)}_{(- 19.2\%)}$ \\

	NLO(LHCb)	& 	& 19.9		& $^{+ 2.6}_{- 2.7}$ & \hspace{-0.82cm} & $^{(+ 13.3\%)}_{(- 13.7\%)}$	& $^{+ 1.4}_{- 1.1}$ & \hspace{-0.82cm} & $^{(+ 6.9\%)}_{(- 5.5\%)}$		& $^{+ 0.3}_{- 0.3}$ & \hspace{-0.82cm} & $^{(+ 1.6\%)}_{(- 1.6\%)}$ 	& $^{+ 1.0}_{- 0.9}$ & \hspace{-0.82cm} & $^{(+ 4.9\%)}_{(- 4.8\%)}$		& $^{+ 4.3}_{- 4.2}$ & \hspace{-0.82cm} & $^{(+ 21.9\%)}_{(- 21.1\%)}$ \\
	\hline
	
		NNLO$^*$(inc.)	& 	& 177.2		& $^{+ 4.8}_{- 4.2}$ & \hspace{-0.82cm} & $^{(+ 2.7\%)}_{(- 2.3\%)}$	& $^{+ 4.0}_{- 6.4}$ & \hspace{-0.82cm} & $^{(+ 2.3\%)}_{(- 3.6\%)}$		& $^{+ 3.0}_{- 3.0}$ & \hspace{-0.82cm} & $^{(+ 1.7\%)}_{(- 1.7\%)}$		& $^{+ 8.1}_{- 7.8}$ & \hspace{-0.82cm} & $^{(+ 4.6\%)}_{(- 4.4\%)}$		& $^{+ 14.3}_{- 14.7}$ & \hspace{-0.82cm} & $^{(+ 8.1\%)}_{(- 8.3\%)}$ \\

	NLO(inc.)	& \hspace{-0.22cm} HERA \hspace{-0.22cm}	& 136.1		& $^{+ 15.6}_{- 16.3}$ & \hspace{-0.82cm} & $^{(+ 11.5\%)}_{(- 12.0\%)}$	& $^{+ 3.9}_{- 3.4}$ & \hspace{-0.82cm} & $^{(+ 2.9\%)}_{(- 2.5\%)}$		& $^{+ 1.3}_{- 1.3}$ & \hspace{-0.82cm} & $^{(+ 1.0\%)}_{(- 1.0\%)}$ 	& $^{+ 6.2}_{- 6.1}$ & \hspace{-0.82cm} & $^{(+ 4.6\%)}_{(- 4.5\%)}$		& $^{+ 23.1}_{- 23.3}$ & \hspace{-0.82cm} & $^{(+ 16.9\%)}_{(- 17.1\%)}$ \\

	NLO(LHCb) & 	& 16.9		& $^{+ 2.1}_{- 2.0}$ & \hspace{-0.82cm} & $^{(+ 12.3\%)}_{(- 12.0\%)}$	& $^{+ 0.5}_{- 0.3}$ & \hspace{-0.82cm} & $^{(+ 2.9\%)}_{(- 1.6\%)}$		& $^{+ 0.2}_{- 0.2}$ & \hspace{-0.82cm} & $^{(+ 1.2\%)}_{(- 1.2\%)}$ 	& $^{+ 0.8}_{- 0.8}$ & \hspace{-0.82cm} & $^{(+ 4.8\%)}_{(- 4.7\%)}$		& $^{+ 3.0}_{- 2.9}$ & \hspace{-0.82cm} & $^{(+ 18.0\%)}_{(- 17.1\%)}$ \\
	\hline
	
	NNLO$^*$(inc.)	 & 	& 172.0		& $^{+ 4.4}_{- 5.8}$ & \hspace{-0.82cm} & $^{(+ 2.6\%)}_{(- 3.4\%)}$	& $^{+ 4.7}_{- 4.7}$ & \hspace{-0.82cm} & $^{(+ 2.7\%)}_{(- 2.7\%)}$		& $^{+ 2.9}_{- 2.9}$ & \hspace{-0.82cm} & $^{(+ 1.7\%)}_{(- 1.7\%)}$ 	& $^{+ 8.0}_{- 7.7}$ & \hspace{-0.82cm} & $^{(+ 4.6\%)}_{(- 4.4\%)}$		& $^{+ 14.1}_{- 15.2}$ & \hspace{-0.82cm} & $^{(+ 8.2\%)}_{(- 8.9\%)}$ \\
	
	NLO(inc.)	& \hspace{-0.22cm} MSTW \hspace{-0.22cm}	& 158.4		& $^{+ 19.6}_{- 21.2}$ & \hspace{-0.82cm} & $^{(+ 12.4\%)}_{(- 13.4\%)}$	& $^{+ 4.0}_{- 5.5}$ & \hspace{-0.82cm} & $^{(+ 2.6\%)}_{(- 3.4\%)}$		& $^{+ 2.1}_{- 2.1}$ & \hspace{-0.82cm} & $^{(+ 1.3\%)}_{(- 1.3\%)}$ 	& $^{+ 7.2}_{- 7.0}$ & \hspace{-0.82cm} & $^{(+ 4.6\%)}_{(- 4.5\%)}$		& $^{+ 28.1}_{- 30.4}$ & \hspace{-0.82cm} & $^{(+ 17.7\%)}_{(- 19.2\%)}$ \\

	NLO(LHCb) 	& 	& 20.8		& $^{+ 2.9}_{- 2.9}$ & \hspace{-0.82cm} & $^{(+ 13.9\%)}_{(- 14.2\%)}$	& $^{+ 0.7}_{- 0.9}$ & \hspace{-0.82cm} & $^{(+ 3.2\%)}_{(- 4.2\%)}$		& $^{+ 0.3}_{- 0.3}$ & \hspace{-0.82cm} & $^{(+ 1.5\%)}_{(- 1.5\%)}$ 	& $^{+ 1.0}_{- 1.0}$ & \hspace{-0.82cm} & $^{(+ 4.8\%)}_{(- 4.8\%)}$		& $^{+ 4.1}_{- 4.3}$ & \hspace{-0.82cm} & $^{(+ 19.9\%)}_{(- 20.8\%)}$ \\
	\hline
	
	NNLO$^*$(inc.)	 & 	& 172.7		& $^{+ 4.6}_{- 6.0}$ & \hspace{-0.82cm} & $^{(+ 2.7\%)}_{(- 3.5\%)}$	& $^{+ 5.2}_{- 5.2}$ & \hspace{-0.82cm} & $^{(+ 3.0\%)}_{(- 3.0\%)}$		& $^{+ 2.7}_{- 2.7}$ & \hspace{-0.82cm} & $^{(+ 1.6\%)}_{(- 1.6\%)}$ 	& $^{+ 8.0}_{- 7.8}$ & \hspace{-0.82cm} & $^{(+ 4.6\%)}_{(- 4.5\%)}$		& $^{+ 14.5}_{- 15.8}$ & \hspace{-0.82cm} & $^{(+ 8.4\%)}_{(- 9.1\%)}$ \\	
	
	NLO(inc.)	& \hspace{-0.42cm} NNPDF \hspace{-0.42cm}	& 158.7		& $^{+ 19.6}_{- 20.2}$ & \hspace{-0.82cm} & $^{(+ 12.4\%)}_{(- 12.7\%)}$	& $^{+ 4.0}_{- 4.0}$ & \hspace{-0.82cm} & $^{(+ 2.5\%)}_{(- 2.5\%)}$		& $^{+ 2.3}_{- 2.3}$ & \hspace{-0.82cm} & $^{(+ 1.5\%)}_{(- 1.5\%)}$ 	& $^{+ 7.3}_{- 7.1}$ & \hspace{-0.82cm} & $^{(+ 4.6\%)}_{(- 4.5\%)}$		& $^{+ 27.2}_{- 28.5}$ & \hspace{-0.82cm} & $^{(+ 17.8\%)}_{(- 18.1\%)}$ \\
	
	NLO(LHCb)	& 	& 20.2		& $^{+ 2.8}_{- 2.7}$ & \hspace{-0.82cm} & $^{(+ 14.0\%)}_{(- 13.3\%)}$	& $^{+ 0.7}_{- 0.7}$ & \hspace{-0.82cm} & $^{(+ 3.4\%)}_{(- 3.4\%)}$		& $^{+ 0.4}_{- 0.4}$ & \hspace{-0.82cm} & $^{(+ 1.8\%)}_{(- 1.8\%)}$ 	& $^{+ 1.0}_{- 0.9}$ & \hspace{-0.82cm} & $^{(+ 4.9\%)}_{(- 4.8\%)}$		& $^{+ 4.1}_{- 3.9}$ & \hspace{-0.82cm} & $^{(+ 20.2\%)}_{(- 19.4\%)}$ \\

	\end{tabular}
	\caption{Summary of inclusive (inc.) and differential (LHCb) cross-sections at NNLO+NLLL (NNLO$^*$) and NLO accuracy and associated theoretical uncertainties at 7~TeV, for PDF sets as described in the text.}
	\label{Sigma7}
\end{footnotesize}
\end{table}

\begin{table}[htdp]
\begin{footnotesize}
\centering
	\begin{tabular}{r | c | c | r c r | r c r | r c r | r c r | r c r }
	Order 	& PDF 	&$\sigma$(pb) 	& \multicolumn{3}{ c | }{$\delta_{\mathrm{scale}}$ (pb)}	& \multicolumn{3}{ c | }{$\delta_{\mathrm{PDF}}$ (pb)}	& \multicolumn{3}{ c | }{$\delta_{\alpha_s}$ (pb)} & \multicolumn{3}{ c | }{$\delta_{m_t}$ (pb)}	& \multicolumn{3}{ c }{$\delta_{\mathrm{total}}$ (pb)}\\ \hline
	
	NNLO$^*$(inc.)	&	& 832.0	&  $^{+ 18.7}_{- 27.4}$ & \hspace{-0.82cm} & $^{(+ 2.2\%)}_{(- 3.3\%)}$	& $^{+ 25.1}_{- 25.1}$ & \hspace{-0.82cm} & $^{(+ 3.0\%)}_{(- 3.0\%)}$		& $^{+ 0.0}_{- 0.0}$ & \hspace{-0.82cm} & $^{(+ 0.0\%)}_{(- 0.0\%)}$ 	& $^{+ 34.9}_{- 33.7}$ & \hspace{-0.82cm} & $^{(+ 4.2\%)}_{(- 4.1\%)}$		& $^{+ 61.7}_{- 69.7}$ & \hspace{-0.82cm} & $^{(+ 7.4\%)}_{(- 8.4\%)}$ \\

	NLO(inc.)	& \hspace{-0.22cm} ABM \hspace{-0.22cm}	& 771.9	& $^{+ 91.0}_{- 92.4}$ & \hspace{-0.82cm} & $^{(+ 11.8\%)}_{(- 12.0\%)}$	& $^{+ 9.4}_{- 9.4}$ & \hspace{-0.82cm} & $^{(+ 1.2\%)}_{(- 1.2\%)}$		& $^{+ 0.0}_{- 0.0}$ & \hspace{-0.82cm} & $^{(+ 0.0\%)}_{(- 0.0\%)}$ 	& $^{+ 32.3}_{- 31.9}$ & \hspace{-0.82cm} & $^{(+ 4.2\%)}_{(- 4.1\%)}$		& $^{+ 124.7}_{- 125.7}$ & \hspace{-0.82cm} & $^{(+ 16.1\%)}_{(- 16.3\%)}$ \\
	
	NLO(LHCb) 	& 	& 117.2	& $^{+ 14.5}_{- 14.1}$ & \hspace{-0.82cm} & $^{(+ 12.3\%)}_{(- 12.0\%)}$	& $^{+ 2.0}_{- 2.0}$ & \hspace{-0.82cm} & $^{(+ 1.7\%)}_{(- 1.7\%)}$		& $^{+ 0.0}_{- 0.0}$ & \hspace{-0.82cm} & $^{(+ 0.0\%)}_{(- 0.0\%)}$ 	& $^{+ 5.2}_{- 5.1}$ & \hspace{-0.82cm} & $^{(+ 4.4\%)}_{(- 4.3\%)}$		& $^{+ 20.0}_{- 19.5}$ & \hspace{-0.82cm} & $^{(+ 17.1\%)}_{(- 16.7\%)}$ \\ \hline

	NNLO$^*$(inc.)	&	& 952.8	&  $^{+ 23.3}_{- 34.5}$ & \hspace{-0.82cm} & $^{(+ 2.4\%)}_{(- 3.6\%)}$	& $^{+ 22.4}_{- 19.9}$ & \hspace{-0.82cm} & $^{(+ 2.3\%)}_{(- 2.1\%)}$		& $^{+ 14.0}_{- 14.0}$ & \hspace{-0.82cm} & $^{(+ 1.5\%)}_{(- 1.5\%)}$ 	& $^{+ 39.2}_{- 37.8}$ & \hspace{-0.82cm} & $^{(+ 4.1\%)}_{(- 4.0\%)}$		& $^{+ 70.6}_{- 79.5}$ & \hspace{-0.82cm} & $^{(+ 7.4\%)}_{(- 8.3\%)}$ \\

	NLO(inc.)	& \hspace{-0.22cm} CT10 \hspace{-0.22cm}	& 832.6	& $^{+ 97.0}_{- 96.7}$ & \hspace{-0.82cm} & $^{(+ 11.7\%)}_{(- 11.6\%)}$	& $^{+ 19.6}_{- 20.2}$ & \hspace{-0.82cm} & $^{(+ 2.4\%)}_{(- 2.4\%)}$		& $^{+ 9.2}_{- 9.2}$ & \hspace{-0.82cm} & $^{(+ 1.1\%)}_{(- 1.1\%)}$ 	& $^{+ 34.0}_{- 33.3}$ & \hspace{-0.82cm} & $^{(+ 4.1\%)}_{(- 4.0\%)}$		& $^{+ 137.4}_{- 136.6}$ & \hspace{-0.82cm} & $^{(+ 16.5\%)}_{(- 16.4\%)}$ \\
	
	NLO(LHCb) 	& 	& 137.0	& $^{+ 16.7}_{- 16.4}$ & \hspace{-0.82cm} & $^{(+ 12.2\%)}_{(- 12.0\%)}$	& $^{+ 5.0}_{- 4.6}$ & \hspace{-0.82cm} & $^{(+ 3.6\%)}_{(- 3.4\%)}$		& $^{+ 1.8}_{- 1.8}$ & \hspace{-0.82cm} & $^{(+ 1.3\%)}_{(- 1.3\%)}$ 	& $^{+ 5.9}_{- 5.8}$ & \hspace{-0.82cm} & $^{(+ 4.3\%)}_{(- 4.2\%)}$		& $^{+ 24.7}_{- 24.0}$ & \hspace{-0.82cm} & $^{(+ 18.0\%)}_{(- 17.5\%)}$ \\ \hline
	
		NNLO$^*$(inc.) &	& 970.5	&  $^{+ 22.1}_{- 22.0}$ & \hspace{-0.82cm} & $^{(+ 2.3\%)}_{(- 2.3\%)}$	& $^{+ 15.7}_{- 25.7}$ & \hspace{-0.82cm} & $^{(+ 1.6\%)}_{(- 2.6\%)}$		& $^{+ 12.8}_{- 12.8}$ & \hspace{-0.82cm} & $^{(+ 1.3\%)}_{(- 1.3\%)}$ 	& $^{+ 39.6}_{- 38.4}$ & \hspace{-0.82cm} & $^{(+ 4.1\%)}_{(- 4.0\%)}$		& $^{+ 66.6}_{- 70.0}$ & \hspace{-0.82cm} & $^{(+ 6.9\%)}_{(- 7.2\%)}$ \\

	NLO(inc.)	& \hspace{-0.22cm} HERA \hspace{-0.22cm}	& 804.2	& $^{+ 91.9}_{- 87.6}$ & \hspace{-0.82cm} & $^{(+ 11.4\%)}_{(- 10.9\%)}$	& $^{+ 16.1}_{- 21.9}$ & \hspace{-0.82cm} & $^{(+ 2.0\%)}_{(- 2.7\%)}$		& $^{+ 5.3}_{- 5.3}$ & \hspace{-0.82cm} & $^{(+ 0.7\%)}_{(- 0.7\%)}$ 	& $^{+ 33.4}_{- 32.4}$ & \hspace{-0.82cm} & $^{(+ 4.1\%)}_{(- 4.0\%)}$		& $^{+ 129.3}_{- 127.1}$ & \hspace{-0.82cm} & $^{(+ 16.1\%)}_{(- 15.8\%)}$ \\
	
	NLO(LHCb)	& 	& 124.7	& $^{+ 14.8}_{- 13.7}$ & \hspace{-0.82cm} & $^{(+ 11.8\%)}_{(- 11.0\%)}$	& $^{+ 3.0}_{- 3.0}$ & \hspace{-0.82cm} & $^{(+ 2.4\%)}_{(- 2.4\%)}$		& $^{+ 1.1}_{- 1.1}$ & \hspace{-0.82cm} & $^{(+ 0.9\%)}_{(- 0.9\%)}$ 	& $^{+ 5.5}_{- 5.3}$ & \hspace{-0.82cm} & $^{(+ 4.4\%)}_{(- 4.3\%)}$		& $^{+ 21.1}_{- 19.9}$ & \hspace{-0.82cm} & $^{(+ 16.9\%)}_{(- 15.9\%)}$ \\ \hline
	
	NNLO$^*$(inc.)	&	& 953.6	&  $^{+ 22.7}_{- 33.9}$ & \hspace{-0.82cm} & $^{(+ 2.4\%)}_{(- 3.6\%)}$	& $^{+ 16.2}_{- 17.8}$ & \hspace{-0.82cm} & $^{(+ 1.7\%)}_{(- 1.9\%)}$		& $^{+ 12.8}_{- 12.8}$ & \hspace{-0.82cm} & $^{(+ 1.3\%)}_{(- 1.3\%)}$ 	& $^{+ 39.1}_{- 37.9}$ & \hspace{-0.82cm} & $^{(+ 4.1\%)}_{(- 4.0\%)}$		& $^{+ 66.9}_{- 77.7}$ & \hspace{-0.82cm} & $^{(+ 7.0\%)}_{(- 8.1\%)}$ \\

	NLO(inc.)	& \hspace{-0.22cm} MSTW \hspace{-0.22cm}	& 885.6	& $^{+ 107.2}_{- 105.7}$ & \hspace{-0.82cm} & $^{(+ 12.1\%)}_{(- 11.9\%)}$	& $^{+ 16.0}_{- 19.4}$ & \hspace{-0.82cm} & $^{(+ 1.8\%)}_{(- 2.2\%)}$		& $^{+ 10.1}_{- 10.1}$ & \hspace{-0.82cm} & $^{(+ 1.1\%)}_{(- 1.1\%)}$ 	& $^{+ 36.2}_{- 35.3}$ & \hspace{-0.82cm} & $^{(+ 4.1\%)}_{(- 4.0\%)}$		& $^{+ 148.1}_{- 147.3}$ & \hspace{-0.82cm} & $^{(+ 16.7\%)}_{(- 16.6\%)}$ \\
	
	NLO(LHCb)	& 	& 144.4		& $^{+ 18.6}_{- 17.8}$ & \hspace{-0.82cm} & $^{(+ 12.8\%)}_{(- 12.3\%)}$	& $^{+ 3.5}_{- 3.9}$ & \hspace{-0.82cm} & $^{(+ 2.4\%)}_{(- 2.7\%)}$		& $^{+ 1.9}_{- 1.9}$ & \hspace{-0.82cm} & $^{(+ 1.3\%)}_{(- 1.3\%)}$ 	& $^{+ 6.2}_{- 6.1}$ & \hspace{-0.82cm} & $^{(+ 4.3\%)}_{(- 4.2\%)}$		& $^{+ 25.9}_{- 25.2}$ & \hspace{-0.82cm} & $^{(+ 18.0\%)}_{(- 17.5\%)}$ \\ \hline
	
	NNLO$^*$(inc.) &	& 977.5	&  $^{+ 23.6}_{- 35.4}$ & \hspace{-0.82cm} & $^{(+ 2.4\%)}_{(- 3.6\%)}$	& $^{+ 16.4}_{- 16.4}$ & \hspace{-0.82cm} & $^{(+ 1.7\%)}_{(- 1.7\%)}$		& $^{+ 12.2}_{- 12.2}$ & \hspace{-0.82cm} & $^{(+ 1.3\%)}_{(- 1.3\%)}$ 	& $^{+ 40.4}_{- 39.1}$ & \hspace{-0.82cm} & $^{(+ 4.1\%)}_{(- 4.0\%)}$		& $^{+ 68.9}_{- 80.0}$ & \hspace{-0.82cm} & $^{(+ 7.0\%)}_{(- 8.1\%)}$ \\

	NLO(inc.) & \hspace{-0.42cm} NNPDF \hspace{-0.42cm}	& 894.5	& $^{+ 107.6}_{- 101.0}$ & \hspace{-0.82cm} & $^{(+ 12.0\%)}_{(- 11.3\%)}$	& $^{+ 12.8}_{- 12.8}$ & \hspace{-0.82cm} & $^{(+ 1.4\%)}_{(- 1.4\%)}$		& $^{+ 9.9}_{- 9.9}$ & \hspace{-0.82cm} & $^{(+ 1.1\%)}_{(- 1.1\%)}$ 	& $^{+ 36.6}_{- 35.8}$ & \hspace{-0.82cm} & $^{(+ 4.1\%)}_{(- 4.0\%)}$		& $^{+ 147.6}_{- 140.3}$ & \hspace{-0.82cm} & $^{(+ 16.5\%)}_{(- 15.7\%)}$ \\
	
	NLO(LHCb) 	& 	& 142.5	& $^{+ 18.1}_{- 16.6}$ & \hspace{-0.82cm} & $^{(+ 12.7\%)}_{(- 11.7\%)}$	& $^{+ 3.0}_{- 3.0}$ & \hspace{-0.82cm} & $^{(+ 2.1\%)}_{(- 2.1\%)}$		& $^{+ 2.0}_{- 2.0}$ & \hspace{-0.82cm} & $^{(+ 1.4\%)}_{(- 1.4\%)}$ 	& $^{+ 6.2}_{- 6.1}$ & \hspace{-0.82cm} & $^{(+ 4.4\%)}_{(- 4.3\%)}$		& $^{+ 25.2}_{- 23.7}$ & \hspace{-0.82cm} & $^{(+ 17.7\%)}_{(- 16.6\%)}$ \\ \hline

	\end{tabular}
	\caption{Summary of inclusive (inc.) and differential (LHCb) cross-sections at NNLO+NLLL (NNLO$^*$) and NLO accuracy and associated theoretical uncertainties at 14~TeV, for PDF sets as described in the text.}
	\label{Sigma14}
\end{footnotesize}
\end{table}

The enhanced sensitivity of measurements at high pseudorapidity can be seen by comparing the relative uncertainties for the inclusive and differential LHCb cross-sections. This comparison is done by taking the ratio of their relative uncertainties,
\begin{equation} 
\delta^{\mathrm{ratio}}_X = \frac{\delta^{\mathrm{LHCb}}_X}{\delta^{NLO}_X},
\end{equation}

\noindent which highlights the sensitivity of measurements at LHCb to PDF uncertainties, in particular to those sets provided by NNPDF and CT10. The results are summarised in Tables~\ref{SigmaRatio7} and \ref{SigmaRatio14} for 7 and 14 TeV respectively.

\begin{table}[htdp]
\begin{small}
\centering
	\begin{tabular}{r | c | c | c | c | c }
	PDF 	& $\delta^{\mathrm{ratio}}_{\mathrm{scale}}$	& $\delta^{\mathrm{ratio}}_{\mathrm{PDF}}$	& $\delta^{\mathrm{ratio}}_{\alpha_s}$ & $\delta^{\mathrm{ratio}}_{m_t}$	& $\delta^{\mathrm{ratio}}_{\mathrm{total}}$\\ \hline
	ABM		& $^{+ 1.05}_{- 1.00}$	& $^{+ 1.11}_{- 1.11}$	& $^{+ 0.00}_{- 0.00}$	& $^{+ 1.06}_{- 1.06}$	& $^{+ 1.05}_{- 1.02}$	\\
	CT10	& $^{+ 1.12}_{- 1.06}$	& $^{+ 1.56}_{- 1.30}$	& $^{+ 1.23}_{- 1.23}$	& $^{+ 1.07}_{- 1.07}$	& $^{+ 1.19}_{- 1.10}$	\\
	HERA	& $^{+ 1.07}_{- 1.01}$	& $^{+ 1.01}_{- 0.65}$	& $^{+ 1.25}_{- 1.25}$	& $^{+ 1.05}_{- 1.06}$	& $^{+ 1.06}_{- 1.00}$	\\  
	MSTW	& $^{+ 1.12}_{- 1.06}$	& $^{+ 1.27}_{- 1.23}$	& $^{+ 1.13}_{- 1.13}$	& $^{+ 1.06}_{- 1.08}$	& $^{+ 1.12}_{- 1.08}$	\\ 
	NNPDF	& $^{+ 1.13}_{- 1.05}$	& $^{+ 1.34}_{- 1.34}$	& $^{+ 1.21}_{- 1.21}$	& $^{+ 1.07}_{- 1.07}$	& $^{+ 1.13}_{- 1.08}$	\\ 
	\end{tabular}
	\caption{Ratio of relative uncertainties at 7~TeV between LHCb/inclusive cross-sections at NLO.}
	\label{SigmaRatio7}
\end{small}
\end{table}

\begin{table}[htdp]
\begin{small}
\centering
	\begin{tabular}{r | c | c | c | c | c }
	PDF 	& $\delta^{\mathrm{ratio}}_{\mathrm{scale}}$	& $\delta^{\mathrm{ratio}}_{\mathrm{PDF}}$	& $\delta^{\mathrm{ratio}}_{\alpha_s}$ & $\delta^{\mathrm{ratio}}_{m_t}$	& $\delta^{\mathrm{ratio}}_{\mathrm{total}}$\\ \hline
	ABM		& $^{+ 1.05}_{- 1.00}$	& $^{+ 1.40}_{- 1.40}$	& $^{+ 0.00}_{- 0.00}$	& $^{+ 1.05}_{- 1.05}$	& $^{+ 1.06}_{- 1.02}$	\\
	CT10	& $^{+ 1.05}_{- 1.03}$	& $^{+ 1.55}_{- 1.40}$	& $^{+ 1.20}_{- 1.20}$	& $^{+ 1.06}_{- 1.05}$	& $^{+ 1.09}_{- 1.07}$	\\
	HERA	& $^{+ 1.04}_{- 1.01}$	& $^{+ 1.19}_{- 0.90}$	& $^{+ 1.33}_{- 1.33}$	& $^{+ 1.07}_{- 1.06}$	& $^{+ 1.05}_{- 1.01}$	\\
	MSTW	& $^{+ 1.06}_{- 1.03}$	& $^{+ 1.35}_{- 1.23}$	& $^{+ 1.13}_{- 1.13}$	& $^{+ 1.05}_{- 1.06}$	& $^{+ 1.07}_{- 1.05}$	\\
	NNPDF	& $^{+ 1.05}_{- 1.03}$	& $^{+ 1.45}_{- 1.45}$	& $^{+ 1.27}_{- 1.27}$	& $^{+ 1.07}_{- 1.07}$	& $^{+ 1.07}_{- 1.06}$	\\
	\end{tabular}
	\caption{Ratio of relative uncertainties at 14~TeV LHCb/inclusive cross-sections at NLO}
	\label{SigmaRatio14}
\end{small}
\end{table}

It is noted that the central value prediction from ABM is substantially lower than the other predictions for differential and inclusive NLO, and NNLO results. At NNLO this can be understood from both a lower value for $\alpha_s(M_Z)$ and a softer gluon PDF at large-$x$~\cite{Ball:2012wy,Czakon:2013tha}. At NLO, even for identical best fit value $\alpha_s(M_Z)$, the prediction from ABM is substantially lower than CT10 as shown in Fig.~\ref{fig:alphas}. In fact, the discrepancy between the central value of ABM and the other predictions is enhanced at high rapidity as a result of the soft large-$x$ gluon PDF. The predictions from different eigenvectors were found to be very stable, with the exception of members 10 and 13, resulting in small PDF uncertainty. Although the PDF uncertainty is small, including LHCb $t\bar{t}$ data in a PDF fit will impact the central value of the gluon PDF in the large-$x$ region.

\begin{figure}
\centering
\begin{tabular}{cc}
\includegraphics[width=.48\linewidth]{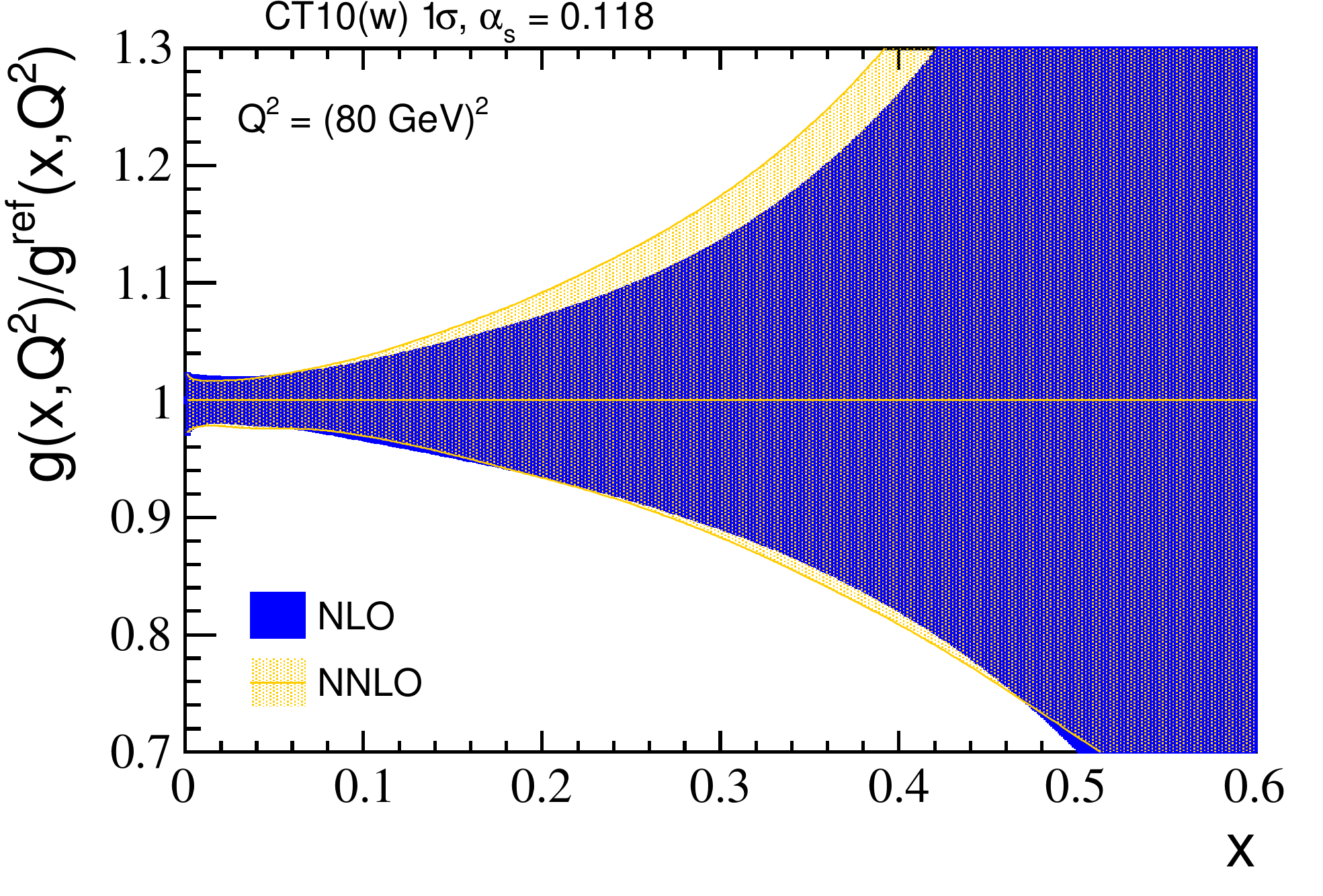} &
\includegraphics[width=.48\linewidth]{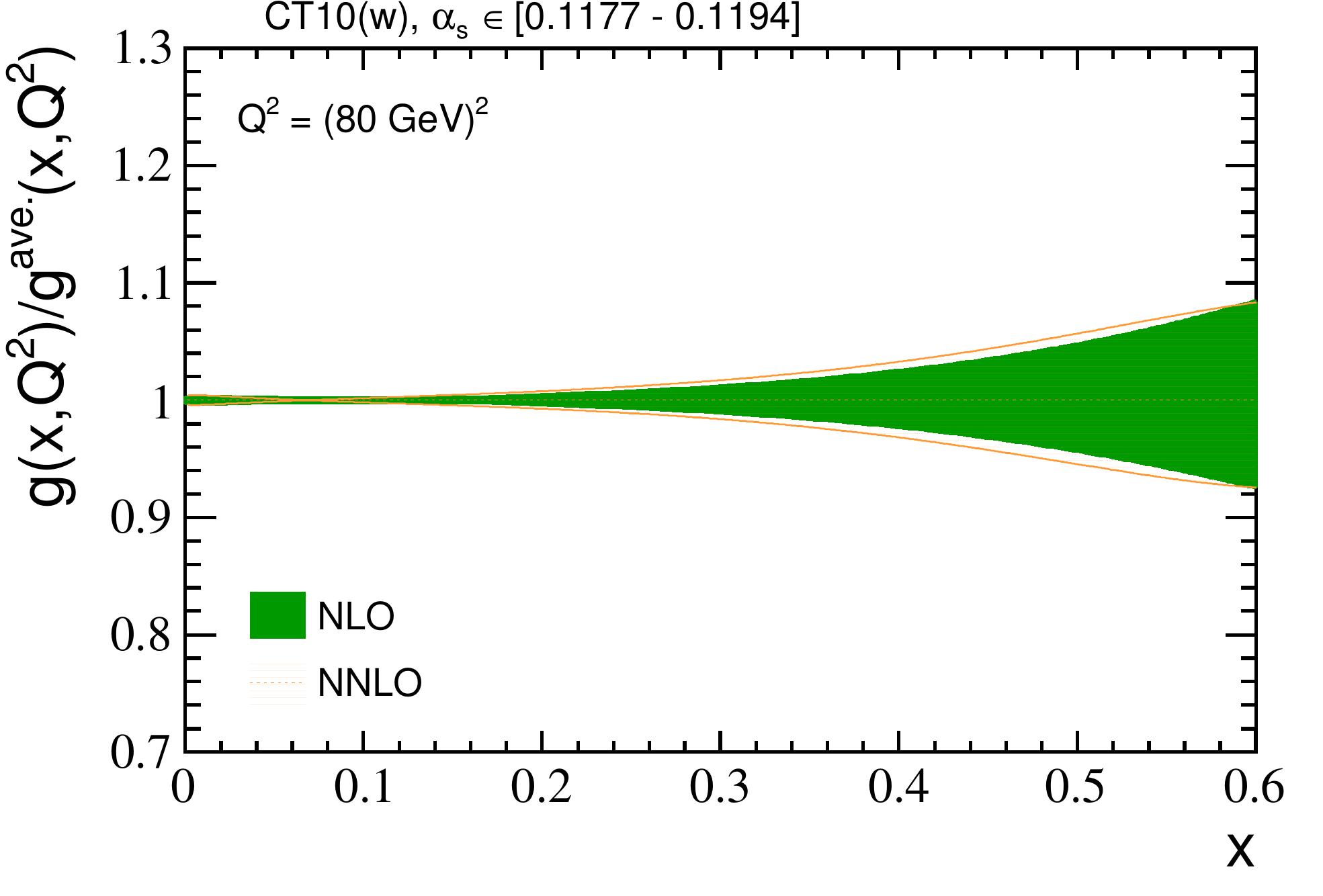} \\
\end{tabular}
\caption{Relative uncertainty on the gluon PDF for CT10 NLO, NNLO sets for PDF and $\alpha_s$ variations.}\label{gPDF}
\end{figure}

At NLO the contribution from the scale variation to the total uncertainty is dominant. However, given the recent theoretical advances in pair production predictions, it is clear that a cross-section measurement in the forward region can be used to constrain the gluon PDF description at high-$x$. It is expected that the observed large ratio of the relative PDF uncertainties between inclusive and LHCb measurements is still present at NNLO. This can be seen by comparing the relative uncertainty on the gluon PDF as function of $x$ for both CT10 NLO and NNLO sets for $\delta_{\mathrm{PDF}}$ (left) and $\delta_{\alpha_s}$ (right) as shown in Fig.~\ref{gPDF}. The uncertainties at NLO and NNLO are of comparable size.

\section{Constraining the gluon PDF}
\begin{figure}[h]
\centering
\begin{tabular}{c}
\includegraphics[width=.65\linewidth]{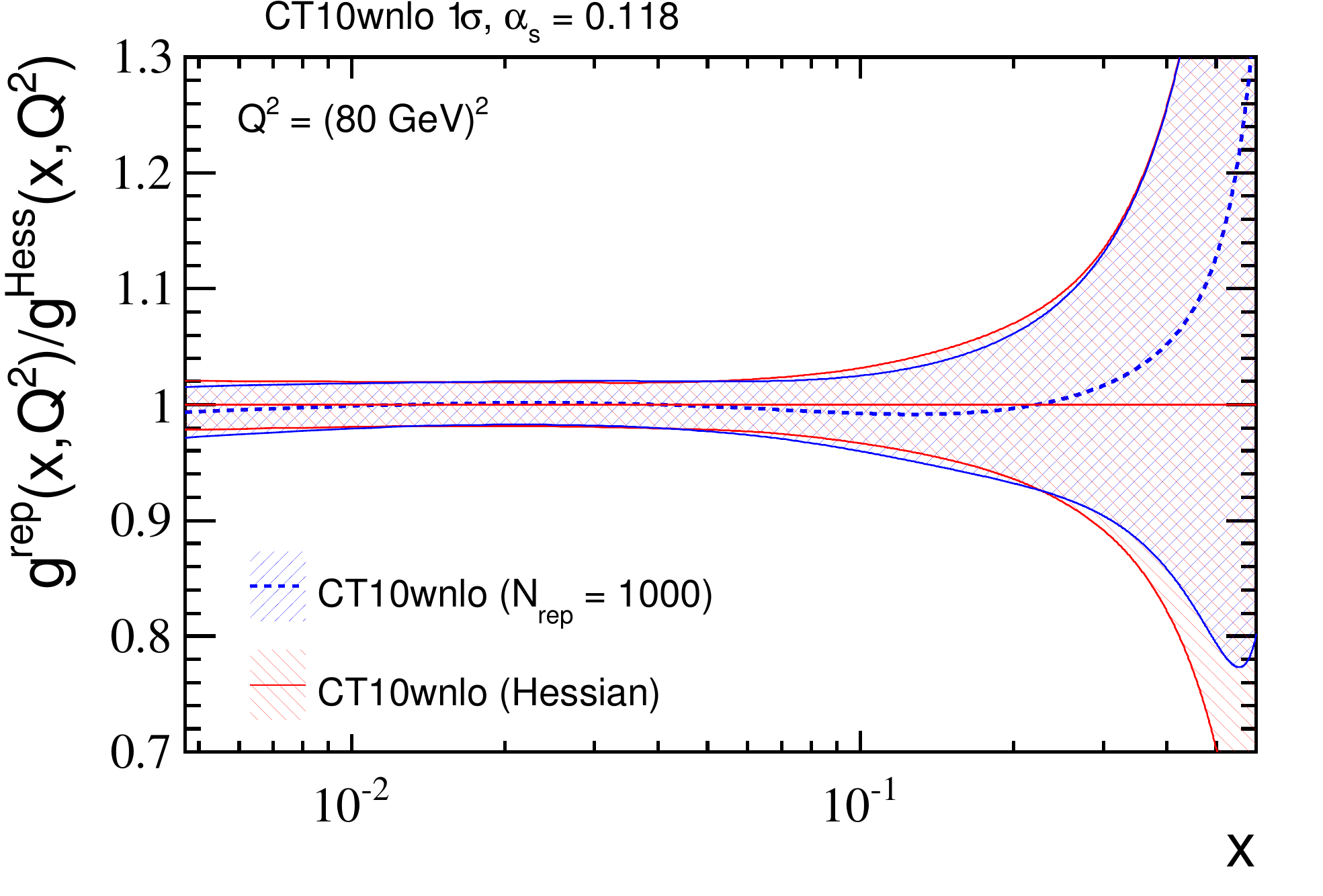}
\end{tabular}
\caption{Relative uncertainty on CT10wnlo gluon PDF, $g(x,Q^2)$, for replica and Hessian set with respect to the central Hessian member.}
\label{fig:Hessian}
\end{figure}

Due to the high statistical precision expected within 1 year of running (5~fb$^{-1}$) at 14~TeV, a differential measurement in bins of pseudorapidity across the entire LHCb acceptance is viable. To demonstrate the potential power of such a measurement on constraining the gluon PDF, we apply a reweighting to the CT10 and NNPDF sets based on a hypothetical measurement of $\sigma^{\mathrm{LHCb}}$. This is done following the prescriptions of Ref.~\cite{Ball:2010gb,DeLorenzi:2010zt,Ball:2011gg,Watt:2012tq} where a Bayesian method based on statistical inference is used. The procedure is easily performed for the NNPDF Monte Carlo sets, while for CT10 (the Hessian set) it is necessary to first generate a set of random PDFs from the eigenvector set. This is done working in the basis of observables, $\{X_0(\mathcal{S}_0),X^-_1(\mathcal{S}^-_1),X^+_1(\mathcal{S}^+_1),...X^-_N(\mathcal{S}^-_N),X^+_N(\mathcal{S}^+_N)\}$, spanning the $N$ eigenvectors. Hypothetical and random observables are generated as:

\beq
\bar{X}_0 = \frac{1}{N_{\mathrm{rep}}}\sum^{N_{\mathrm{rep}}}_{k=1}X_0(\mathcal{S}_0)[1+R_{k0}],\hspace{0.2cm} \hspace{0.2cm} X(\mathcal{S}_k) = X(\mathcal{S}_0)+\sum^N_{j=1}[X(\mathcal{S}_j^{\pm})-X(\mathcal{S}_0)]|R_{kj}|
\eeq

\noindent where $R_{kj}$ is a random gaussian-distributed number with zero mean and variance of one. The choice of negative or positive displacements $S_j^-$ or $S_j^+$ depends on the sign of $R_{kj}$. For the generated CT10 and NNPDF sets studied, the number of replicas are 1000 and 100 respectively. This procedure is applied to the evolved gluon PDF $g(x,Q^2)$ for CT10 and then compared to the Hessian result in Fig.~\ref{fig:Hessian}, where the relative uncertainty for the replica and Hessian set is plotted with respect to the Hessian central value. The difference between the two sets occurs for large $x$ where the PDF uncertainties are most asymmetric (see also Fig.~\ref{gPDF}). It is re-assuring that the two parameterisations are in very good agreement.

By storing the set of random numbers $R_{kj}$ generated in producing the replica set, it is possible to then generate an equivalent set of observables at the level of $\sigma^{\mathrm{LHCb}}$. From these sets of random observables $\sigma^{\mathrm{LHCb}}(\mathcal{S}_k)$ a reweighting can be performed by computing the $\chi^2_k$ with respect to $\bar{\sigma}^{\mathrm{LHCb}}_0$, assuming an experimental uncertainty in the range 4-8\%. The relevant formulas are:

\begin{equation}
\langle X \rangle_{\mathrm{old}} = \frac{1}{N_{\mathrm{rep}}} \sum_{k=1}^{N_{\mathrm{rep}}} X(\mathcal{S}_k) \mathrm{, \hspace{0.5cm}} \langle X \rangle_{\mathrm{new}} = \frac{1}{N_{\mathrm{rep}}} \sum_{k=1}^{N_{\mathrm{rep}}} w_k(\chi^2_k) X(\mathcal{S}_k) \mathrm{,}
\end{equation}

\noindent where the weights are computed as

\beq
w_k(\chi^2_k) = \frac{ W_k(\chi^2_k) }{\frac{1}{N_{\mathrm{rep}}} \sum_{j=1}^{N_{\mathrm{rep}}} W_j(\chi^2_k)} \mathrm{, \hspace{0.5cm}} W_k(\chi^2_k) = (\chi^2_k)^{\frac{1}{2}(N_{\mathrm{pts.}}-1)} \mathrm{exp}(-\frac{1}{2}\chi^2_k) \mathrm{,}
\eeq

\noindent and the dominator fixing the normalisation is,

\beq
\sum_{k=1}^{N_{\mathrm{rep}}} w_k(\chi^2_k) = N_{\mathrm{rep}}.
\eeq

\noindent 
After applying this reweighting technique, the number of effective remaining replicas can be found after calculating the Shannon entropy as,
\beq
N_{\mathrm{eff}} = \mathrm{exp}\left(\frac{1}{N_{\mathrm{rep}}}\sum_{k=1}^{N_{\mathrm{rep}}}w_k\mathrm{log}(N_{\mathrm{rep}}/w_k)\right).
\eeq
The effective number of replicas after having applied this reweighting technique to the random NNPDF ($N_{\mathrm{rep}} =$ 100) and CT10 ($N_{\mathrm{rep}} =$ 1000) sets for different experimental uncertainties are provided in Table~\ref{neff}.

\begin{table}[htdp]
\begin{small}
\centering
	\begin{tabular}{r | c | c | c | c }
	PDF 		& 	$\bar{\sigma}_0^{\mathrm{LHCb}}$(pb)	& Exp. uncertainty	& $N_{\mathrm{rep}}$	& $N_{\mathrm{eff}}$ \\ \hline
			& 							& 4\%			& 1000		& 942		\\
	CT10	& 	137.3					& 6\%			& 1000		& 983		\\
			& 							& 8\%			& 1000		& 994		\\ \hline
			& 							& 4\%			& 100		& 97			\\
	NNPDF	& 	145.1					& 6\%			& 100		& 99			\\
			& 							& 8\%			& 100		& 100		\\			
	\end{tabular}
	\caption{Effective replicas after reweighting with the inclusion of an LHCb semi-inclusive measurement, the associated experimental  uncertainty is within the range 4-8\%.}
	\label{neff}
\end{small}
\end{table}

\begin{figure}
\centering
\begin{tabular}{cc}
\includegraphics[width=.48\linewidth]{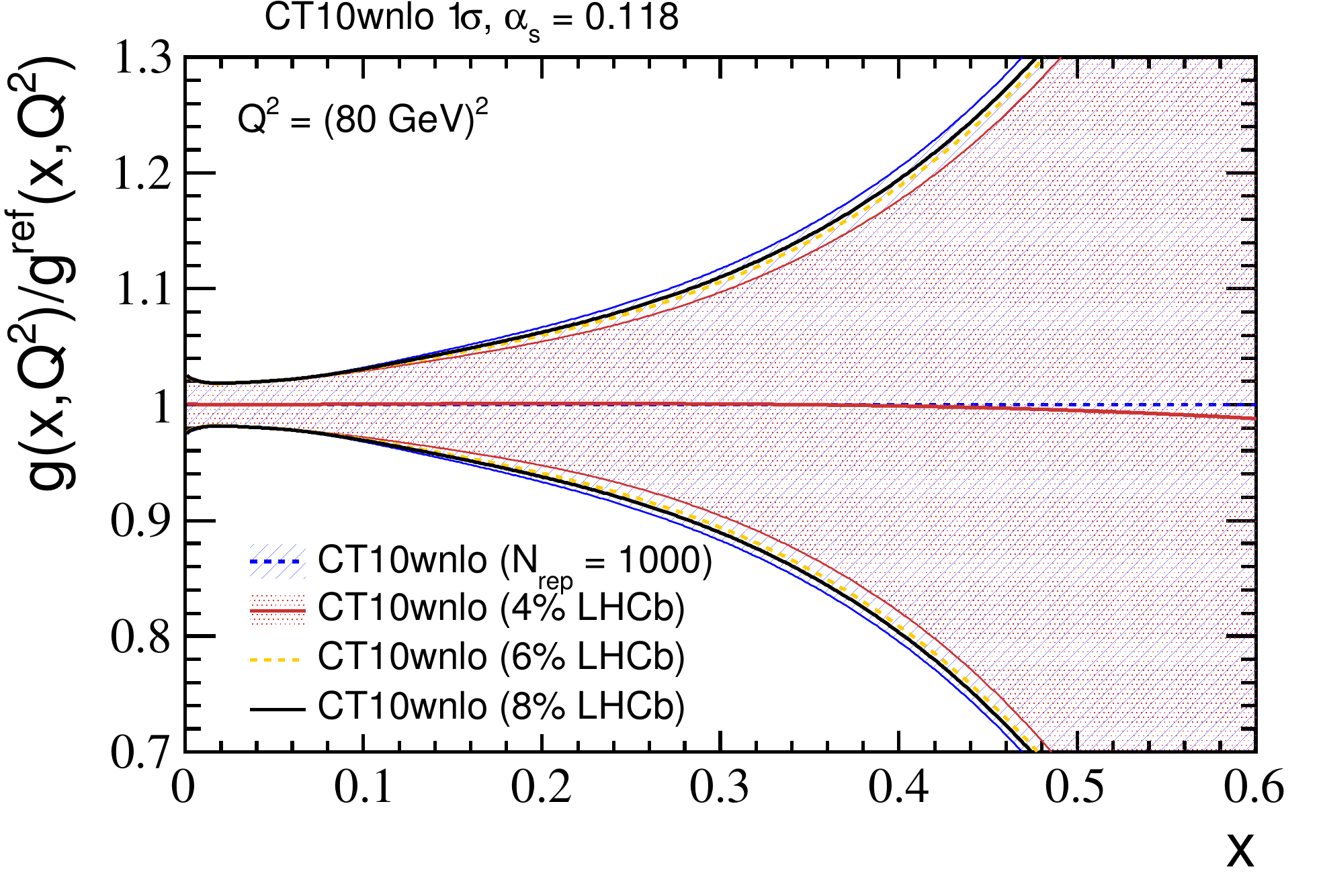} &
\includegraphics[width=.48\linewidth]{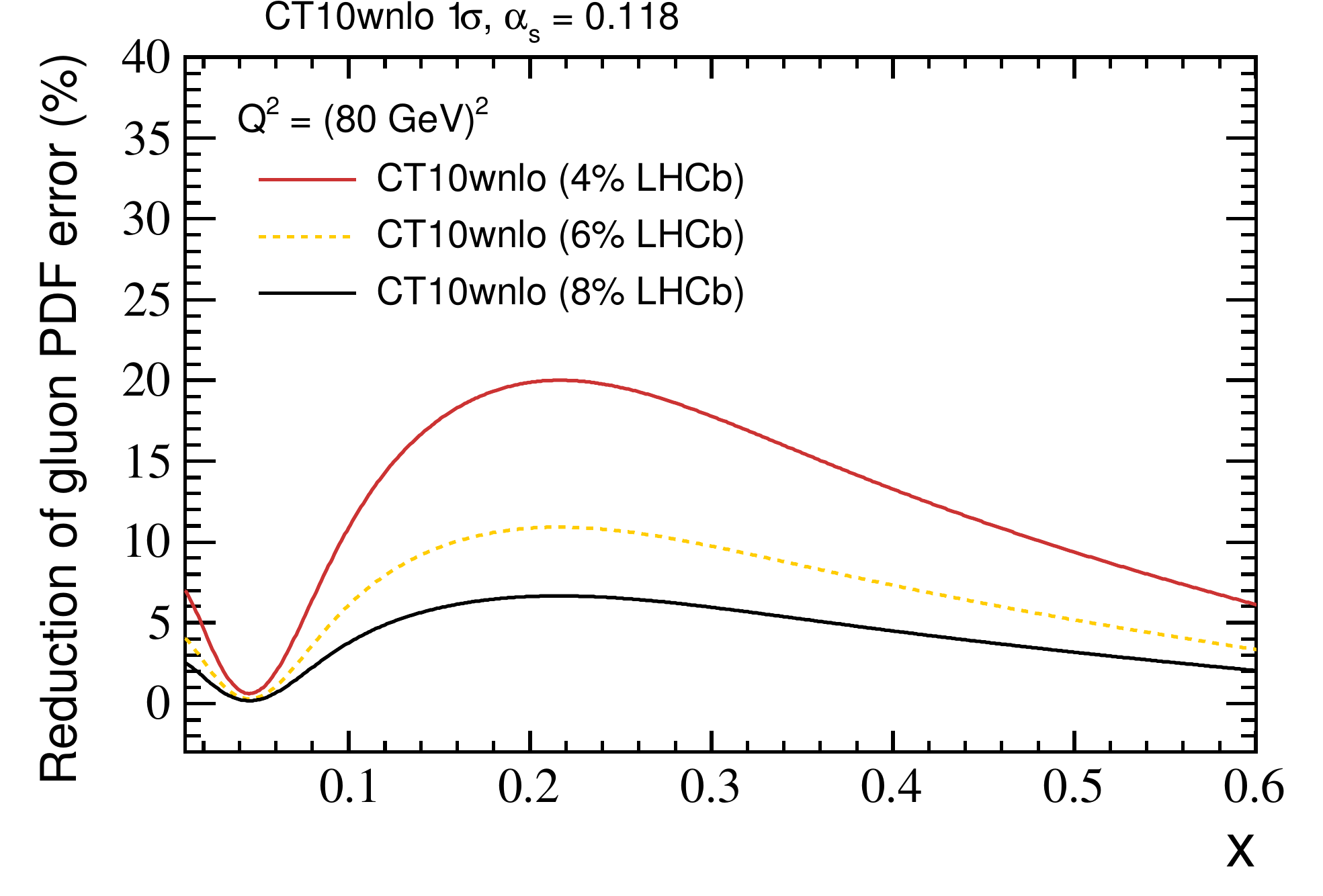} \\
\includegraphics[width=.48\linewidth]{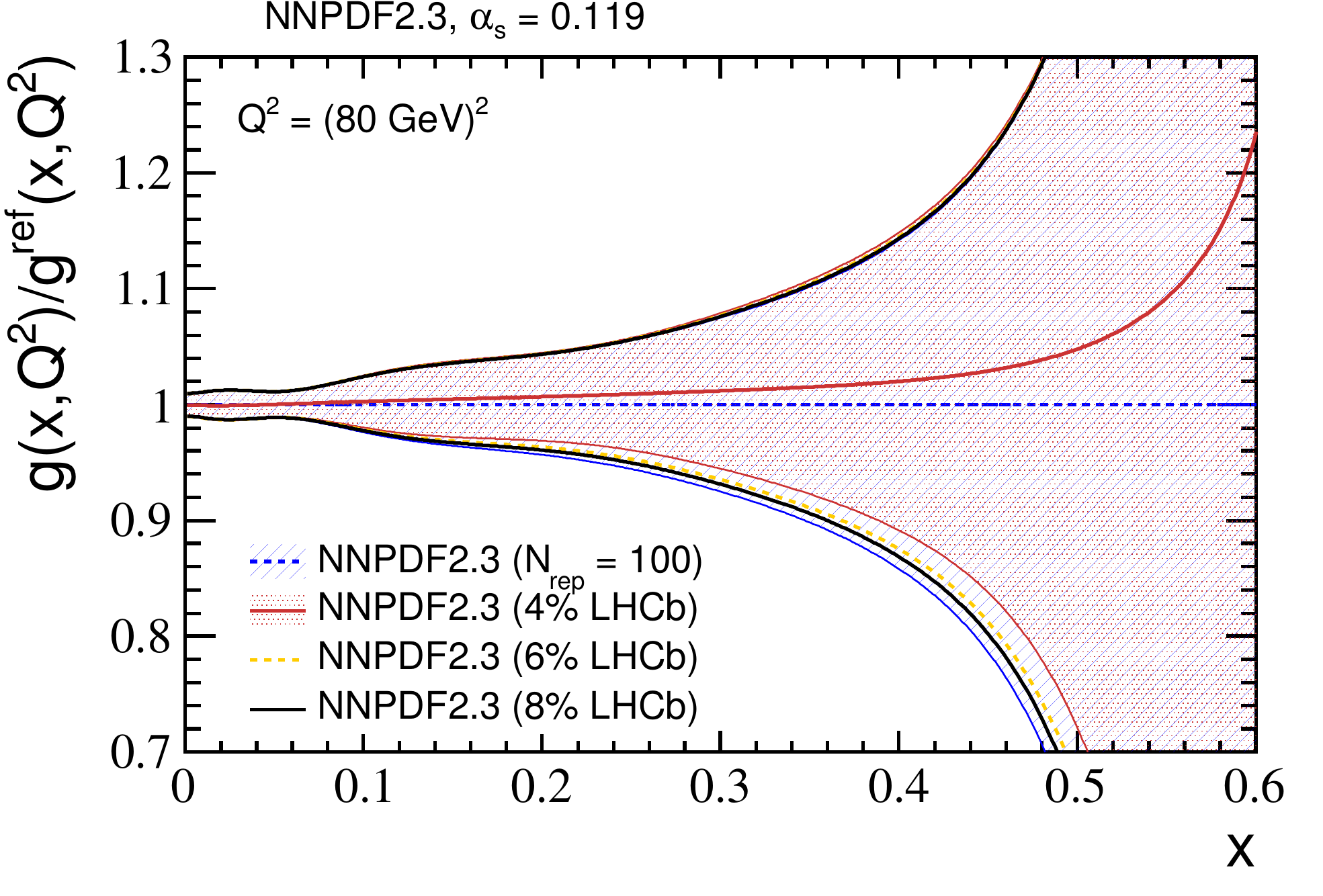} &
\includegraphics[width=.48\linewidth]{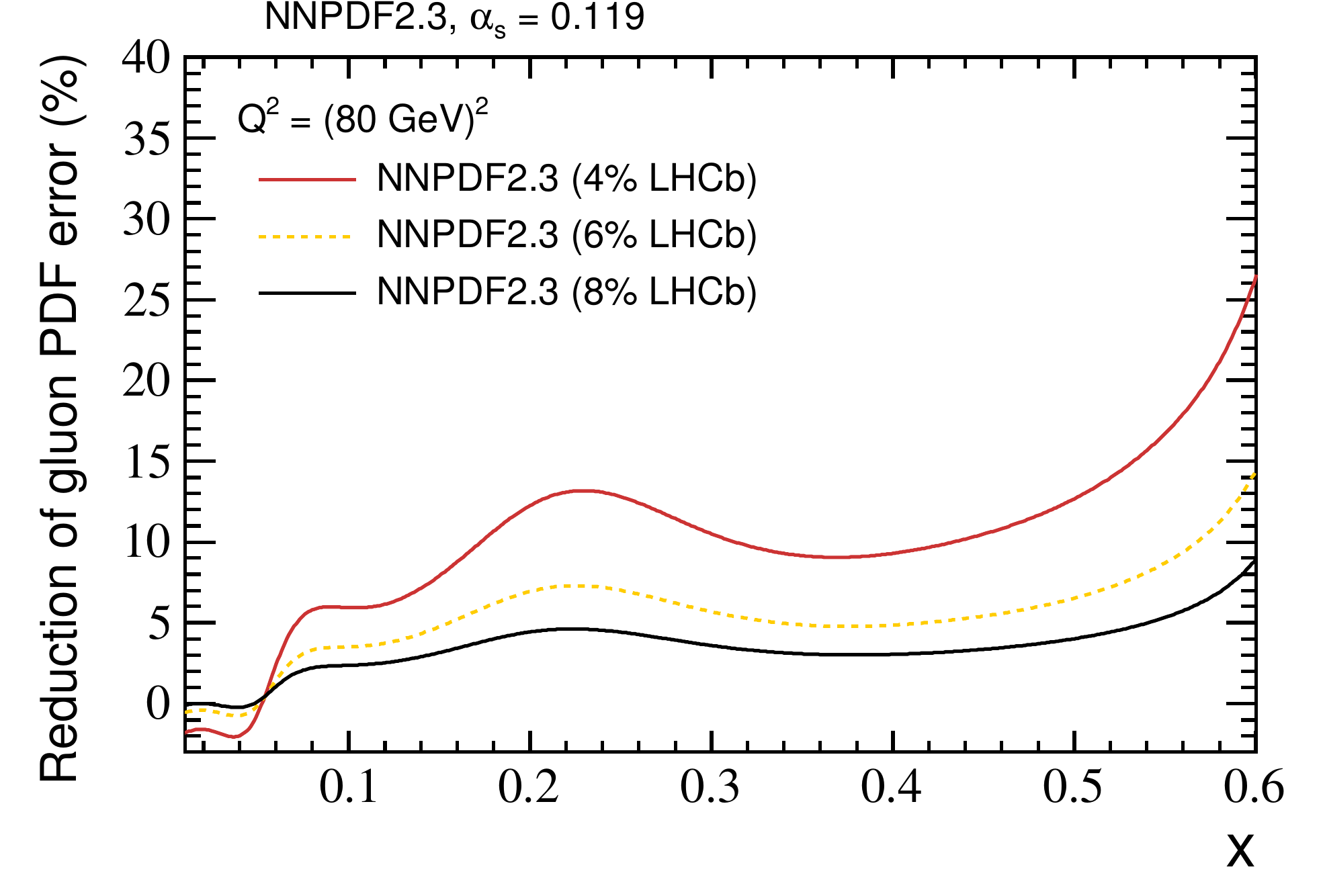} \\
\end{tabular}
\caption{Potential constraint (left) on gluon PDF for CT10wnlo (upper) NNPDF2.3 (lower) with the inclusion of an LHCb semi-inclusive measurement with an associated uncertainty of 4-8\%. The corresponding reduction of the PDF uncertainty is also plotted for assumed uncertainties of 4, 6, 8\%(right).}\label{fig:gluonPDF}
\end{figure}

The effect of this reweighting on the evolved gluon PDF is presented in Fig. \ref{fig:gluonPDF} for CT10 (upper) and NNPDF (lower) replica sets. The reduction of the relative uncertainty of the gluon PDF with respect to the average of the unweighted PDF, $g^{\mathrm{ref}}(x,Q^2$), is plotted (Fig. \ref{fig:gluonPDF}, left) assuming experimental uncertainties of 4, 6, and 8\% of the pseudoata $\bar{\sigma}_0^{\mathrm{LHCb}}$. The reduction of the gluon PDF uncertainty for the same range of experimental uncertainties are also plotted (right).

\begin{figure}
\centering
\begin{tabular}{cc}
\includegraphics[width=.48\linewidth]{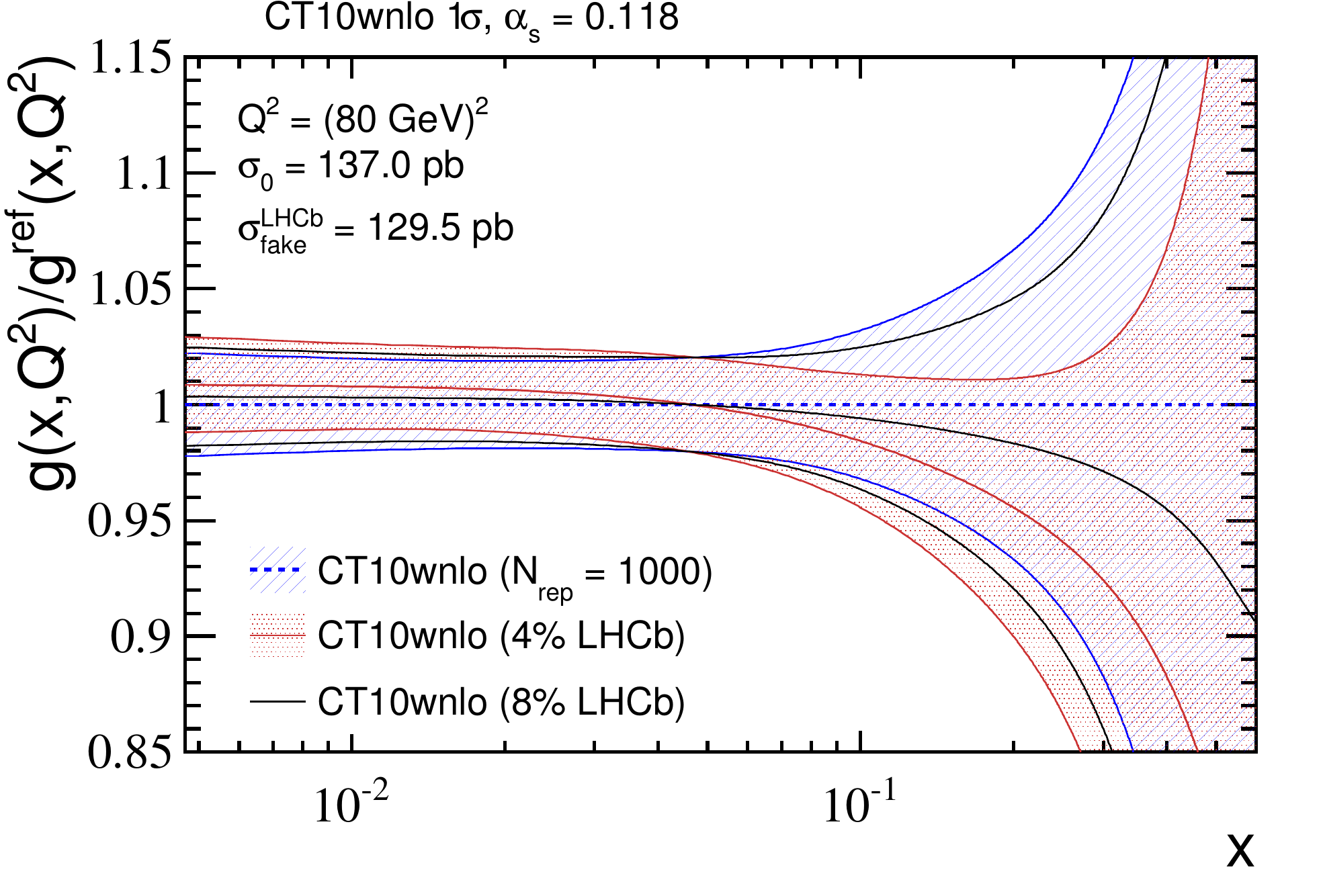} &
\includegraphics[width=.48\linewidth]{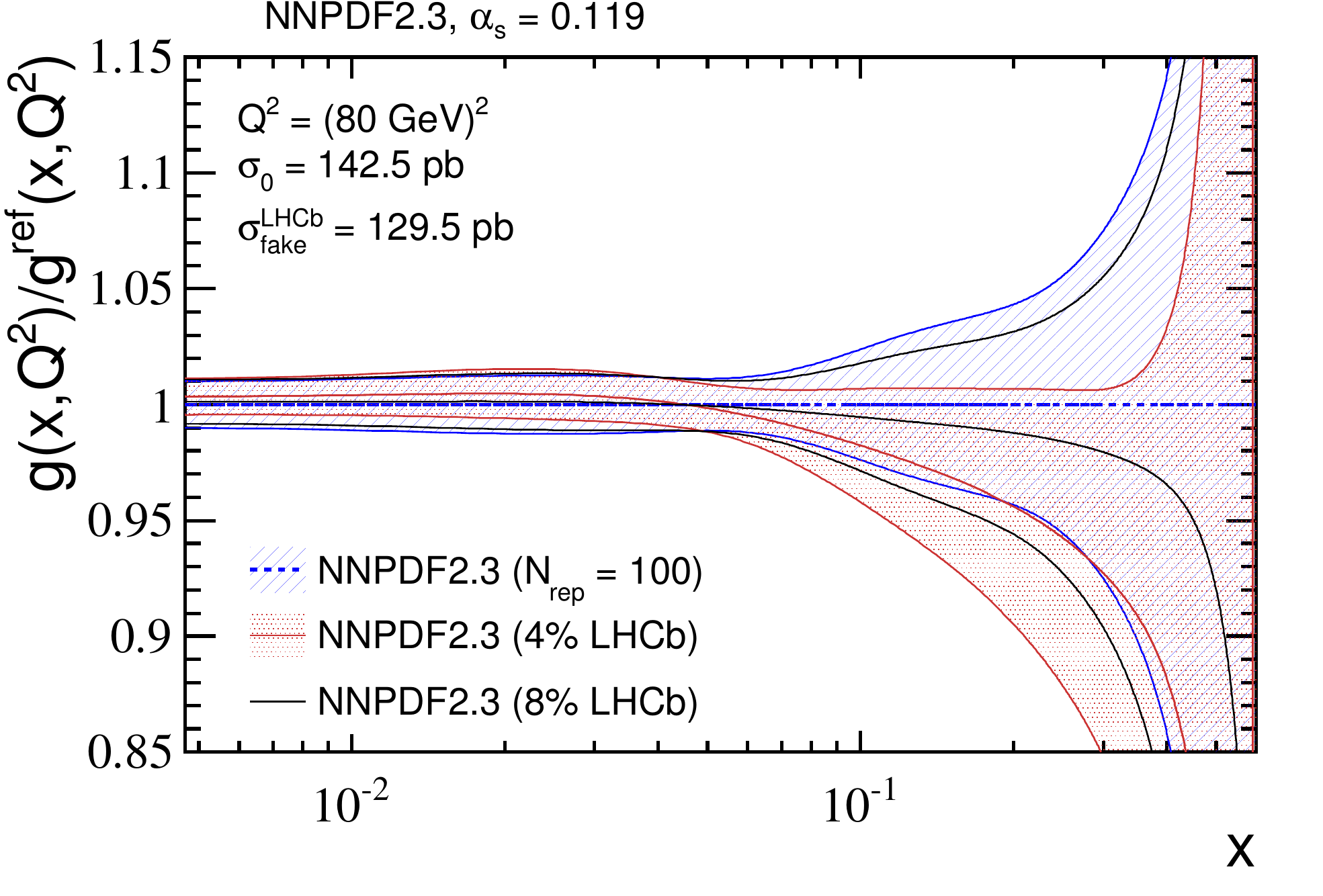} \\
\end{tabular}
\caption{Potential shift in central value of the evolved gluon PDF for CT10wnlo (left) and NNPDF2.3 (right) after reweighting with respect to pseudodata generated from HERA1.5. The 1$\sigma$ CL uncertainties are also included.}\label{fig:gluonShift}
\end{figure}

The largest sensitivity lies within the range of $0.1 < x < 0.3$ for 14 TeV pseudodata. The experimental precision achievable at LHCb will therefore have a large impact on future PDF fits within this range. The choice of generating pseudodata from an observable generated from the  central value PDF set is useful for demonstrating the reduction in uncertainty, however the impact on the central value is clearly small by construction. To demonstrate the impact of a measurement at LHCb on the central value, pseudodata is generated from the HERA central value, $\bar{\sigma}_0^{\mathrm{LHCb}} = \sigma_{\mathrm{fake}}^{\mathrm{LHCb}} =$ 129.5~pb,  and then used to reweight the CT10 and NNPDF replica sets. The results are plotted in Fig.~\ref{fig:gluonShift} and the effective number of replicas are also computed and included in Table~\ref{neff2}.

\begin{table}[htdp]
\begin{small}
\centering
	\begin{tabular}{r | c | c | c | c }
	PDF 		& 	$\bar{\sigma}_0^{\mathrm{LHCb}}$(pb)	& Exp. uncertainty	& $N_{\mathrm{rep}}$	& $N_{\mathrm{eff}}$ \\ \hline
			& 							& 4\%			& 1000		& 720		\\
	CT10	& 	129.5					& 6\%			& 1000		& 895		\\
			& 							& 8\%			& 1000		& 956		\\ \hline
			& 							& 4\%			& 100		& 53			\\
	NNPDF	& 	129.5					& 6\%			& 100		& 85			\\
			& 							& 8\%			& 100		& 95		\\			
	\end{tabular}
	\caption{Effective replicas after reweighting with the inclusion of an LHCb semi-inclusive measurement generated from the HERA1.5 central value, the associated experimental uncertainty is within the range 4-8\%.}
	\label{neff2}
\end{small}
\end{table}

For convenience of the PDF collaborations, we list the eigenvectors (and their directions) for all studied asymmetric Hessian sets which have a substanial impact on replicas with large $\chi^2_k$ values. Given that the pseudodata values are centred on the observable calculated from the central Hessian member ($\sigma_0^{\mathrm{LHCb}}$), this can be quantified as
\beq
\Delta X_j^{\pm} = \frac{ X(\mathcal{S}_j^{\pm})-X(\mathcal{S}_0) }{X(\mathcal{S}_0)}.
\eeq

\begin{figure}
\centering
\begin{tabular}{cc}
\includegraphics[width=.48\linewidth]{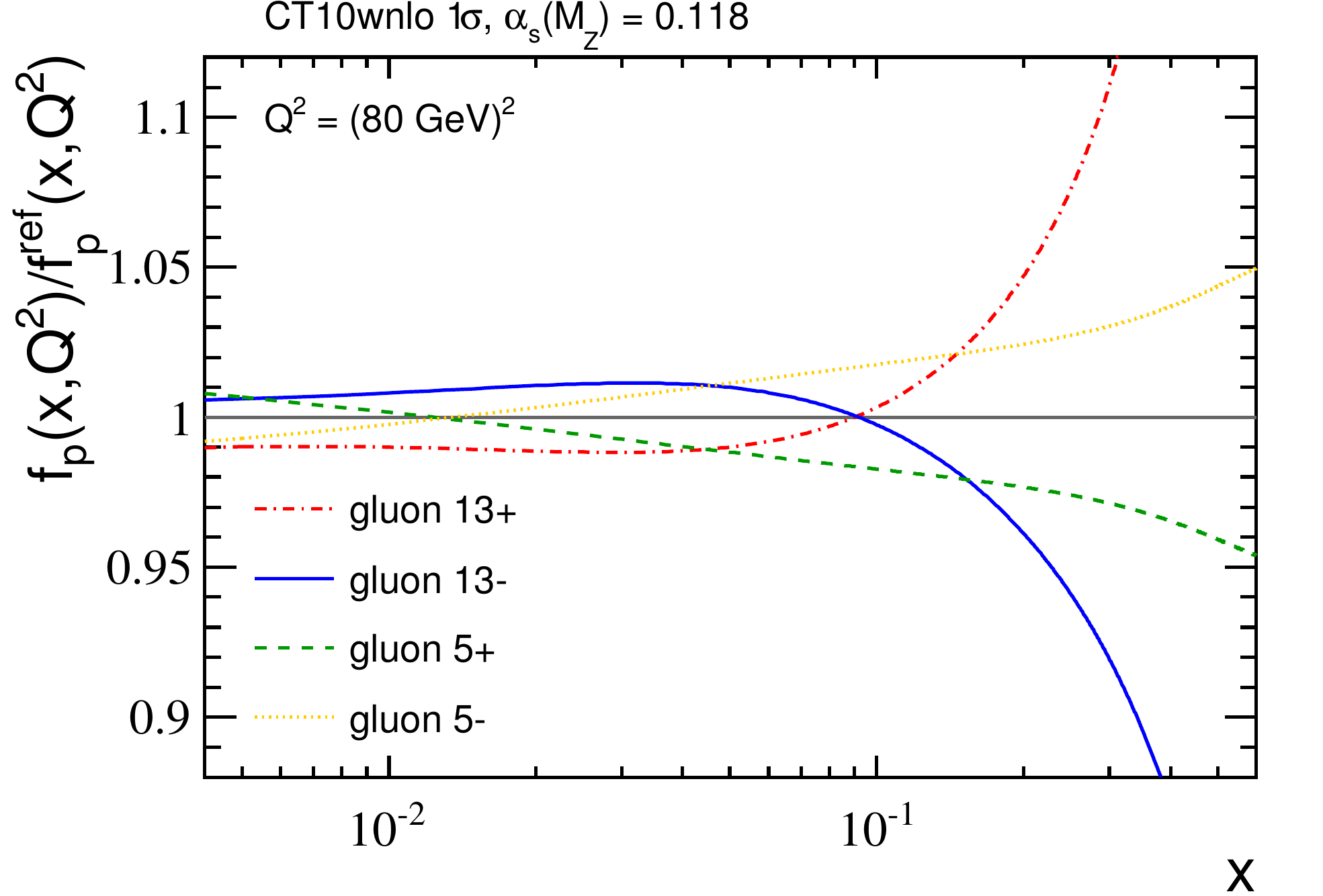} &
\includegraphics[width=.48\linewidth]{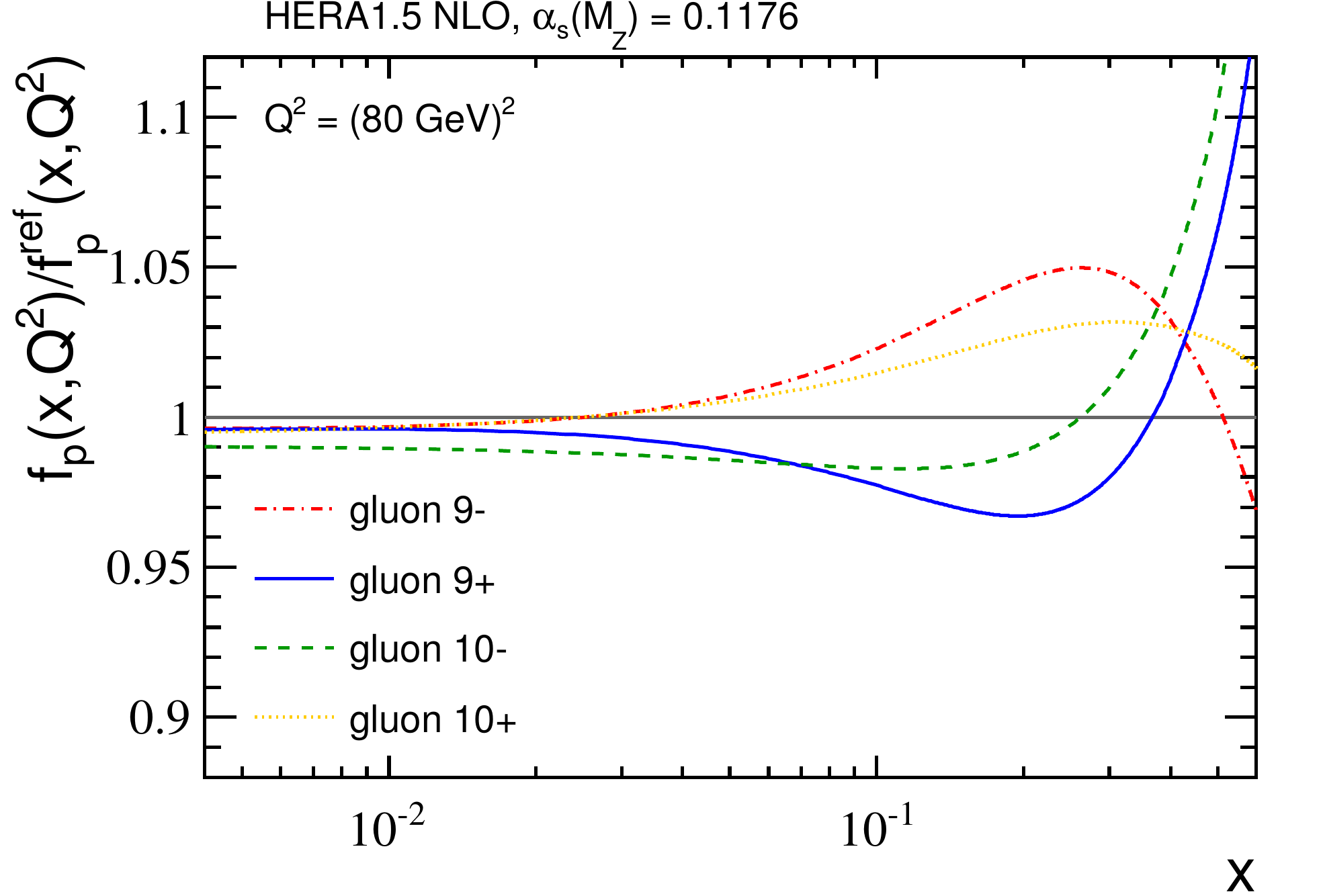} \\
\includegraphics[width=.48\linewidth]{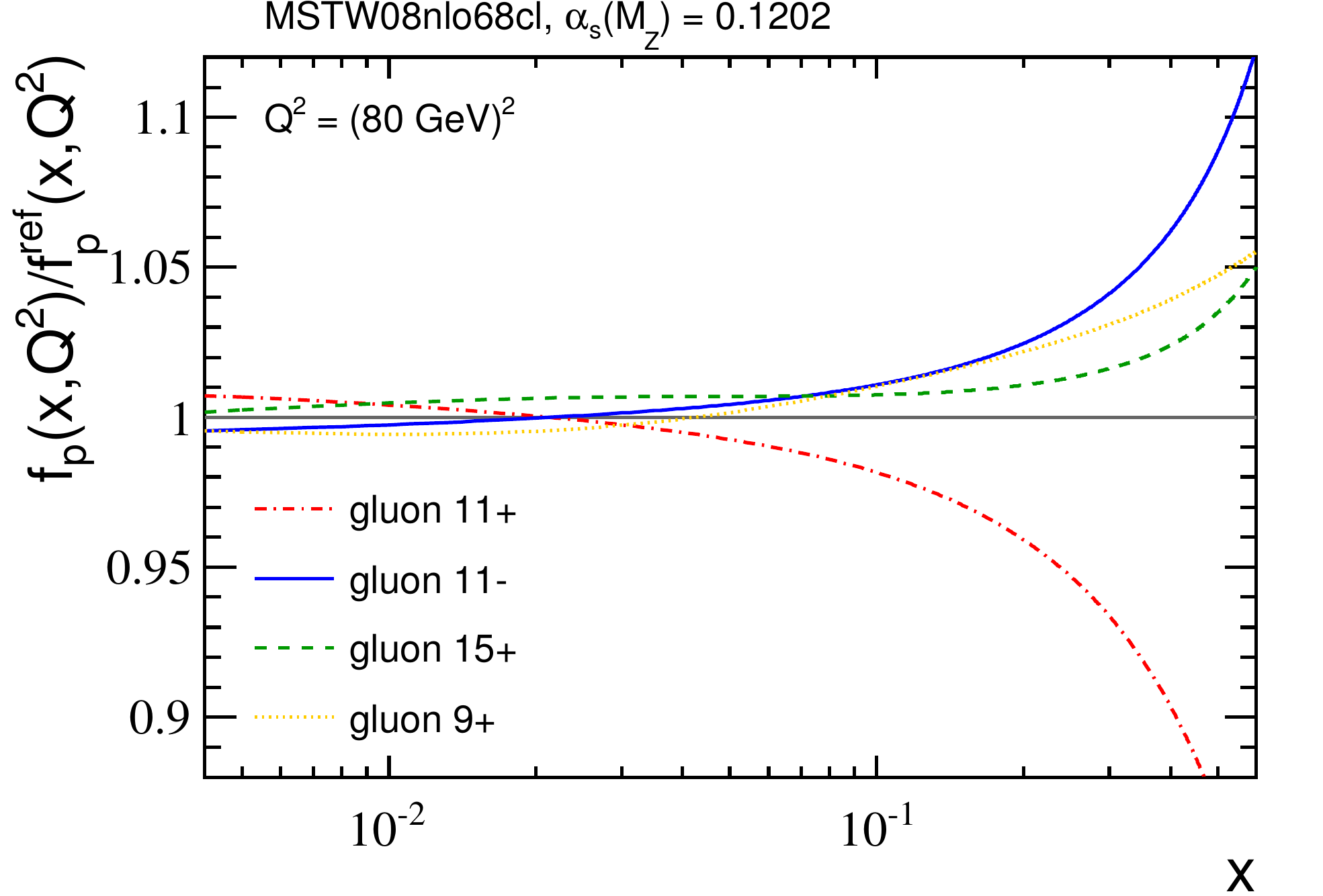} &
\includegraphics[width=.48\linewidth]{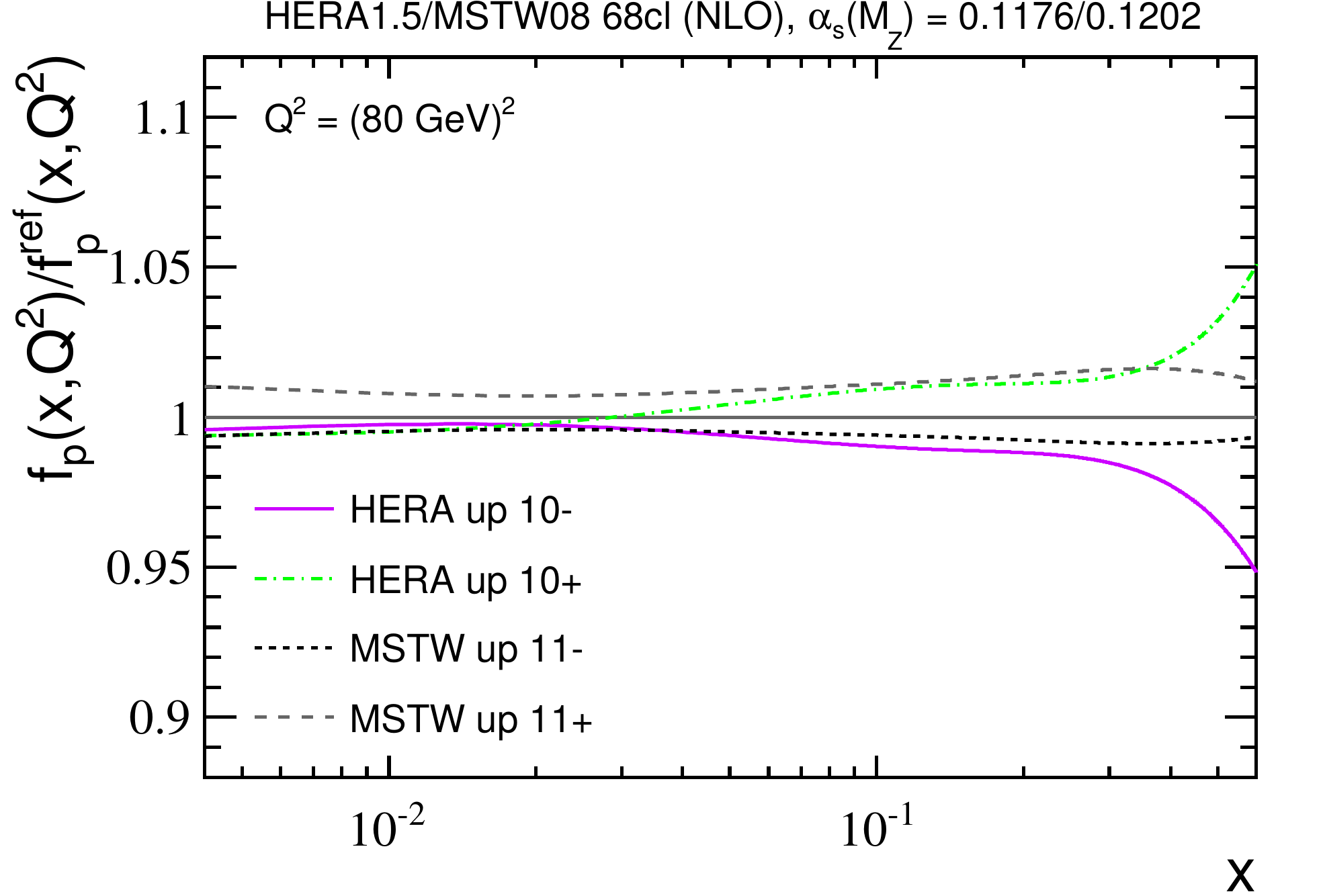} \\
\end{tabular}
\caption{Ratio of evolved quark, gluon PDFs ($f_p(x,Q^2 )$) with respect to their corresponding central value for selected members - as described in the text.}\label{fig:partonPDF}
\end{figure}

\noindent These members and their deviations with respect to the central value are presented in Table \ref{tab:members}. The ratio of the evolved gluon and up quark PDFs with respect to the central value for these members are also presented in Fig.~\ref{fig:partonPDF}. The deviations at high-$x$ are found to be largest for the gluon PDF, with the exception of the valence content for a few eigenvectors, demonstrating the dominance of the gluon PDF uncertainties on the observable $\sigma^{\mathrm{LHCb}}$.

\begin{table}[htdp]
\begin{small}
\centering
	\begin{tabular}{ r c | r c | r c }
	 			 \multicolumn{2}{ c | }{CT10}	&  \multicolumn{2}{ c | }{HERA}		&  \multicolumn{2}{ c  }{MSTW}\\
				 $\mathcal{S}_j^{\pm}$ & $\Delta \sigma_j^{\mathrm{LHCb}}$(\%) & $\mathcal{S}_j^{\pm}$ & $\Delta  \sigma_j^{\mathrm{LHCb}}$(\%) & $\mathcal{S}_j^{\pm}$ & $\Delta  \sigma_j^{\mathrm{LHCb}}$(\%) \\ \hline
				 13$+$		& $+$2.47	& 9$-$		&	$+$1.92		& 11$+$		&	$-$2.13			\\
				 13$-$		& $-$1.71	& 9$+$		&	$-$1.84		& 11$-$		&	$+$1.41		\\  
				 5$+$ 		& $-$1.55	& 10$-$		&	$-$1.39		& 15$+$ 		&	$+$1.17		\\ 
				 5$-$			& $+$1.53	& 10$+$		&	$+$1.26		& 9$+$		&	$+$0.97		\\
				 26$+$		& $-$1.39	& 4$-$		&	$-$0.55		& 12$-$		&	$-$0.85			\\
				 24$+$		& $+$1.23	& 4$+$		&	$+$0.52		& 9$-$		&	$-$0.82			\\
	\end{tabular}
	\caption{Eigenvectors (and directions) with a strong impact on replicas with large $\chi^2_k$ values with respect to a hypothetical $\bar{\sigma}_0^{\mathrm{LHCb}}$ at 14~TeV.}
	\label{tab:members}
\end{small}
\end{table}

\begin{figure}
\centering
\begin{tabular}{cc}
\includegraphics[width=.48\linewidth]{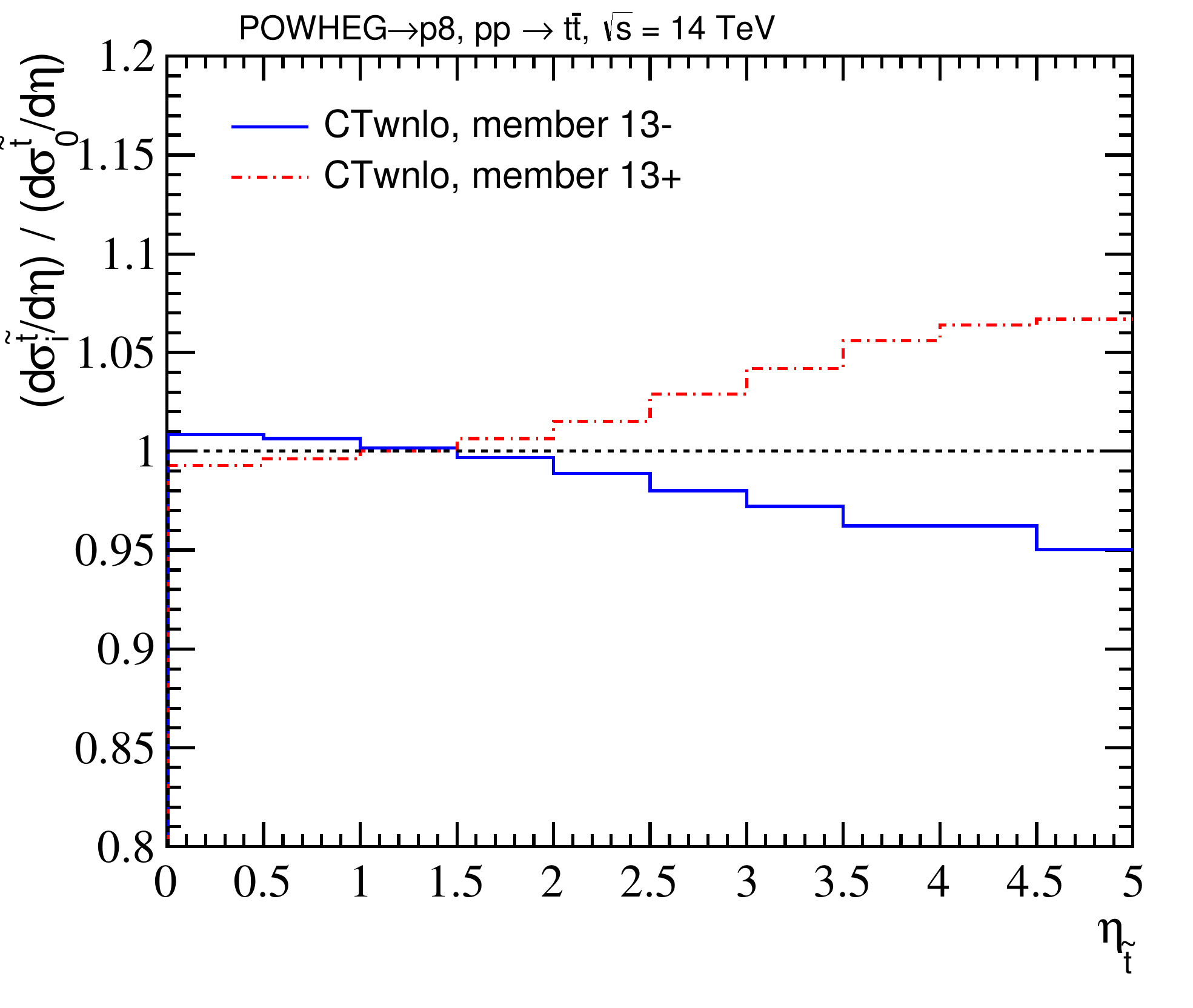} &
\includegraphics[width=.48\linewidth]{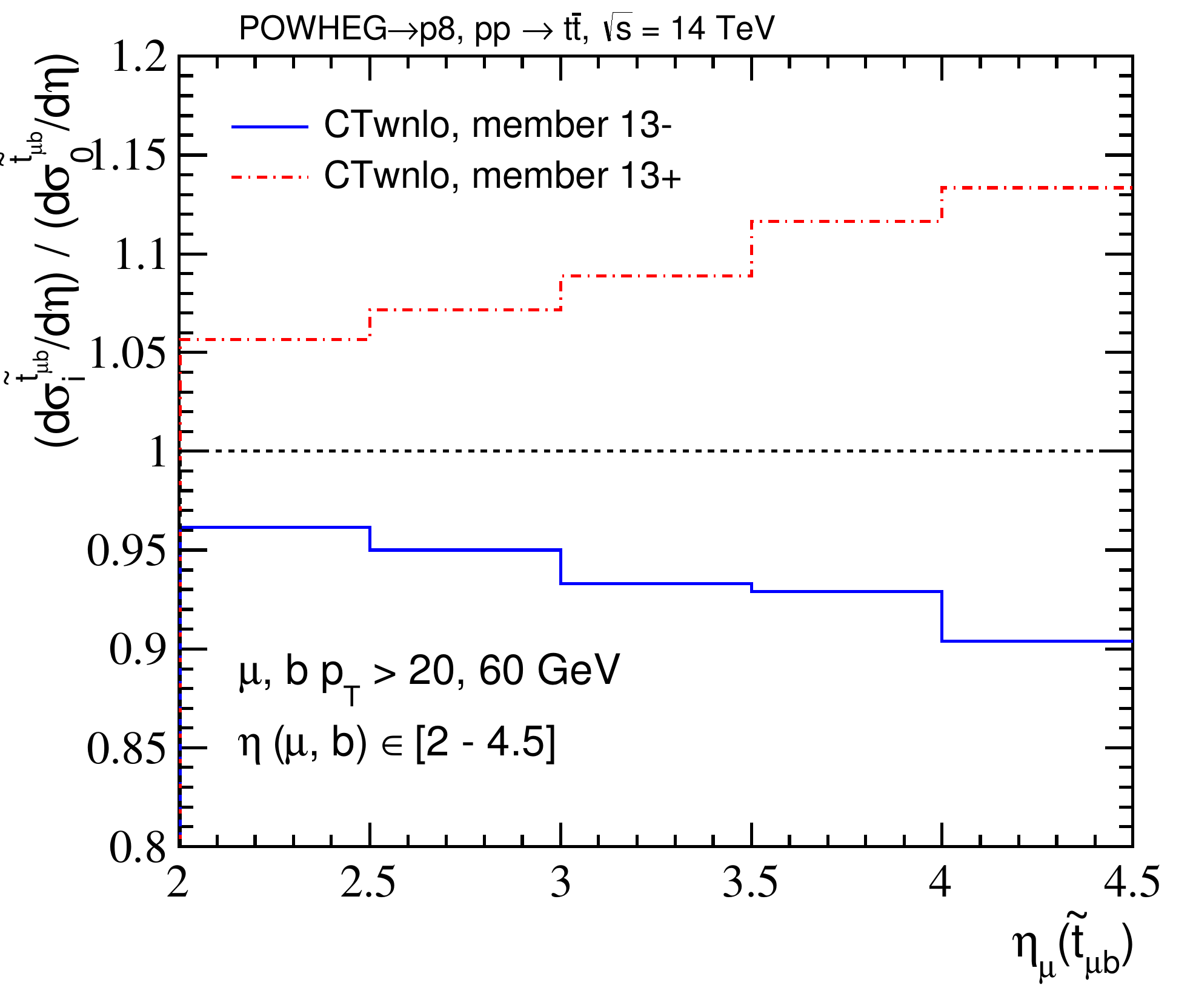} \\
\end{tabular}
\caption{Ratio of pseudotop (left) and $\tilde{t}_{\mu b}$ (right) differential cross-sections for CT10wnlo eigenvectors 13$-$ and 13$+$ with respect to the central value.}\label{toplep14}
\end{figure}

\begin{figure}
\centering
\begin{tabular}{cc}
\includegraphics[width=.48\linewidth]{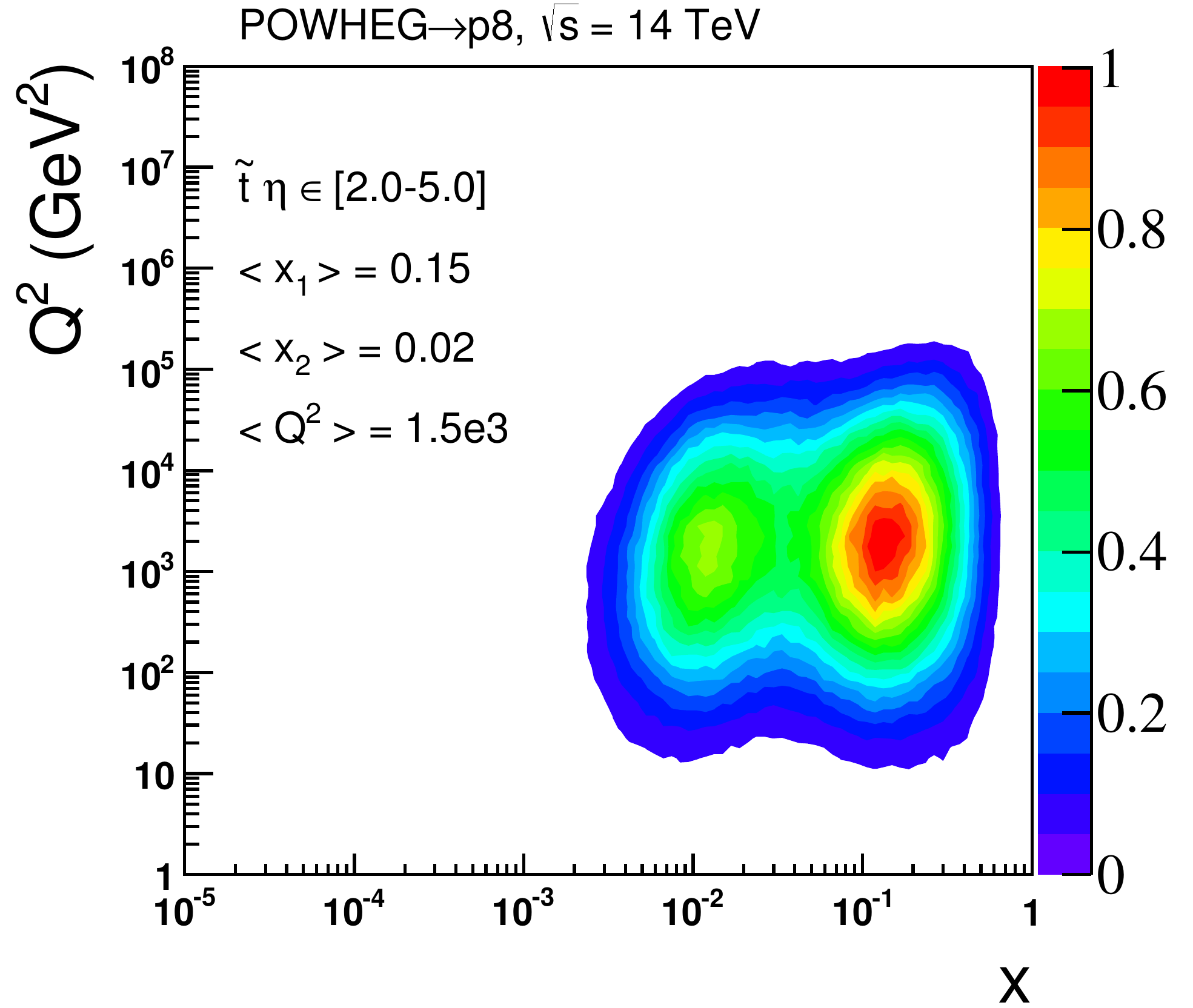} &
\includegraphics[width=.48\linewidth]{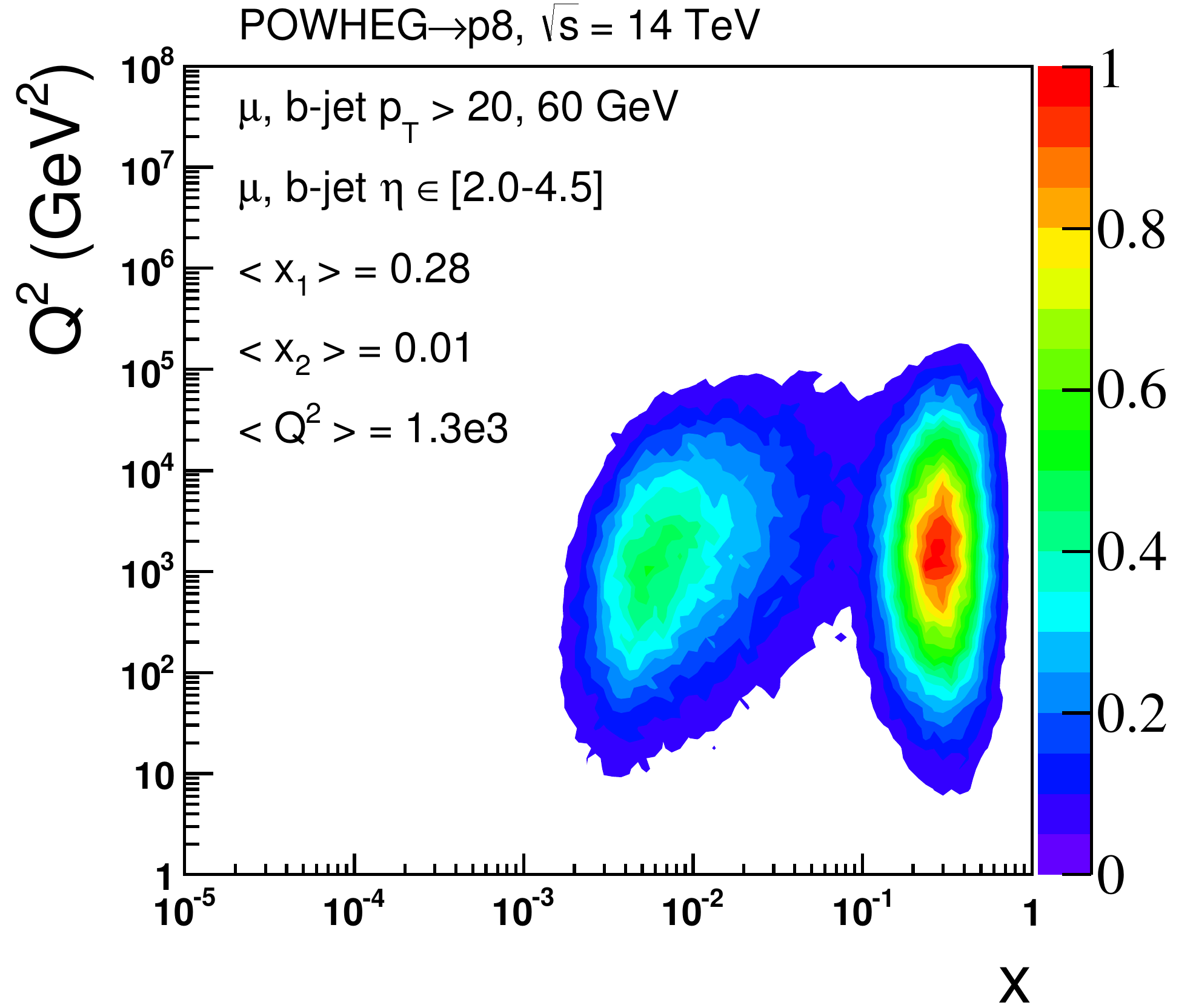} \\
\end{tabular}
\caption{Event momentum scale, $Q^2$, with respect to incoming parton momentum fraction $x$ for a pseudotop within the LHCb acceptance (left) and pseudotop final state $\tilde{t}_{\mu b}$ passing analysis cuts (right).}\label{xq2}
\end{figure}

These particular eigenvectors are only similar to the list obtained from calculating the inclusive $t\bar{t}$ cross-section. This is due to partial cancellation across the entire pseudorapidity region for the cross-section prediction from some eigenvectors, as shown in Fig.~\ref{toplep14} where the ratio of pseudotop (left) differential cross-sections for CT10 eigenvectors 13$-$ and 13$+$ are plotted with respect to the central value. This cancellation is a consequence of the evolved gluon PDF for eigenvectors 13$-$ and 13$+$ passing a crossing point with respect to the central value at $x$ = 0.1, which dominates central $t\bar{t}$ production at 14~TeV - see Fig.~\ref{fig:partonPDF}. Also plotted in Fig.~\ref{toplep14} (right) is the ratio of differential cross-sections for the $\tilde{t}_{\mu b}$ final state passing all analysis cuts discussed in the previous section, again for eigenvectors 13$-$ and 13$+$ where the deviation from the central value is larger. This demonstrates that the analysis of the impact of a measurement $\sigma^{\mathrm{LHCb}}$ on the gluon PDF is an underestimate as more information is contained in a binned differential cross-section. In fact, the kinematic cuts applied in the analysis, which are required to improve the signal/background ratio, for the $b$-jet and muon of $p_T >$ 60, 20~GeV select harder events which are produced from higher $x_1$ incoming partons improving the constraints at yet higher $x$. This is demonstrated in Fig.~\ref{xq2} where the incoming parton momentum fraction $x$ is plotted against the event momentum scale squared ($Q^2$). The left plot corresponds to events where a parton level top is within the LHCb acceptance, and the right plot to events passing the full analysis cuts. As a larger fraction of events are at high-$x_1$ ($<x_1 > =$ 0.28) after applying analysis cuts, this increases the sensitivity within this region which can be seen by comparing the bin-by-bin deviation in Fig.~\ref{toplep14}. Fully quantifying the sensitivity after applying analysis cuts will require a full study of NLO+PS for all eigenvector members as well as knowledge of cuts which will be eventually used in the analysis.

\section{Application to the charge asymmetry}
Improvements to the gluon PDF description at high-$x$ are useful for reducing uncertainties in Standard Model (SM), such as Higgs production, as well as Beyond-SM (BSM) physics processes which are often swamped by $t\bar{t}$ backgrounds.  Another interesting application of an improved high-$x$ gluon PDF is to the prediction of the $t\bar{t}$ charge asymmetry, diluted by symmetric $gg$-scattering.

There is tension between NLO predictions and the observed charge asymmetry from foward-backward measurements with the full TeVatron data sets, Ref.~\cite{Aaltonen:2012it,Aaltonen:2013vaf,Abazov:2013wxa,d0:preliminary}, where the measured asymmetries are larger than expected. Although the same behaviour is not seen in the current LHC forward-central measurements, Ref.~\cite{ATLAS-CONF-2013-078,Chatrchyan:2012cxa,CMS-PAS-TOP-12-033} , it is difficult for any conclusion to be made as the combined uncertainties on the LHC measurements are of comparable size to the theoretical predictions. The small asymmetry prediction at the LHC, in comparison to the TeVatron, is a result of the large $gg$-dilution present in multi-TeV $pp$ collisions as well as the redefinition of asymmetry variables required as the initial state is symmetric.

The proposal of Ref.~\cite{Kuhn:1998kw}, and specifically to LHCb in Ref.\cite{Kagan:2011yx},  was to measure the production rate of $t$/$\bar{t}$ from pair production in the high pseudorapidity bins at the LHC as,

\beq
A^{t\bar{t}} = \left( \frac{d\sigma^{t}/d\eta-d\sigma^{\bar{t}}/d\eta} {\mathrm{2} d\sigma^{\tilde{t}}_{\mathrm{LO}}/d\eta}\right).
\eeq

\noindent Due to the reduction in the dilution from $gg$-scattering, the asymmetry grows substantially with increasing pseudorapidity. With LHCb data sets of 5, 50~fb$^{-1}$ at 14~TeV, the number of $t\bar{t}\rightarrow \mu b$ events passing the analysis cuts of Section~\ref{LHCbanalysis} beyond $\eta =3.2$ are $\mathcal{O}(1k, 10k)$ respectively. Therefore, an asymmetry measurement with these data sets will also be systematically dominated. We have already demonstrated the sensitivity of cross-section measurements at LHCb to the high-$x$ gluon uncertainties, meaning that the associated PDF systematic for the asymmetry is also large in comparison to central measurements.

To demonstrate how a cross-section measurement at LHCb impacts the PDF uncertainty of $A^{t\bar{t}}$, we perform a reweighting of the observables $A^{t\bar{t}}(\mathcal{S}_k)$ generated from CT10 and NNPDF replica sets based on the assumption of a cross-section measurement $\sigma^{\mathrm{LHCb}}_{\mathrm{fake}}$. Note that in this case, the reweighting of both CT10 and NNPDF sets is done assuming the same cross-section, where as in the previous section this was not the case. The predictions from the replica sets are combined in the following way,

\beq
\begin{aligned}
X_{\mathrm{central}} &= 0.5\cdot (\mathrm{max}(X_1+\delta X_1,X_2+\delta X_2)+\mathrm{min}(X_1-\delta X_1,X_2-\delta X_2) )\\
\delta X &= 0.5\cdot (\mathrm{max}(X_1+\delta X_1,X_2+\delta X_2)-\mathrm{min}(X_1-\delta X_1,X_2-\delta X_2) ).
\end{aligned}
\eeq

\begin{figure}[h]
\centering
\begin{tabular}{c}
\includegraphics[width=.655\linewidth]{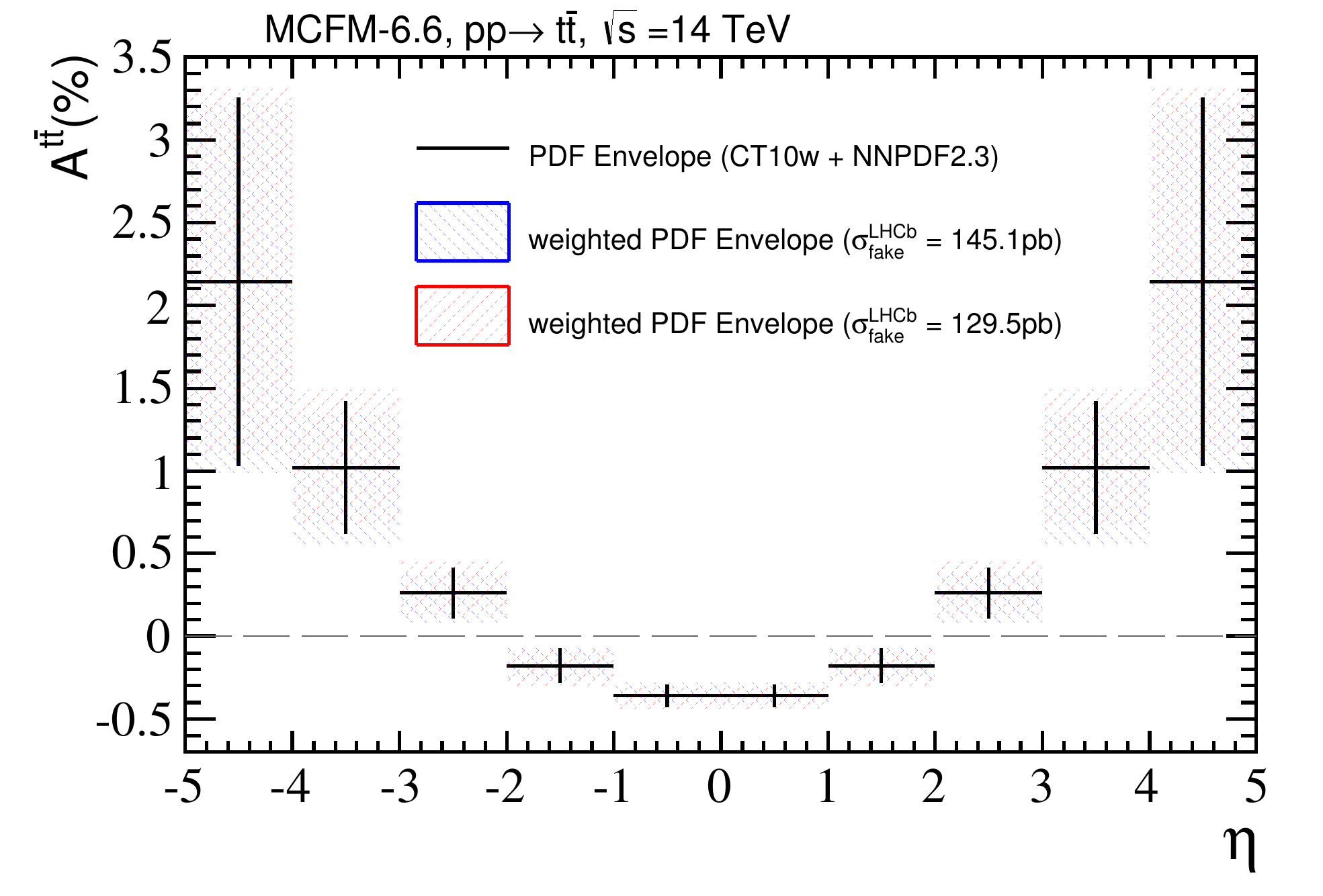}
\end{tabular}
\caption{Differential asymmetry $A^{t\bar{t}}$ with respect to pseudorapidity. The effect of reweighting after including hypothetical measurements of $\sigma^{\mathrm{LHCb}}$ is also included - see text for details.}
\label{fig:Asymmetry}
\end{figure}

\noindent $X_{1,2}$ correspond to the observable calculated from CT10 and NNPDF replica sets. A reweighting is performed for both CT10 and NNPDF sets assuming $\sigma^{\mathrm{LHCb}}_{\mathrm{fake}} =$ 145.1~pb and then also for $\sigma^{\mathrm{LHCb}}_{\mathrm{fake}} =$129.5~pb, with an associated experimental uncertainty of 4\%. The predictions for the asymmetry before and after these reweightings are shown in Fig.~\ref{fig:Asymmetry}.

\noindent The assumed cross-section of $\sigma^{\mathrm{LHCb}}_{\mathrm{fake}} =$ 145.1~pb results in a decrease in magnitude of the asymmetry, while for the smaller cross-section $\sigma^{\mathrm{LHCb}}_{\mathrm{fake}} =$ 129.5~pb the opposite behaviour is observed, accounting respectively for and increase and decrease in the gluon PDF at high-$x$ required to account for the assumed cross-section. The asymmetry expectation and associated uncertainty is provided in Table~\ref{tab:asymshift} where the relative shifts in the asymmetry after reweighting are also included. The change to the overall relative uncertainty on the asymmetry is found to be negligible for the given choices of pseudodata cross-sections. The largest shift to the central value is in the region 2.0 $< \eta <$ 3.0, which is the region where the contribution to $\sigma^{\mathrm{LHCb}}$ is largest - see Fig. \ref{diffsigma}.

\begin{table}[htdp]
\begin{footnotesize}
\centering
	\begin{tabular}{ r | c | c | c | c | c }
	 				&  0.0 $<\eta<$ 1.0	& 1.0 $< \eta <$ 2.0 	& 2.0 $< \eta <$ 3.0 	&  3.0 $< \eta <$ 4.0 & 4.0 $< \eta <$ 5.0\\ \hline
		$A^{t\bar{t}}$(\%)$\pm \delta A^{t\bar{t}}$	& $-$0.36$\pm$0.07	& $-$0.18$\pm$0.11	& 0.26$\pm$0.16	& 1.02$\pm$0.40	& 2.14$\pm$1.11	\\ \hline
		shift $A^{t\bar{t}}$ ($\sigma^{\mathrm{LHCb}}_{\mathrm{fake}} =$ 145.1~pb)	&  $-$3.07\%		& $-$7.34\%			& $-$10.82\%			& $-$7.24\%			&	$-$1.89\%		\\
		shift $A^{t\bar{t}}$ ($\sigma^{\mathrm{LHCb}}_{\mathrm{fake}} =$ 129.5~pb)	& $+$2.92\%	&  $+$7.40\%		& $+$10.75\%			& $+$6.55\%		&	$+$2.70\%		\\
	\end{tabular}
	\caption{Summary of $A^{t\bar{t}}$ with respect to pseudorapidity and the relative shift of this asymmetry after reweighting assuming a cross-section measurement within the LHCb acceptance.}
	\label{tab:asymshift}
\end{footnotesize}
\end{table}

\section{Discussion}
The chosen experimental uncertainty range of 4-8\% is an estimate of the systematic reach of future measurements at LHCb, it is expected that the largest uncertainties arise from; background/signal modelling, $b$-tagging mis-tag/efficiency and luminosity. Current cross-section results from ATLAS~\cite{ATLAS-CONF-2013-097} have already achieved a total relative uncertainty below 5\%, it is therefore not unreasonable to expect similar precision from measurements with the upgraded LHCb detector. Especially given that the same simulation technology is available to LHCb (currently NLO$\rightarrow$parton shower) and that the luminosity uncertainty at LHCb~\cite{Aaij:2011er} is of similar size to the result~\cite{ATLAS-CONF-2013-097}. It is therefore expected that future $t\bar{t}$ cross-section measurements at LHCb have the potential to reduce uncertainties on  the high-$x$ gluon PDF by up to 20\%. A direct application of such an improvement is to better predict the charge asymmetry within the LHCb acceptance. Given that the prediction of the $t\bar{t}$ charge asymmetry is dependent on the high-$x$ gluon PDF, comparisons between BSM scenarios and perturbative QCD will rely on such improvements.

The analysis strategy presented for top reconstruction relies on isolating the charged lepton in the decay $t \rightarrow (W \rightarrow \mu \nu_\mu) b$ as well as any additional jets in the event, the main motivation for this is to reduce the QCD $b\bar{b}$ background. However, this also removes highly boosted top quark decays in which the top decays products are very close together. There are many BSM scenarios which contain top partners as a solution to the hierarchy problem, for example~\cite{Cheng:2005as,Agashe:2006hk}. If these new particles are kinematically accessible at the LHC, and they decay via top quarks - which is often the case - then boosted top quarks are an interesting signal for BSM~\cite{Matsumoto:2006ws,Belyaev:2006jh,Freitas:2006vy,Meade:2006dw,Thaler:2008ju,Nojiri:2008ir}. In the case of boosted top quarks at LHCb, an investigation into the separation power of very energetic fat jets and top decays should be undertaken. Semi-leptonic decays may be promising due to the excellent impact parameter resolution for charged leptons. 

We conclude that a cross-section measurement with the current 7~TeV data set is statistically limited. However, given the larger data set available at 8~TeV (2~fb$^{-1}$) and an increase in $\sigma_{t\bar{t}}\cdot Acc$, statistical precision of 6\% is achievable in the highly populated bins (see. Fig~\ref{plot7}). It is also worth investigating the precision achievable in the electron channel, which could further improve statistics. At 14~TeV the impact of a cross-section measurement on the gluon PDF ultimately depends on the experimental precision. Measurements of the background cross-sections such as $W$($b$)-jets will be a necessary ingredient to achieving high precision.

\section{Acknowledgements}
We are grateful to Amanda Cooper-Sarkar, Ulrich Haisch and Victor Coco for many useful suggestions and in particular Juan Rojo for advice which helped improve the study. The research of R.G. is supported by an STFC Postgraduate Studentship.

\bibliographystyle{BIB}
\bibliography{arXiv}

\ifx\mcitethebibliography\mciteundefinedmacro
\PackageError{LHCb.bst}{mciteplus.sty has not been loaded}
{This bibstyle requires the use of the mciteplus package.}\fi
\providecommand{\href}[2]{#2}
\begin{mcitethebibliography}{10}
\mciteSetBstSublistMode{n}
\mciteSetBstMaxWidthForm{subitem}{\alph{mcitesubitemcount})}
\mciteSetBstSublistLabelBeginEnd{\mcitemaxwidthsubitemform\space}
{\relax}{\relax}

\bibitem{Kagan:2011yx}
A.~L. Kagan, J.~F. Kamenik, G.~Perez, and S.~Stone,
  \ifthenelse{\boolean{articletitles}}{{\it {Top LHCb Physics}},
  }{}\href{http://dx.doi.org/10.1103/PhysRevLett.107.082003}{Phys.\ Rev.\
  Lett.\  {\bf 107} (2011) 082003}, \href{http://arxiv.org/abs/1103.3747}{{\tt
  arXiv:1103.3747}}\relax
\mciteBstWouldAddEndPuncttrue
\mciteSetBstMidEndSepPunct{\mcitedefaultmidpunct}
{\mcitedefaultendpunct}{\mcitedefaultseppunct}\relax
\EndOfBibitem
\bibitem{Kuhn:1998kw}
J.~H. Kuhn and G.~Rodrigo, \ifthenelse{\boolean{articletitles}}{{\it {Charge
  asymmetry of heavy quarks at hadron colliders}},
  }{}\href{http://dx.doi.org/10.1103/PhysRevD.59.054017}{Phys.\ Rev.\  {\bf
  D59} (1999) 054017}, \href{http://arxiv.org/abs/hep-ph/9807420}{{\tt
  arXiv:hep-ph/9807420}}\relax
\mciteBstWouldAddEndPuncttrue
\mciteSetBstMidEndSepPunct{\mcitedefaultmidpunct}
{\mcitedefaultendpunct}{\mcitedefaultseppunct}\relax
\EndOfBibitem
\bibitem{Czakon:2013goa}
M.~Czakon, P.~Fiedler, and A.~Mitov, \ifthenelse{\boolean{articletitles}}{{\it
  {The total top quark pair production cross-section at hadron colliders
  through O(alpha\_S\^4)}},
  }{}\href{http://dx.doi.org/10.1103/PhysRevLett.110.252004}{Phys.\ Rev.\
  Lett.\  {\bf 110} (2013) 252004}, \href{http://arxiv.org/abs/1303.6254}{{\tt
  arXiv:1303.6254}}\relax
\mciteBstWouldAddEndPuncttrue
\mciteSetBstMidEndSepPunct{\mcitedefaultmidpunct}
{\mcitedefaultendpunct}{\mcitedefaultseppunct}\relax
\EndOfBibitem
\bibitem{Baernreuther:2012ws}
P.~Baernreuther, M.~Czakon, and A.~Mitov,
  \ifthenelse{\boolean{articletitles}}{{\it {Percent Level Precision Physics at
  the Tevatron: First Genuine NNLO QCD Corrections to $q \bar{q} \to t \bar{t}
  + X$}}, }{}\href{http://dx.doi.org/10.1103/PhysRevLett.109.132001}{Phys.\
  Rev.\ Lett.\  {\bf 109} (2012) 132001},
  \href{http://arxiv.org/abs/1204.5201}{{\tt arXiv:1204.5201}}\relax
\mciteBstWouldAddEndPuncttrue
\mciteSetBstMidEndSepPunct{\mcitedefaultmidpunct}
{\mcitedefaultendpunct}{\mcitedefaultseppunct}\relax
\EndOfBibitem
\bibitem{Czakon:2012zr}
M.~Czakon and A.~Mitov, \ifthenelse{\boolean{articletitles}}{{\it {NNLO
  corrections to top-pair production at hadron colliders: the all-fermionic
  scattering channels}},
  }{}\href{http://dx.doi.org/10.1007/JHEP12(2012)054}{JHEP {\bf 1212} (2012)
  054}, \href{http://arxiv.org/abs/1207.0236}{{\tt arXiv:1207.0236}}\relax
\mciteBstWouldAddEndPuncttrue
\mciteSetBstMidEndSepPunct{\mcitedefaultmidpunct}
{\mcitedefaultendpunct}{\mcitedefaultseppunct}\relax
\EndOfBibitem
\bibitem{Czakon:2012pz}
M.~Czakon and A.~Mitov, \ifthenelse{\boolean{articletitles}}{{\it {NNLO
  corrections to top pair production at hadron colliders: the quark-gluon
  reaction}}, }{}\href{http://dx.doi.org/10.1007/JHEP01(2013)080}{JHEP {\bf
  1301} (2013) 080}, \href{http://arxiv.org/abs/1210.6832}{{\tt
  arXiv:1210.6832}}\relax
\mciteBstWouldAddEndPuncttrue
\mciteSetBstMidEndSepPunct{\mcitedefaultmidpunct}
{\mcitedefaultendpunct}{\mcitedefaultseppunct}\relax
\EndOfBibitem
\bibitem{Czakon:2009zw}
M.~Czakon, A.~Mitov, and G.~F. Sterman,
  \ifthenelse{\boolean{articletitles}}{{\it {Threshold Resummation for Top-Pair
  Hadroproduction to Next-to-Next-to-Leading Log}},
  }{}\href{http://dx.doi.org/10.1103/PhysRevD.80.074017}{Phys.\ Rev.\  {\bf
  D80} (2009) 074017}, \href{http://arxiv.org/abs/0907.1790}{{\tt
  arXiv:0907.1790}}\relax
\mciteBstWouldAddEndPuncttrue
\mciteSetBstMidEndSepPunct{\mcitedefaultmidpunct}
{\mcitedefaultendpunct}{\mcitedefaultseppunct}\relax
\EndOfBibitem
\bibitem{Beneke:2009rj}
M.~Beneke, P.~Falgari, and C.~Schwinn,
  \ifthenelse{\boolean{articletitles}}{{\it {Soft radiation in heavy-particle
  pair production: All-order colour structure and two-loop anomalous
  dimension}},
  }{}\href{http://dx.doi.org/10.1016/j.nuclphysb.2009.11.004}{Nucl.\ Phys.\
  {\bf B828} (2010) 69}, \href{http://arxiv.org/abs/0907.1443}{{\tt
  arXiv:0907.1443}}\relax
\mciteBstWouldAddEndPuncttrue
\mciteSetBstMidEndSepPunct{\mcitedefaultmidpunct}
{\mcitedefaultendpunct}{\mcitedefaultseppunct}\relax
\EndOfBibitem
\bibitem{Cacciari:2011hy}
M.~Cacciari {\em et~al.}, \ifthenelse{\boolean{articletitles}}{{\it {Top-pair
  production at hadron colliders with next-to-next-to-leading logarithmic
  soft-gluon resummation}},
  }{}\href{http://dx.doi.org/10.1016/j.physletb.2012.03.013}{Phys.\ Lett.\
  {\bf B710} (2012) 612}, \href{http://arxiv.org/abs/1111.5869}{{\tt
  arXiv:1111.5869}}\relax
\mciteBstWouldAddEndPuncttrue
\mciteSetBstMidEndSepPunct{\mcitedefaultmidpunct}
{\mcitedefaultendpunct}{\mcitedefaultseppunct}\relax
\EndOfBibitem
\bibitem{Czakon:2013tha}
M.~Czakon, M.~L. Mangano, A.~Mitov, and J.~Rojo,
  \ifthenelse{\boolean{articletitles}}{{\it {Constraints on the gluon PDF from
  top quark pair production at hadron colliders}},
  }{}\href{http://dx.doi.org/10.1007/JHEP07(2013)167}{JHEP {\bf 1307} (2013)
  167}, \href{http://arxiv.org/abs/1303.7215}{{\tt arXiv:1303.7215}}\relax
\mciteBstWouldAddEndPuncttrue
\mciteSetBstMidEndSepPunct{\mcitedefaultmidpunct}
{\mcitedefaultendpunct}{\mcitedefaultseppunct}\relax
\EndOfBibitem
\bibitem{Kidonakis:2010dk}
N.~Kidonakis, \ifthenelse{\boolean{articletitles}}{{\it
  {Next-to-next-to-leading soft-gluon corrections for the top quark cross
  section and transverse momentum distribution}},
  }{}\href{http://dx.doi.org/10.1103/PhysRevD.82.114030}{Phys.\ Rev.\  {\bf
  D82} (2010) 114030}, \href{http://arxiv.org/abs/1009.4935}{{\tt
  arXiv:1009.4935}}\relax
\mciteBstWouldAddEndPuncttrue
\mciteSetBstMidEndSepPunct{\mcitedefaultmidpunct}
{\mcitedefaultendpunct}{\mcitedefaultseppunct}\relax
\EndOfBibitem
\bibitem{Ahrens:2011mw}
V.~Ahrens {\em et~al.}, \ifthenelse{\boolean{articletitles}}{{\it {RG-improved
  single-particle inclusive cross sections and forward-backward asymmetry in
  $t\bar t$ production at hadron colliders}},
  }{}\href{http://dx.doi.org/10.1007/JHEP09(2011)070}{JHEP {\bf 1109} (2011)
  070}, \href{http://arxiv.org/abs/1103.0550}{{\tt arXiv:1103.0550}}\relax
\mciteBstWouldAddEndPuncttrue
\mciteSetBstMidEndSepPunct{\mcitedefaultmidpunct}
{\mcitedefaultendpunct}{\mcitedefaultseppunct}\relax
\EndOfBibitem
\bibitem{Kidonakis:2011ca}
N.~Kidonakis and B.~D. Pecjak, \ifthenelse{\boolean{articletitles}}{{\it
  {Top-quark production and QCD}},
  }{}\href{http://dx.doi.org/10.1140/epjc/s10052-012-2084-0}{Eur.\ Phys.\ J.\
  {\bf C72} (2012) 2084}, \href{http://arxiv.org/abs/1108.6063}{{\tt
  arXiv:1108.6063}}\relax
\mciteBstWouldAddEndPuncttrue
\mciteSetBstMidEndSepPunct{\mcitedefaultmidpunct}
{\mcitedefaultendpunct}{\mcitedefaultseppunct}\relax
\EndOfBibitem
\bibitem{Beneke:2012wb}
M.~Beneke {\em et~al.}, \ifthenelse{\boolean{articletitles}}{{\it {Inclusive
  Top-Pair Production Phenomenology with TOPIXS}},
  }{}\href{http://dx.doi.org/10.1007/JHEP07(2012)194}{JHEP {\bf 1207} (2012)
  194}, \href{http://arxiv.org/abs/1206.2454}{{\tt arXiv:1206.2454}}\relax
\mciteBstWouldAddEndPuncttrue
\mciteSetBstMidEndSepPunct{\mcitedefaultmidpunct}
{\mcitedefaultendpunct}{\mcitedefaultseppunct}\relax
\EndOfBibitem
\bibitem{Guzzi:2013noa}
M.~Guzzi, K.~Lipka, and S.-O. Moch, \ifthenelse{\boolean{articletitles}}{{\it
  {Top-quark production at the LHC: differential cross section and
  phenomenological applications}},
  }{}\href{http://arxiv.org/abs/1308.1635}{{\tt arXiv:1308.1635}}\relax
\mciteBstWouldAddEndPuncttrue
\mciteSetBstMidEndSepPunct{\mcitedefaultmidpunct}
{\mcitedefaultendpunct}{\mcitedefaultseppunct}\relax
\EndOfBibitem
\bibitem{Gauld:1557385}
R.~Gauld, \ifthenelse{\boolean{articletitles}}{{\it {Measuring top quark
  production asymmetries at LHCb}}, }{} Tech. Rep. LHCb-PUB-2013-009.
  CERN-LHCb-PUB-2013-009, CERN, Geneva, Jun, 2013\relax
\mciteBstWouldAddEndPuncttrue
\mciteSetBstMidEndSepPunct{\mcitedefaultmidpunct}
{\mcitedefaultendpunct}{\mcitedefaultseppunct}\relax
\EndOfBibitem
\bibitem{Nason:2004rx}
P.~Nason, \ifthenelse{\boolean{articletitles}}{{\it {A New method for combining
  NLO QCD with shower Monte Carlo algorithms}},
  }{}\href{http://dx.doi.org/10.1088/1126-6708/2004/11/040}{JHEP {\bf 0411}
  (2004) 040}, \href{http://arxiv.org/abs/hep-ph/0409146}{{\tt
  arXiv:hep-ph/0409146}}\relax
\mciteBstWouldAddEndPuncttrue
\mciteSetBstMidEndSepPunct{\mcitedefaultmidpunct}
{\mcitedefaultendpunct}{\mcitedefaultseppunct}\relax
\EndOfBibitem
\bibitem{Frixione:2007vw}
S.~Frixione, P.~Nason, and C.~Oleari, \ifthenelse{\boolean{articletitles}}{{\it
  {Matching NLO QCD computations with Parton Shower simulations: the POWHEG
  method}}, }{}\href{http://dx.doi.org/10.1088/1126-6708/2007/11/070}{JHEP {\bf
  0711} (2007) 070}, \href{http://arxiv.org/abs/0709.2092}{{\tt
  arXiv:0709.2092}}\relax
\mciteBstWouldAddEndPuncttrue
\mciteSetBstMidEndSepPunct{\mcitedefaultmidpunct}
{\mcitedefaultendpunct}{\mcitedefaultseppunct}\relax
\EndOfBibitem
\bibitem{Alioli:2010xd}
S.~Alioli, P.~Nason, C.~Oleari, and E.~Re,
  \ifthenelse{\boolean{articletitles}}{{\it {A general framework for
  implementing NLO calculations in shower Monte Carlo programs: the POWHEG
  BOX}}, }{}\href{http://dx.doi.org/10.1007/JHEP06(2010)043}{JHEP {\bf 1006}
  (2010) 043}, \href{http://arxiv.org/abs/1002.2581}{{\tt
  arXiv:1002.2581}}\relax
\mciteBstWouldAddEndPuncttrue
\mciteSetBstMidEndSepPunct{\mcitedefaultmidpunct}
{\mcitedefaultendpunct}{\mcitedefaultseppunct}\relax
\EndOfBibitem
\bibitem{Frixione:2007nw}
S.~Frixione, P.~Nason, and G.~Ridolfi,
  \ifthenelse{\boolean{articletitles}}{{\it {A Positive-weight
  next-to-leading-order Monte Carlo for heavy flavour hadroproduction}},
  }{}\href{http://dx.doi.org/10.1088/1126-6708/2007/09/126}{JHEP {\bf 0709}
  (2007) 126}, \href{http://arxiv.org/abs/0707.3088}{{\tt
  arXiv:0707.3088}}\relax
\mciteBstWouldAddEndPuncttrue
\mciteSetBstMidEndSepPunct{\mcitedefaultmidpunct}
{\mcitedefaultendpunct}{\mcitedefaultseppunct}\relax
\EndOfBibitem
\bibitem{Alioli:2010qp}
S.~Alioli, P.~Nason, C.~Oleari, and E.~Re,
  \ifthenelse{\boolean{articletitles}}{{\it {Vector boson plus one jet
  production in POWHEG}},
  }{}\href{http://dx.doi.org/10.1007/JHEP01(2011)095}{JHEP {\bf 1101} (2011)
  095}, \href{http://arxiv.org/abs/1009.5594}{{\tt arXiv:1009.5594}}\relax
\mciteBstWouldAddEndPuncttrue
\mciteSetBstMidEndSepPunct{\mcitedefaultmidpunct}
{\mcitedefaultendpunct}{\mcitedefaultseppunct}\relax
\EndOfBibitem
\bibitem{Alioli:2009je}
S.~Alioli, P.~Nason, C.~Oleari, and E.~Re,
  \ifthenelse{\boolean{articletitles}}{{\it {NLO single-top production matched
  with shower in POWHEG: s- and t-channel contributions}},
  }{}\href{http://dx.doi.org/10.1007/JHEP02(2010)011,
  10.1088/1126-6708/2009/09/111}{JHEP {\bf 0909} (2009) 111},
  \href{http://arxiv.org/abs/0907.4076}{{\tt arXiv:0907.4076}}\relax
\mciteBstWouldAddEndPuncttrue
\mciteSetBstMidEndSepPunct{\mcitedefaultmidpunct}
{\mcitedefaultendpunct}{\mcitedefaultseppunct}\relax
\EndOfBibitem
\bibitem{Frederix:2012dh}
R.~Frederix, E.~Re, and P.~Torrielli, \ifthenelse{\boolean{articletitles}}{{\it
  {Single-top t-channel hadroproduction in the four-flavour scheme with POWHEG
  and aMC@NLO}}, }{}\href{http://dx.doi.org/10.1007/JHEP09(2012)130}{JHEP {\bf
  1209} (2012) 130}, \href{http://arxiv.org/abs/1207.5391}{{\tt
  arXiv:1207.5391}}\relax
\mciteBstWouldAddEndPuncttrue
\mciteSetBstMidEndSepPunct{\mcitedefaultmidpunct}
{\mcitedefaultendpunct}{\mcitedefaultseppunct}\relax
\EndOfBibitem
\bibitem{Oleari:2011ey}
C.~Oleari and L.~Reina, \ifthenelse{\boolean{articletitles}}{{\it {W +- b
  $\bar{b}$ production in POWHEG}},
  }{}\href{http://dx.doi.org/10.1007/JHEP11(2011)040,
  10.1007/JHEP08(2011)061}{JHEP {\bf 1108} (2011) 061},
  \href{http://arxiv.org/abs/1105.4488}{{\tt arXiv:1105.4488}}\relax
\mciteBstWouldAddEndPuncttrue
\mciteSetBstMidEndSepPunct{\mcitedefaultmidpunct}
{\mcitedefaultendpunct}{\mcitedefaultseppunct}\relax
\EndOfBibitem
\bibitem{Lai:2010vv}
H.-L. Lai {\em et~al.}, \ifthenelse{\boolean{articletitles}}{{\it {New parton
  distributions for collider physics}},
  }{}\href{http://dx.doi.org/10.1103/PhysRevD.82.074024}{Phys.\ Rev.\  {\bf
  D82} (2010) 074024}, \href{http://arxiv.org/abs/1007.2241}{{\tt
  arXiv:1007.2241}}\relax
\mciteBstWouldAddEndPuncttrue
\mciteSetBstMidEndSepPunct{\mcitedefaultmidpunct}
{\mcitedefaultendpunct}{\mcitedefaultseppunct}\relax
\EndOfBibitem
\bibitem{Sjostrand:2007gs}
T.~Sjostrand, S.~Mrenna, and P.~Z. Skands,
  \ifthenelse{\boolean{articletitles}}{{\it {A Brief Introduction to PYTHIA
  8.1}}, }{}\href{http://dx.doi.org/10.1016/j.cpc.2008.01.036}{Comput.\ Phys.\
  Commun.\  {\bf 178} (2008) 852}, \href{http://arxiv.org/abs/0710.3820}{{\tt
  arXiv:0710.3820}}\relax
\mciteBstWouldAddEndPuncttrue
\mciteSetBstMidEndSepPunct{\mcitedefaultmidpunct}
{\mcitedefaultendpunct}{\mcitedefaultseppunct}\relax
\EndOfBibitem
\bibitem{Alwall:2011uj}
J.~Alwall {\em et~al.}, \ifthenelse{\boolean{articletitles}}{{\it {MadGraph 5 :
  Going Beyond}}, }{}\href{http://dx.doi.org/10.1007/JHEP06(2011)128}{JHEP {\bf
  1106} (2011) 128}, \href{http://arxiv.org/abs/1106.0522}{{\tt
  arXiv:1106.0522}}\relax
\mciteBstWouldAddEndPuncttrue
\mciteSetBstMidEndSepPunct{\mcitedefaultmidpunct}
{\mcitedefaultendpunct}{\mcitedefaultseppunct}\relax
\EndOfBibitem
\bibitem{Pumplin:2002vw}
J.~Pumplin {\em et~al.}, \ifthenelse{\boolean{articletitles}}{{\it {New
  generation of parton distributions with uncertainties from global QCD
  analysis}}, }{}JHEP {\bf 0207} (2002) 012,
  \href{http://arxiv.org/abs/hep-ph/0201195}{{\tt arXiv:hep-ph/0201195}}\relax
\mciteBstWouldAddEndPuncttrue
\mciteSetBstMidEndSepPunct{\mcitedefaultmidpunct}
{\mcitedefaultendpunct}{\mcitedefaultseppunct}\relax
\EndOfBibitem
\bibitem{Campbell:2010ff}
J.~M. Campbell and R.~Ellis, \ifthenelse{\boolean{articletitles}}{{\it {MCFM
  for the Tevatron and the LHC}},
  }{}\href{http://dx.doi.org/10.1016/j.nuclphysbps.2010.08.011}{Nucl.\ Phys.\
  Proc.\ Suppl.\  {\bf 205-206} (2010) 10},
  \href{http://arxiv.org/abs/1007.3492}{{\tt arXiv:1007.3492}}\relax
\mciteBstWouldAddEndPuncttrue
\mciteSetBstMidEndSepPunct{\mcitedefaultmidpunct}
{\mcitedefaultendpunct}{\mcitedefaultseppunct}\relax
\EndOfBibitem
\bibitem{Cacciari:2011ma}
M.~Cacciari, G.~P. Salam, and G.~Soyez,
  \ifthenelse{\boolean{articletitles}}{{\it {FastJet User Manual}},
  }{}\href{http://dx.doi.org/10.1140/epjc/s10052-012-1896-2}{Eur.\ Phys.\ J.\
  {\bf C72} (2012) 1896}, \href{http://arxiv.org/abs/1111.6097}{{\tt
  arXiv:1111.6097}}\relax
\mciteBstWouldAddEndPuncttrue
\mciteSetBstMidEndSepPunct{\mcitedefaultmidpunct}
{\mcitedefaultendpunct}{\mcitedefaultseppunct}\relax
\EndOfBibitem
\bibitem{Aaij:2012vn}
LHCb Collaboration, R.~Aaij {\em et~al.},
  \ifthenelse{\boolean{articletitles}}{{\it {Inclusive $W$ and $Z$ production
  in the forward region at $\sqrt{s} = 7$ TeV}},
  }{}\href{http://dx.doi.org/10.1007/JHEP06(2012)058}{JHEP {\bf 1206} (2012)
  058}, \href{http://arxiv.org/abs/1204.1620}{{\tt arXiv:1204.1620}}\relax
\mciteBstWouldAddEndPuncttrue
\mciteSetBstMidEndSepPunct{\mcitedefaultmidpunct}
{\mcitedefaultendpunct}{\mcitedefaultseppunct}\relax
\EndOfBibitem
\bibitem{Beringer:1900zz}
Particle Data Group, J.~Beringer {\em et~al.},
  \ifthenelse{\boolean{articletitles}}{{\it {Review of Particle Physics
  (RPP)}}, }{}\href{http://dx.doi.org/10.1103/PhysRevD.86.010001}{Phys.\ Rev.\
  {\bf D86} (2012) 010001}\relax
\mciteBstWouldAddEndPuncttrue
\mciteSetBstMidEndSepPunct{\mcitedefaultmidpunct}
{\mcitedefaultendpunct}{\mcitedefaultseppunct}\relax
\EndOfBibitem
\bibitem{ATLAS-CONF-2013-102}
\ifthenelse{\boolean{articletitles}}{{\it {Combination of ATLAS and CMS results
  on the mass of the top-quark using up to 4.9 fb$^{-1}$ of $\sqrt{s}=7$ TeV
  LHC data}}, }{} Tech. Rep. ATLAS-CONF-2013-102, CERN, Geneva, Sep, 2013\relax
\mciteBstWouldAddEndPuncttrue
\mciteSetBstMidEndSepPunct{\mcitedefaultmidpunct}
{\mcitedefaultendpunct}{\mcitedefaultseppunct}\relax
\EndOfBibitem
\bibitem{Alekhin:2012ig}
S.~Alekhin, J.~Blumlein, and S.~Moch, \ifthenelse{\boolean{articletitles}}{{\it
  {Parton Distribution Functions and Benchmark Cross Sections at NNLO}},
  }{}\href{http://dx.doi.org/10.1103/PhysRevD.86.054009}{Phys.\ Rev.\  {\bf
  D86} (2012) 054009}, \href{http://arxiv.org/abs/1202.2281}{{\tt
  arXiv:1202.2281}}\relax
\mciteBstWouldAddEndPuncttrue
\mciteSetBstMidEndSepPunct{\mcitedefaultmidpunct}
{\mcitedefaultendpunct}{\mcitedefaultseppunct}\relax
\EndOfBibitem
\bibitem{CooperSarkar:2011aa}
ZEUS Collaboration, H1 Collaboration, A.~Cooper-Sarkar,
  \ifthenelse{\boolean{articletitles}}{{\it {PDF Fits at HERA}}, }{}PoS {\bf
  EPS-HEP2011} (2011) 320, \href{http://arxiv.org/abs/1112.2107}{{\tt
  arXiv:1112.2107}}\relax
\mciteBstWouldAddEndPuncttrue
\mciteSetBstMidEndSepPunct{\mcitedefaultmidpunct}
{\mcitedefaultendpunct}{\mcitedefaultseppunct}\relax
\EndOfBibitem
\bibitem{Martin:2009iq}
A.~Martin, W.~Stirling, R.~Thorne, and G.~Watt,
  \ifthenelse{\boolean{articletitles}}{{\it {Parton distributions for the
  LHC}}, }{}\href{http://dx.doi.org/10.1140/epjc/s10052-009-1072-5}{Eur.\
  Phys.\ J.\  {\bf C63} (2009) 189}, \href{http://arxiv.org/abs/0901.0002}{{\tt
  arXiv:0901.0002}}\relax
\mciteBstWouldAddEndPuncttrue
\mciteSetBstMidEndSepPunct{\mcitedefaultmidpunct}
{\mcitedefaultendpunct}{\mcitedefaultseppunct}\relax
\EndOfBibitem
\bibitem{Ball:2011mu}
R.~D. Ball {\em et~al.}, \ifthenelse{\boolean{articletitles}}{{\it {Impact of
  Heavy Quark Masses on Parton Distributions and LHC Phenomenology}},
  }{}\href{http://dx.doi.org/10.1016/j.nuclphysb.2011.03.021}{Nucl.\ Phys.\
  {\bf B849} (2011) 296}, \href{http://arxiv.org/abs/1101.1300}{{\tt
  arXiv:1101.1300}}\relax
\mciteBstWouldAddEndPuncttrue
\mciteSetBstMidEndSepPunct{\mcitedefaultmidpunct}
{\mcitedefaultendpunct}{\mcitedefaultseppunct}\relax
\EndOfBibitem
\bibitem{Whalley:2005nh}
M.~Whalley, D.~Bourilkov, and R.~Group,
  \ifthenelse{\boolean{articletitles}}{{\it {The Les Houches accord PDFs
  (LHAPDF) and LHAGLUE}}, }{}\href{http://arxiv.org/abs/hep-ph/0508110}{{\tt
  arXiv:hep-ph/0508110}}\relax
\mciteBstWouldAddEndPuncttrue
\mciteSetBstMidEndSepPunct{\mcitedefaultmidpunct}
{\mcitedefaultendpunct}{\mcitedefaultseppunct}\relax
\EndOfBibitem
\bibitem{Dittmaier:2011ti}
LHC Higgs Cross Section Working Group, S.~Dittmaier {\em et~al.},
  \ifthenelse{\boolean{articletitles}}{{\it {Handbook of LHC Higgs Cross
  Sections: 1. Inclusive Observables}},
  }{}\href{http://arxiv.org/abs/1101.0593}{{\tt arXiv:1101.0593}}\relax
\mciteBstWouldAddEndPuncttrue
\mciteSetBstMidEndSepPunct{\mcitedefaultmidpunct}
{\mcitedefaultendpunct}{\mcitedefaultseppunct}\relax
\EndOfBibitem
\bibitem{Ball:2012wy}
R.~D. Ball {\em et~al.}, \ifthenelse{\boolean{articletitles}}{{\it {Parton
  Distribution Benchmarking with LHC Data}},
  }{}\href{http://dx.doi.org/10.1007/JHEP04(2013)125}{JHEP {\bf 1304} (2013)
  125}, \href{http://arxiv.org/abs/1211.5142}{{\tt arXiv:1211.5142}}\relax
\mciteBstWouldAddEndPuncttrue
\mciteSetBstMidEndSepPunct{\mcitedefaultmidpunct}
{\mcitedefaultendpunct}{\mcitedefaultseppunct}\relax
\EndOfBibitem
\bibitem{Ball:2010gb}
NNPDF Collaboration, R.~D. Ball {\em et~al.},
  \ifthenelse{\boolean{articletitles}}{{\it {Reweighting NNPDFs: the W lepton
  asymmetry}}, }{}\href{http://dx.doi.org/10.1016/j.nuclphysb.2011.03.017,
  10.1016/j.nuclphysb.2011.10.024, 10.1016/j.nuclphysb.2011.09.011}{Nucl.\
  Phys.\  {\bf B849} (2011) 112}, \href{http://arxiv.org/abs/1012.0836}{{\tt
  arXiv:1012.0836}}\relax
\mciteBstWouldAddEndPuncttrue
\mciteSetBstMidEndSepPunct{\mcitedefaultmidpunct}
{\mcitedefaultendpunct}{\mcitedefaultseppunct}\relax
\EndOfBibitem
\bibitem{DeLorenzi:2010zt}
F.~De~Lorenzi, \ifthenelse{\boolean{articletitles}}{{\it {Parton Distribution
  Function sensitivity studies using electroweak processes at LHCb}},
  }{}\href{http://arxiv.org/abs/1011.4260}{{\tt arXiv:1011.4260}}\relax
\mciteBstWouldAddEndPuncttrue
\mciteSetBstMidEndSepPunct{\mcitedefaultmidpunct}
{\mcitedefaultendpunct}{\mcitedefaultseppunct}\relax
\EndOfBibitem
\bibitem{Ball:2011gg}
R.~D. Ball {\em et~al.}, \ifthenelse{\boolean{articletitles}}{{\it {Reweighting
  and Unweighting of Parton Distributions and the LHC W lepton asymmetry
  data}}, }{}\href{http://dx.doi.org/10.1016/j.nuclphysb.2011.10.018}{Nucl.\
  Phys.\  {\bf B855} (2012) 608}, \href{http://arxiv.org/abs/1108.1758}{{\tt
  arXiv:1108.1758}}\relax
\mciteBstWouldAddEndPuncttrue
\mciteSetBstMidEndSepPunct{\mcitedefaultmidpunct}
{\mcitedefaultendpunct}{\mcitedefaultseppunct}\relax
\EndOfBibitem
\bibitem{Watt:2012tq}
G.~Watt and R.~Thorne, \ifthenelse{\boolean{articletitles}}{{\it {Study of
  Monte Carlo approach to experimental uncertainty propagation with MSTW 2008
  PDFs}}, }{}\href{http://dx.doi.org/10.1007/JHEP08(2012)052}{JHEP {\bf 1208}
  (2012) 052}, \href{http://arxiv.org/abs/1205.4024}{{\tt
  arXiv:1205.4024}}\relax
\mciteBstWouldAddEndPuncttrue
\mciteSetBstMidEndSepPunct{\mcitedefaultmidpunct}
{\mcitedefaultendpunct}{\mcitedefaultseppunct}\relax
\EndOfBibitem
\bibitem{Aaltonen:2012it}
CDF Collaboration, T.~Aaltonen {\em et~al.},
  \ifthenelse{\boolean{articletitles}}{{\it {Measurement of the top quark
  forward-backward production asymmetry and its dependence on event kinematic
  properties}}, }{}\href{http://dx.doi.org/10.1103/PhysRevD.87.092002}{Phys.\
  Rev.\  {\bf D87} (2013) 092002}, \href{http://arxiv.org/abs/1211.1003}{{\tt
  arXiv:1211.1003}}\relax
\mciteBstWouldAddEndPuncttrue
\mciteSetBstMidEndSepPunct{\mcitedefaultmidpunct}
{\mcitedefaultendpunct}{\mcitedefaultseppunct}\relax
\EndOfBibitem
\bibitem{Aaltonen:2013vaf}
CDF Collaboration, T.~A. Aaltonen {\em et~al.},
  \ifthenelse{\boolean{articletitles}}{{\it {Measurement of the leptonic
  asymmetry in ttbar events produced in ppbar collisions at sqrt(s)=1.96 TeV}},
  }{}\href{http://dx.doi.org/10.1103/PhysRevD.88.072003}{Phys.\ Rev.\ D 88,
  {\bf 072003} (2013) }, \href{http://arxiv.org/abs/1308.1120}{{\tt
  arXiv:1308.1120}}\relax
\mciteBstWouldAddEndPuncttrue
\mciteSetBstMidEndSepPunct{\mcitedefaultmidpunct}
{\mcitedefaultendpunct}{\mcitedefaultseppunct}\relax
\EndOfBibitem
\bibitem{Abazov:2013wxa}
D0 Collaboration, V.~M. Abazov {\em et~al.},
  \ifthenelse{\boolean{articletitles}}{{\it {Measurement of the asymmetry in
  angular distributions of leptons produced in dilepton ttbar final states in
  $p\bar p$ collisions at $\sqrt{s}=1.96$ TeV}},
  }{}\href{http://arxiv.org/abs/1308.6690}{{\tt arXiv:1308.6690}}\relax
\mciteBstWouldAddEndPuncttrue
\mciteSetBstMidEndSepPunct{\mcitedefaultmidpunct}
{\mcitedefaultendpunct}{\mcitedefaultseppunct}\relax
\EndOfBibitem
\bibitem{d0:preliminary}
D0 Collaboration, V.~M. Abazov {\em et~al.},
  \ifthenelse{\boolean{articletitles}}{{\it {Forward-backward asymmetry in the
  distribution of leptons in $t\bar{t}$ events}}, }{}
  http://www-d0.fnal.gov/Run2Physics/WWW/results/prelim/TOP/T98/T98.pdf,
  2013\relax
\mciteBstWouldAddEndPuncttrue
\mciteSetBstMidEndSepPunct{\mcitedefaultmidpunct}
{\mcitedefaultendpunct}{\mcitedefaultseppunct}\relax
\EndOfBibitem
\bibitem{ATLAS-CONF-2013-078}
\ifthenelse{\boolean{articletitles}}{{\it {Measurement of the top quark pair
  production charge asymmetry in proton-proton collisions at $sqrt{s}=7$ TeV
  using the ATLAS detector}}, }{} Tech. Rep. ATLAS-CONF-2013-078, CERN, Geneva,
  Jul, 2013\relax
\mciteBstWouldAddEndPuncttrue
\mciteSetBstMidEndSepPunct{\mcitedefaultmidpunct}
{\mcitedefaultendpunct}{\mcitedefaultseppunct}\relax
\EndOfBibitem
\bibitem{Chatrchyan:2012cxa}
CMS Collaboration, S.~Chatrchyan {\em et~al.},
  \ifthenelse{\boolean{articletitles}}{{\it {Inclusive and differential
  measurements of the $t \bar{t}$ charge asymmetry in proton-proton collisions
  at 7 TeV}}, }{}\href{http://dx.doi.org/10.1016/j.physletb.2012.09.028}{Phys.\
  Lett.\  {\bf B717} (2012) 129}, \href{http://arxiv.org/abs/1207.0065}{{\tt
  arXiv:1207.0065}}\relax
\mciteBstWouldAddEndPuncttrue
\mciteSetBstMidEndSepPunct{\mcitedefaultmidpunct}
{\mcitedefaultendpunct}{\mcitedefaultseppunct}\relax
\EndOfBibitem
\bibitem{CMS-PAS-TOP-12-033}
CMS Collaboration, \ifthenelse{\boolean{articletitles}}{{\it {Measurement of
  the ttbar charge asymmetry with lepton+jets events at 8 TeV}}, }{} Tech. Rep.
  CMS-PAS-TOP-12-033, CERN, Geneva, 2013\relax
\mciteBstWouldAddEndPuncttrue
\mciteSetBstMidEndSepPunct{\mcitedefaultmidpunct}
{\mcitedefaultendpunct}{\mcitedefaultseppunct}\relax
\EndOfBibitem
\bibitem{ATLAS-CONF-2013-097}
\ifthenelse{\boolean{articletitles}}{{\it {Measurement of the $t\bar{t}$
  production cross-section in $pp$ collisions at $\sqrt{s}=8$ TeV using $e\mu$
  events with $b$-tagged jets}}, }{} Tech. Rep. ATLAS-CONF-2013-097, CERN,
  Geneva, Sep, 2013\relax
\mciteBstWouldAddEndPuncttrue
\mciteSetBstMidEndSepPunct{\mcitedefaultmidpunct}
{\mcitedefaultendpunct}{\mcitedefaultseppunct}\relax
\EndOfBibitem
\bibitem{Aaij:2011er}
LHCb Collaboration, R.~Aaij {\em et~al.},
  \ifthenelse{\boolean{articletitles}}{{\it {Absolute luminosity measurements
  with the LHCb detector at the LHC}},
  }{}\href{http://dx.doi.org/10.1088/1748-0221/7/01/P01010}{JINST {\bf 7}
  (2012) P01010}, \href{http://arxiv.org/abs/1110.2866}{{\tt
  arXiv:1110.2866}}\relax
\mciteBstWouldAddEndPuncttrue
\mciteSetBstMidEndSepPunct{\mcitedefaultmidpunct}
{\mcitedefaultendpunct}{\mcitedefaultseppunct}\relax
\EndOfBibitem
\bibitem{Cheng:2005as}
H.-C. Cheng, I.~Low, and L.-T. Wang, \ifthenelse{\boolean{articletitles}}{{\it
  {Top partners in little Higgs theories with T-parity}},
  }{}\href{http://dx.doi.org/10.1103/PhysRevD.74.055001}{Phys.\ Rev.\  {\bf
  D74} (2006) 055001}, \href{http://arxiv.org/abs/hep-ph/0510225}{{\tt
  arXiv:hep-ph/0510225}}\relax
\mciteBstWouldAddEndPuncttrue
\mciteSetBstMidEndSepPunct{\mcitedefaultmidpunct}
{\mcitedefaultendpunct}{\mcitedefaultseppunct}\relax
\EndOfBibitem
\bibitem{Agashe:2006hk}
K.~Agashe {\em et~al.}, \ifthenelse{\boolean{articletitles}}{{\it {LHC Signals
  from Warped Extra Dimensions}},
  }{}\href{http://dx.doi.org/10.1103/PhysRevD.77.015003}{Phys.\ Rev.\  {\bf
  D77} (2008) 015003}, \href{http://arxiv.org/abs/hep-ph/0612015}{{\tt
  arXiv:hep-ph/0612015}}\relax
\mciteBstWouldAddEndPuncttrue
\mciteSetBstMidEndSepPunct{\mcitedefaultmidpunct}
{\mcitedefaultendpunct}{\mcitedefaultseppunct}\relax
\EndOfBibitem
\bibitem{Matsumoto:2006ws}
S.~Matsumoto, M.~M. Nojiri, and D.~Nomura,
  \ifthenelse{\boolean{articletitles}}{{\it {Hunting for the Top Partner in the
  Littlest Higgs Model with T-parity at the CERN LHC}},
  }{}\href{http://dx.doi.org/10.1103/PhysRevD.75.055006}{Phys.\ Rev.\  {\bf
  D75} (2007) 055006}, \href{http://arxiv.org/abs/hep-ph/0612249}{{\tt
  arXiv:hep-ph/0612249}}\relax
\mciteBstWouldAddEndPuncttrue
\mciteSetBstMidEndSepPunct{\mcitedefaultmidpunct}
{\mcitedefaultendpunct}{\mcitedefaultseppunct}\relax
\EndOfBibitem
\bibitem{Belyaev:2006jh}
A.~Belyaev, C.-R. Chen, K.~Tobe, and C.-P. Yuan,
  \ifthenelse{\boolean{articletitles}}{{\it {Phenomenology of littlest Higgs
  model with $T^-$ parity: including effects of $T^-$ odd fermions}},
  }{}\href{http://dx.doi.org/10.1103/PhysRevD.74.115020}{Phys.\ Rev.\  {\bf
  D74} (2006) 115020}, \href{http://arxiv.org/abs/hep-ph/0609179}{{\tt
  arXiv:hep-ph/0609179}}\relax
\mciteBstWouldAddEndPuncttrue
\mciteSetBstMidEndSepPunct{\mcitedefaultmidpunct}
{\mcitedefaultendpunct}{\mcitedefaultseppunct}\relax
\EndOfBibitem
\bibitem{Freitas:2006vy}
A.~Freitas and D.~Wyler, \ifthenelse{\boolean{articletitles}}{{\it
  {Phenomenology of mirror fermions in the littlest Higgs model with
  T-parity}}, }{}\href{http://dx.doi.org/10.1088/1126-6708/2006/11/061}{JHEP
  {\bf 0611} (2006) 061}, \href{http://arxiv.org/abs/hep-ph/0609103}{{\tt
  arXiv:hep-ph/0609103}}\relax
\mciteBstWouldAddEndPuncttrue
\mciteSetBstMidEndSepPunct{\mcitedefaultmidpunct}
{\mcitedefaultendpunct}{\mcitedefaultseppunct}\relax
\EndOfBibitem
\bibitem{Meade:2006dw}
P.~Meade and M.~Reece, \ifthenelse{\boolean{articletitles}}{{\it {Top partners
  at the LHC: Spin and mass measurement}},
  }{}\href{http://dx.doi.org/10.1103/PhysRevD.74.015010}{Phys.\ Rev.\  {\bf
  D74} (2006) 015010}, \href{http://arxiv.org/abs/hep-ph/0601124}{{\tt
  arXiv:hep-ph/0601124}}\relax
\mciteBstWouldAddEndPuncttrue
\mciteSetBstMidEndSepPunct{\mcitedefaultmidpunct}
{\mcitedefaultendpunct}{\mcitedefaultseppunct}\relax
\EndOfBibitem
\bibitem{Thaler:2008ju}
J.~Thaler and L.-T. Wang, \ifthenelse{\boolean{articletitles}}{{\it {Strategies
  to Identify Boosted Tops}},
  }{}\href{http://dx.doi.org/10.1088/1126-6708/2008/07/092}{JHEP {\bf 0807}
  (2008) 092}, \href{http://arxiv.org/abs/0806.0023}{{\tt
  arXiv:0806.0023}}\relax
\mciteBstWouldAddEndPuncttrue
\mciteSetBstMidEndSepPunct{\mcitedefaultmidpunct}
{\mcitedefaultendpunct}{\mcitedefaultseppunct}\relax
\EndOfBibitem
\bibitem{Nojiri:2008ir}
M.~M. Nojiri and M.~Takeuchi, \ifthenelse{\boolean{articletitles}}{{\it {Study
  of the top reconstruction in top-partner events at the LHC}},
  }{}\href{http://dx.doi.org/10.1088/1126-6708/2008/10/025}{JHEP {\bf 0810}
  (2008) 025}, \href{http://arxiv.org/abs/0802.4142}{{\tt
  arXiv:0802.4142}}\relax
\mciteBstWouldAddEndPuncttrue
\mciteSetBstMidEndSepPunct{\mcitedefaultmidpunct}
{\mcitedefaultendpunct}{\mcitedefaultseppunct}\relax
\EndOfBibitem
\end{mcitethebibliography}

\end{document}